\documentclass[onecolumn ]{aastex631}

\shorttitle{Ten extremely low mass ratio contact binaries}
\shortauthors{Li et al.}

\graphicspath{{./}{figures/}}

\begin{document}

\title{Extremely low mass ratio contact binaries - I. the first photometric and spectroscopic investigations of ten systems}

\author{Kai Li}
\correspondingauthor{Kai Li}
\email{kaili@sdu.edu.cn}
\author{Xiang Gao}
\author{Xin-Yi Liu}
\affiliation{Shandong Key Laboratory of Optical Astronomy and Solar-Terrestrial Environment, School of Space Science and Physics, Institute of Space Sciences, Shandong University, Weihai, Shandong, 264209, China}

\author{Xing Gao}
\affiliation{Xinjiang Astronomical Observatory, 150 Science 1-Street, Urumqi 830011, China}
\author{Ling-Zhi Li}
\author{Xu Chen}
\affiliation{Shandong Key Laboratory of Optical Astronomy and Solar-Terrestrial Environment, School of Space Science and Physics, Institute of Space Sciences, Shandong University, Weihai, Shandong, 264209, China}
\author{Guo-You Sun}
\affiliation{Wenzhou Astronomical Association}

\begin{abstract}
The photometric and spectroscopic investigations of ten contact binaries were presented for the first time. It is discovered that the mass ratios of all the ten targets are smaller than 0.15, they are extremely low mass ratio contact binaries. Seven of them are deep contact binaries, two are medium contact binaries, while only one is a shallow contact system. Five of them show O'Connell effect, and a dark spot on one of the two components can lead to a good fit of the asymmetric light curves. The orbital period studies of the ten binaries reveal that they all exhibit long-term period changes, six of them are increasing, while the others are shrinking. The LAMOST spectra were analyzed by the spectral subtraction method, and all the ten targets exhibit excess emissions in the H$_\alpha$ line, indicating chromospheric activity. The evolutionary states of the two components of the ten binaries were studied, and it is found that their evolutionary states are identical to those of the other contact binaries. Based on the study of the relation between orbital angular momentum and total mass, we discovered the ten systems may be at the late evolutionary stage of a contact binary. The initial masses of the two components and the ages of them were obtained. By calculating the instability parameters, we found that the ten contact binaries are relatively stable at present.
\end{abstract}


\section{Introduction} \label{Sec:Intro}
A binary which is composed by two Roche lobe filling component stars is called a contact binary. The two components are usually late type stars and have very close surface effective temperatures. Contact binaries are so common in the Galaxy that one in every 500 stars may be a contact binary \citep{2007MNRAS.382..393R}. This type of binary is thought to be formed by short period detached binaries through angular  momentum loss, and will merge into rapidly rotating single star (e.g., \citealt{ 1988ASIC..241..345G,1994ASPC...56..228B,2002ApJ...575..461E,2003MNRAS.342.1260Q,2006Ap&SS.304...25Q,2017RAA....17...87Q,2005ApJ...629.1055Y,2006AcA....56..347S,2011AcA....61..139S}). Estimations by \cite{2014MNRAS.443.1319K} suggested that the stellar merging event happens roughly once every 10 yr in our Galaxy. However, only one confirmed contact binary merging event has been observed up to now, that is, V1309 Sco \citep{2011A&A...528A.114T}. Theoretical studies have proposed that contact binaries exhibit a low mass ratio cut-off and will merge into fast-rotating single star due to the Darwin instability (e.g., \citealt{1995ApJ...444L..41R,2006MNRAS.369.2001L,2007MNRAS.377.1635A,2009MNRAS.394..501A,2010MNRAS.405.2485J,2021MNRAS.501..229W}).
The research on contact binaries with extremely low mass ratios is very essential to our understanding of the merging process and the low mass ratio limit.

Extremely low mass ratio contact binaries (ELMRCBs) were defined as contact binaries with mass ratios less than 0.15. At present, only a few ELMRCBs were discovered, such as V1187 Her ($q\sim0.044$; \citealt{2019PASP..131e4203C}), VSX J082700.8+462850 ($q\sim0.055$; \citealt{2021ApJ...922..122L}), V857 Her ($q\sim0.065$; \citealt{2005AJ....130.1206Q}), ASAS J083241+2332.4 ($q\sim0.068$; \citealt{2016AJ....151...69S}). We carried out a project on the analysis of ELMRCBs. Statistical and simulation studies have indicated that the photometric mass ratios are consistent with the spectroscopic ones for totally eclipsing contact binaries \citep{2001AJ....122.1007R,2003CoSka..33...38P,2005Ap&SS.296..221T,2021AJ....162...13L}. \cite{2001AJ....122.1007R} determined the relationship between the maximum amplitude, the mass ratio, and the contact degree for contact binaries and found that the smaller the amplitude, the smaller the mass ratio. Therefore, a number of low amplitude totally eclipsing contact binaries were selected from the variable star catalog of All-Sky Automated Survey for SuperNovae (ASAS-SN; \citealt{2014ApJ...788...48S}; \citealt{2018MNRAS.477.3145J}) and have been observed since 2019. This paper presents the first photometric and spectroscopic investigation of ten ELMRCBs. The information of the ten systems is shown in Table \ref{tab:information}.

\begin{deluxetable*}{ccccccc}
\tablecaption{The information of the ten systems from ASAS-SN database\label{tab:information}}
\tablewidth{0pt}
\tablehead{
\colhead{ASAS-SN Name} & \colhead{Hereafter} & \colhead{Other Name} & \colhead{Period} & \colhead{HJD$_0$} & \colhead{V} & \colhead{Amplitude} \\
\colhead{} & \colhead{} & \colhead{} & \colhead{days} &\colhead{2450000+} & \colhead{mag} & \colhead{(mag)}\\
}
\startdata
ASASSN-V J001422.52+415331.1 & J001422 & NSVS 3650324               & 0.3792974 & 9133.40443$\pm$0.00029 & 12.03 & 0.39 \\
ASASSN-V J022733.60+360447.1 & J022733 & 1SWASP J022733.57+360447.5 & 0.413965 & 8842.24954$\pm$0.00028 & 12.01 & 0.30 \\
ASASSN-V J042640.22+370047.7 & J042640 & NSVS 6798913               & 0.3625408 & 8812.33649$\pm$0.00072 & 11.64 & 0.31 \\
ASASSN-V J054950.33+584208.9 & J054950 & V0556 Cam                  & 0.444062 & 9203.06578$\pm$0.00052 & 12.35 & 0.26 \\
ASASSN-V J071924.64+415705.4 & J071924 & NSVS 4701980               & 0.355763 & 8869.23882$\pm$0.00046 & 11.97 & 0.33 \\
ASASSN-V J093458.49+290543.0 & J093458 & NSVS 7480723               & 0.299294 & 8901.19512$\pm$0.00021 & 11.67 & 0.32 \\
ASASSN-V J110658.46+511200.1 & J110658 & CSS J110658.4+511201       & 0.400968 & 9280.02520$\pm$0.00046 & 12.37 & 0.33 \\
ASASSN-V J115742.23+074820.3 & J115742 & NSVS 10368868              & 0.2871599 & 8914.35747$\pm$0.00041 & 12.31 & 0.29 \\
ASASSN-V J153433.52+122516.7 & J153433 & ASAS J153433+1225.3        & 0.3300572 & 9339.28454$\pm$0.00041 & 12.33 & 0.26 \\
ASASSN-V J233332.90+180429.9 & J233332 & CSS J233332.9+180430       & 0.6287837 & 9482.33063$\pm$0.00168 & 14.84 & 0.30 \\
\enddata
\end{deluxetable*}

\section{Observations} \label{sec:observations}
\subsection{Photometic Observations\label{sec:Photometric}}
Four meter-class telescopes were applied to observe the ten ELMRCBs. The information of the four meter-class telescopes is as follows: (1) the 60cm Ningbo Bureau of Education and Xinjiang Observatory Telescope (NEXT), this telescope is equipped with an FLI PL23042 CCD, resulting a field of view of $22^{'}\times22^{'}$; (2) the Tsinghua-NAOC 80 cm Telescope (TNT; \citealt{2012RAA....12.1585H}), the PIXIS 1300B CCD was used for this telescope, providing a field of view of $11.4^{'}\times11.1^{'}$  (3) the 85 cm telescope at the Xinglong Station of National Astronomical Observatories (XL85), this telescope used an Andor DZ936 CCD, and its field of view is $32^{'}\times32^{'}$; (4) the Weihai Observatory 1.0-m telescope of Shandong University (WHOT; \citealt{2014RAA....14..719H}), the PIXIS 2048B CCD was equipped on this telescope, and the field of view is $12^{'}\times12^{'}$.
The observational details of the ten systems are listed in Table \ref{tab:observation}. All the CCD images were reduced with IRAF\footnote{IRAF (\url{http://iraf.noao.edu/}) is distributed by the National Optical Astronomy Observatories, which are operated by the Association of Universities for Research in Astronomy, Inc., under cooperative agreement with the National Science Foundation.}.
The reduction processes include the bias and flat corrections, and the aperture photometry for the target, the comparison and check stars. Then, the differential magnitudes between the target and the comparison star and the comparison and the check stars were obtained. Two sets of complete multi-color light curves were obtained for J071924 and J110658, while one set of complete multi-color light curves were derived for the others.

\begin{deluxetable*}{ccccccc}
\tablecaption{The observation log of the ten binaries\label{tab:observation}}
\tablewidth{0pt}
\tablehead{
\colhead{Star} & \colhead{UT Date} & \colhead{Exposures} & \colhead{Comparison$^a$} & \colhead{Check$^a$} & \colhead{Mean error$^b$} & \colhead{Telescope} \\
\colhead{} & \colhead{yyyymmdd} & \colhead{s} & \colhead{2MASS} &\colhead{2MASS} & \colhead{mag} & \colhead{}\\
	}
\startdata
J001422 & 2020 Oct 07 & V22 R16 I22 & 00141914+4153316 & 00142542+4153196 & V0.006 R0.006 I0.007 & NEXT \\
	    & 2020 Oct 10 & V22 R16 I22 & 00141914+4153316 & 00142542+4153196 & V0.005 R0.004 I0.005 & NEXT \\
	    & 2020 Oct 11 & V22 R16 I22 & 00141914+4153316 & 00142542+4153196 & V0.005 R0.005 I0.005 & NEXT \\
     	& 2020 Oct 16 & V22 R16 I22 & 00141914+4153316 & 00142542+4153196 & V0.006 R0.007 I0.006 & NEXT \\
J022733 & 2019 Dec 16 & B30 V18 R14 I17 & 02274280+3605074 & 02272128+3605376 & B0.010 V0.007 R0.009 I0.010 & NEXT \\
	    & 2019 Dec 18 & B40 V23 R18 I22 & 02274280+3605074 & 02272128+3605376 & B0.007 V0.005 R0.007 I0.007 & NEXT \\
	    & 2019 Dec 24 & B40 V23 R18 I22 & 02274280+3605074 & 02272128+3605376 & B0.007 V0.006 R0.005 I0.007 & NEXT \\
        & 2019 Dec 25 & B40 V23 R18 I22 & 02274280+3605074 & 02272128+3605376 & B0.005 V0.005 R0.004 I0.005 & NEXT \\
J042640 & 2019 Nov 20 & B40 V20 R12 I12 & 04261396+3656554 & 04255488+3658220 & B0.010 V0.009 R0.008 I0.008 & XL85 \\
	    & 2019 Nov 24 & B20 V13 R9 I12 & 04261396+3656554 & 04255488+3658220 & B0.011 V0.007 R0.007 I0.007 & NEXT \\
J054950 & 2020 Dec 19 & V25 R18 I25 & 05493600+5843185 & 05491452+5841295 & V0.004 R0.004 I0.004 & NEXT \\
	    & 2020 Dec 20 & V25 R18 I25 & 05493600+5843185 & 05491452+5841295 & V0.008 R0.006 I0.005 & NEXT \\
	    & 2020 Dec 21 & V25 R18 I25 & 05493600+5843185 & 05491452+5841295 & V0.004 R0.004 I0.004 & NEXT \\
J071924 & 2020 Jan 20 & B40 V25 R18 I18 & 07194801+4153253 & 07194782+4153136 & B0.006 V0.006 R0.005 I0.005 & TNT \\
	    & 2020 Feb 17 & B40 V25 R18 I25 & 07194801+4153253 & 07194782+4153136 & B0.009 V0.005 R0.006 I0.008 & NEXT \\
	    & 2020 Feb 20 & B40 V25 R18 I25 & 07194801+4153253 & 07194782+4153136 & B0.007 V0.006 R0.006 I0.007 & NEXT \\
J093458 & 2020 Feb 21 & B30 V18 R13 I19 & 09345181+2859250 & 09345877+2910184 & B0.009 V0.008 R0.008 I0.011 & NEXT \\
	    & 2020 Feb 22 & B30 V18 R13 I19 & 09345181+2859250 & 09345877+2910184 & B0.013 V0.009 R0.006 I0.006 & NEXT \\
	    & 2020 Feb 24 & B30 V18 R13 I19 & 09345181+2859250 & 09345877+2910184 & B0.006 V0.008 R0.005 I0.005 & NEXT \\
J110658 & 2020 Mar 15 & B60 V40 R30 I40 & 11064921+5113466 & 11065838+5107046 & B0.012 V0.009 R0.005 I0.008 & NEXT \\
	    & 2020 Mar 16 & B60 V40 R30 I40 & 11064921+5113466 & 11065838+5107046 & B0.006 V0.005 R0.005 I0.006 & NEXT \\
	    & 2021 Mar 06 & B70 V40 R30 I30 & 11064921+5113466 & 11065838+5107046 & B0.006 V0.004 R0.005 I0.007 & XL85 \\
J115742 & 2020 Feb 26 & B80 V40 R25 I20 & 11572845+0754048 & 11574055+0746086 & B0.013 V0.016 R0.015 I0.021 & WHOT \\
	    & 2020 Mar 03 & B60 V40 R30 I40 & 11572845+0754048 & 11574055+0746086 & B0.007 V0.008 R0.010 I0.009 & NEXT \\
	    & 2020 Mar 05 & B60 V40 R30 I40 & 11572845+0754048 & 11574055+0746086 & B0.011 V0.010 R0.012 I0.013 & NEXT \\
J153433 & 2020 May 26 & V33 R20 I30 & 15343953+1224491 & 15344196+1225290 & V0.007 R0.007 I0.007 & NEXT \\
	    & 2021 May 02 & g23 r20 i25 & 15343953+1224491 & 15344196+1225290 & g0.005 r0.005 i0.006 & NEXT \\
	    & 2021 May 04 & g23 r20 i25 & 15343953+1224491 & 15344196+1225290 & g0.006 r0.005 i0.006 & NEXT \\
	    & 2021 May 06 & g23 r20 i25 & 15343953+1224491 & 15344196+1225290 & g0.006 r0.005 i0.006 & NEXT \\
	    & 2021 May 13 & g23 r20 i25 & 15343953+1224491 & 15344196+1225290 & g0.007 r0.006 i0.007 & NEXT \\
J233332 & 2021 Sep 07 & g110 r95 i130 & 23331817+1803150 & 23334886+1806115 & g0.007 r0.006 i0.010 & NEXT \\
	    & 2021 Sep 08 & g110 r95 i130 & 23331817+1803150 & 23334886+1806115 & g0.007 r0.008 i0.008 & NEXT \\
	    & 2021 Sep 17 & g110 r95 i130 & 23331817+1803150 & 23334886+1806115 & g0.008 r0.007 i0.007 & NEXT \\
	    & 2021 Sep 24 & g110 r95 i130 & 23331817+1803150 & 23334886+1806115 & g0.010 r0.009 i0.013 & NEXT \\
	    & 2021 Sep 26 & g110 r95 i130 & 23331817+1803150 & 23334886+1806115 & g0.008 r0.009 i0.006 & NEXT \\
\enddata
\tablecomments{$^a$ The names of the comparison and check stars were determined from 2MASS catalog \citep{2003yCat.2246....0C}.\\
$^b$ Mean errors were estimated by the standard deviation of the magnitudes between the comparison and check stars. }
\end{deluxetable*}

\subsection{Spectroscopic Observations\label{sec:Spectroscopic}}
All the ten systems have been observed as part of the The Large sky Area Multi-Object Fibre Spectroscopic Telescope (LAMOST) low-resolution spectroscopic survey \citep{2012RAA....12.1197C,2015RAA....15.1095L}. LAMOST is a 4-m reflecting Schmidt telescope with a 5$^\circ$ field of view. It has 4000 fibers mounted on its focal plane to make a very high spectral acquisition rate. The spectral resolution of the low-resolution spectroscopic survey is $R\sim1800$ covering the wavelength range of about $3700\sim9000$ {\AA}. Table \ref{tab:LAMOST} displays an overview of the observations and the atmospheric stellar parameters of our ten targets. The atmospheric surface parameters were derived from LAMOST spectra by the spectrum fitting based on the LAMOST Stellar Parameter pipeline (LASP, \citealt{2011RAA....11..924W,2014IAUS..306..340W,2015RAA....15.1095L}). Although LASP used the single-star model, the difference between the determined temperature and the brighter primary component's temperature is almost less than 200K, no matter what the mass ratio of the binary is \citep{2019ApJS..244...43Z}. he systematic error of temperatures derived for single stars with LASP is estimated to be 81 K for a spectrum with a signal noise ratio in the g band (SNR$_g$) of 10 \citep{2020ApJS..251...27W}. And an additional error on LASP temperatures of primary stars of the binary systems resulting from the procedure used may be up to 200 K.

\begin{deluxetable*}{cccccccccc}
\tablecaption{The stellar parameters of the ten binaries obtaned by LAMOST (DR8)\label{tab:LAMOST}}
\tablewidth{0pt}
\tablehead{
\colhead{Star} &\colhead{UT Date}&\colhead{HJD}&\colhead{Exposure}&\colhead{SNR} &\colhead{Sp} &\colhead{T$_{eff}^*$} &\colhead{log g} &\colhead{[Fe/H]}& \colhead{H$_\alpha$}\\
& & \colhead{2450000+}& \colhead{s}  & \colhead{g} &  & \colhead{K} & \colhead{dex} & \colhead{dex}& \colhead{${\AA}$}\\
}
\startdata
J001422 & 2013 Nov 14 & 6611.0283  &1800 & 279 & F0 & 6568($\pm$9)   & 4.125($\pm$0.013) & $-$0.213($\pm$0.007)& 0.294($\pm$0.006) \\
J022733 & 2017 Jan 06 & 7759.9430  &1800 & 174 & F5 & 6477($\pm$9)   & 4.250($\pm$0.015) & $-$0.098($\pm$0.008)& 0.252($\pm$0.003) \\
J042640 & 2013 Feb 02 & 6325.9977  &1200 & 2   & $-$& $-$            &  $-$              & $-$                 &  $-$              \\
	    & 2013 Oct 04 & 6570.3390  &1800 & 55  & F7 & 5936($\pm$41)  & 3.838($\pm$0.067) & $-$0.361($\pm$0.039)& 0.285($\pm$0.021) \\
	    & 2013 Oct 04 & 6570.3778  &1800 & 89  & F6 & 6054($\pm$18)  & 4.094($\pm$0.030) & $-$0.289($\pm$0.017)& 0.223($\pm$0.011) \\
        & 2015 Jan 12 & 7035.1000  &1800 & 148 & F6 & 6115($\pm$14)  & 4.025($\pm$0.023) & $-$0.228($\pm$0.012)& 0.368($\pm$0.037) \\
J054950 & 2013 Feb 18 & 6342.0501  &1200 & 71  & F0 & 6988($\pm$30)  & 4.093($\pm$0.049) & $-$0.212($\pm$0.014)& 0.239($\pm$0.019) \\
	    & 2013 Dec 27 & 6654.1851  &1800 & 131 & F0 & 6987($\pm$15)  & 4.143($\pm$0.025) & $-$0.223($\pm$0.028)& 0.223($\pm$0.009) \\
J071924 & 2017 Oct 13 & 8040.3942  &600  & 87  & G2 & 5916($\pm$31)  & 4.006($\pm$0.051) &    0.040($\pm$0.029)& 0.433($\pm$0.016) \\
J093458 & 2011 Dec 11 & 5907.3259  &1800 & 90  & F2 & 5789($\pm$18)  & 4.033($\pm$0.030) & $-$0.397($\pm$0.017)& 0.558($\pm$0.021) \\
	    & 2011 Dec 26 & 5922.2895  &2700 & 133 & G2 & 5742($\pm$17)  & 4.211($\pm$0.028) & $-$0.356($\pm$0.016)& 0.626($\pm$0.026) \\
	    & 2016 Mar 17 & 7465.0494  &1800 & 39  & G2 & 5886($\pm$51)  & 4.138($\pm$0.084) & $-$0.310($\pm$0.049)& 0.545($\pm$0.029) \\
	    & 2017 Mar 17 & 7830.1195  &1800 & 65  & F9 & 5863($\pm$37)  & 4.168($\pm$0.061) & $-$0.283($\pm$0.035)& 0.465($\pm$0.035) \\
J110658 & 2013 Dec 21 & 7387.3738  &1800 & 203 & F5 & 6252($\pm$12)  & 4.097($\pm$0.018) & $-$0.141($\pm$0.010)& 0.372($\pm$0.022) \\
	    & 2015 Dec 30 & 6648.3706  &1800 & 233 & F6 & 6255($\pm$11)  & 4.120($\pm$0.017) & $-$0.121($\pm$0.009)& 0.323($\pm$0.030) \\
	    & 2016 Jan 28 & 7416.3105  &1800 & 121 & F7 & 6195($\pm$20)  & 4.077($\pm$0.034) & $-$0.119($\pm$0.019)& 0.200($\pm$0.018) \\
J115742 & 2014 May 22 & 6800.0747  &1200 & 56  & F5 & 5997($\pm$47)  & 4.196($\pm$0.077) & $-$0.793($\pm$0.045)& 0.202($\pm$0.013) \\
J153433 & 2015 Feb 13 & 7067.4228  &1800 & 14  & F6 & 6077($\pm$366) & 4.085($\pm$0.575) & $-$0.483($\pm$0.344)& 0.638($\pm$0.065) \\
J233332 & 2014 Oct 13 & 6944.1204  &4500 & 73  & F0 & 6728($\pm$27)  & 4.177($\pm$0.045) & $-$0.365($\pm$0.026)& 0.301($\pm$0.026) \\
\enddata
\tablecomments{$^*$ An additional error on LASP temperatures of primary stars of the binary systems resulting from the procedure used may be up to 200 K.}
\end{deluxetable*}

\section{Light-curve investigation} \label{sec:LC}
The new observed multiple color light curves of the ten contact binaries were investigated by the 2013 version of the W-D code \citep{1971ApJ...166..605W,1979ApJ...234.1054W,1990ApJ...356..613W}. At first, we have to derive the effective temperature of the primary component for each system. The temperature obtained by LAMOST was assumed to be the primary effective temperature during the following study (the average value was used when a target was observed more than one time). Based on the primary effective temperature, the gravity darkening and bolometric albedo coefficients can be fixed at $g_1=g_2=0.32$ and $A_1=A_2=0.5$ \citep{1924MNRAS..84..665V}. The limb-darkening coefficients were interpolated from \cite{1993AJ....106.2096V}'s table due to a square root law (ld=-3).

\begin{figure}
\epsscale{0.48}
\vspace{-5cm}
\plotone{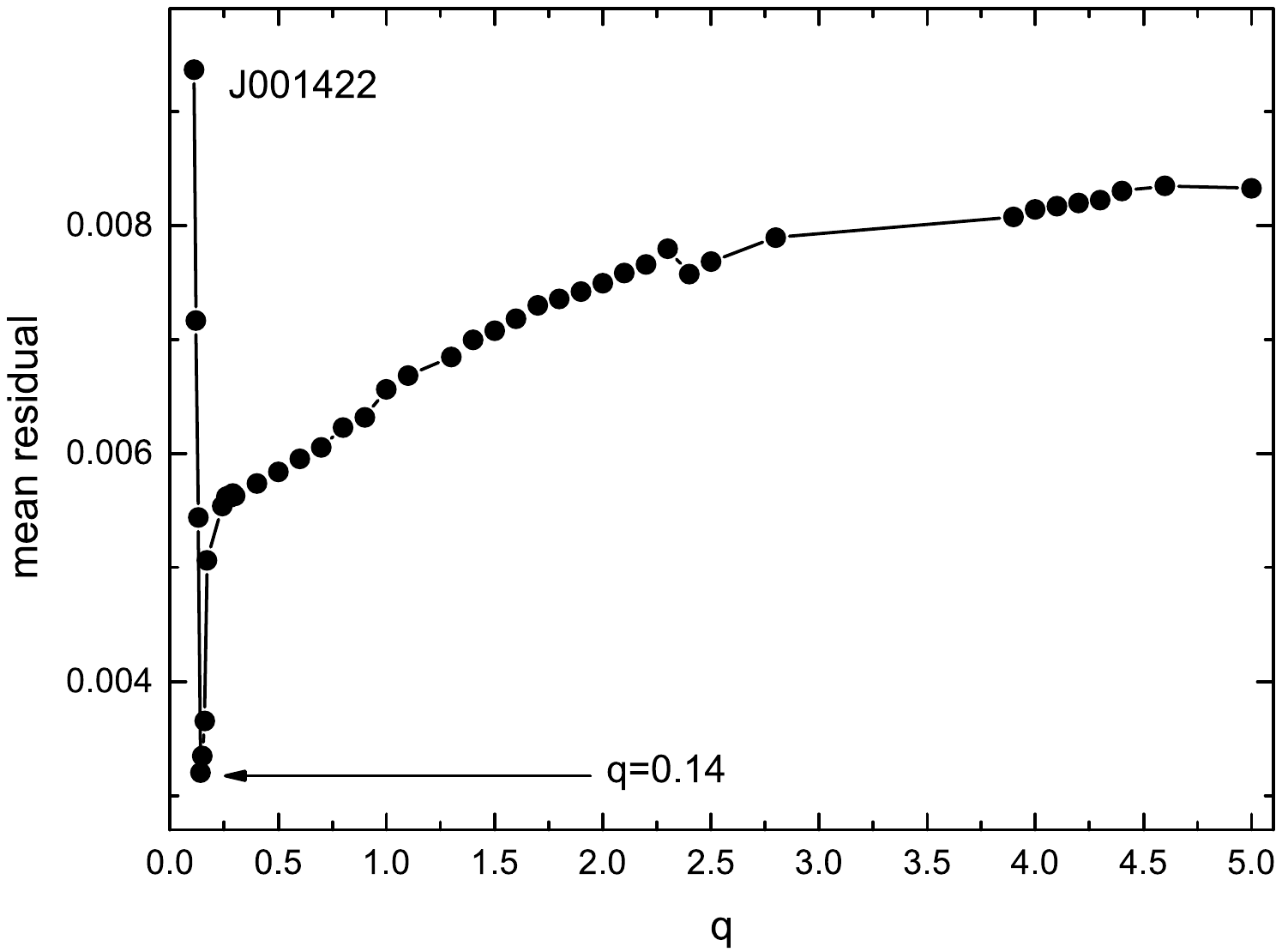}\hspace{-2.3cm}
\plotone{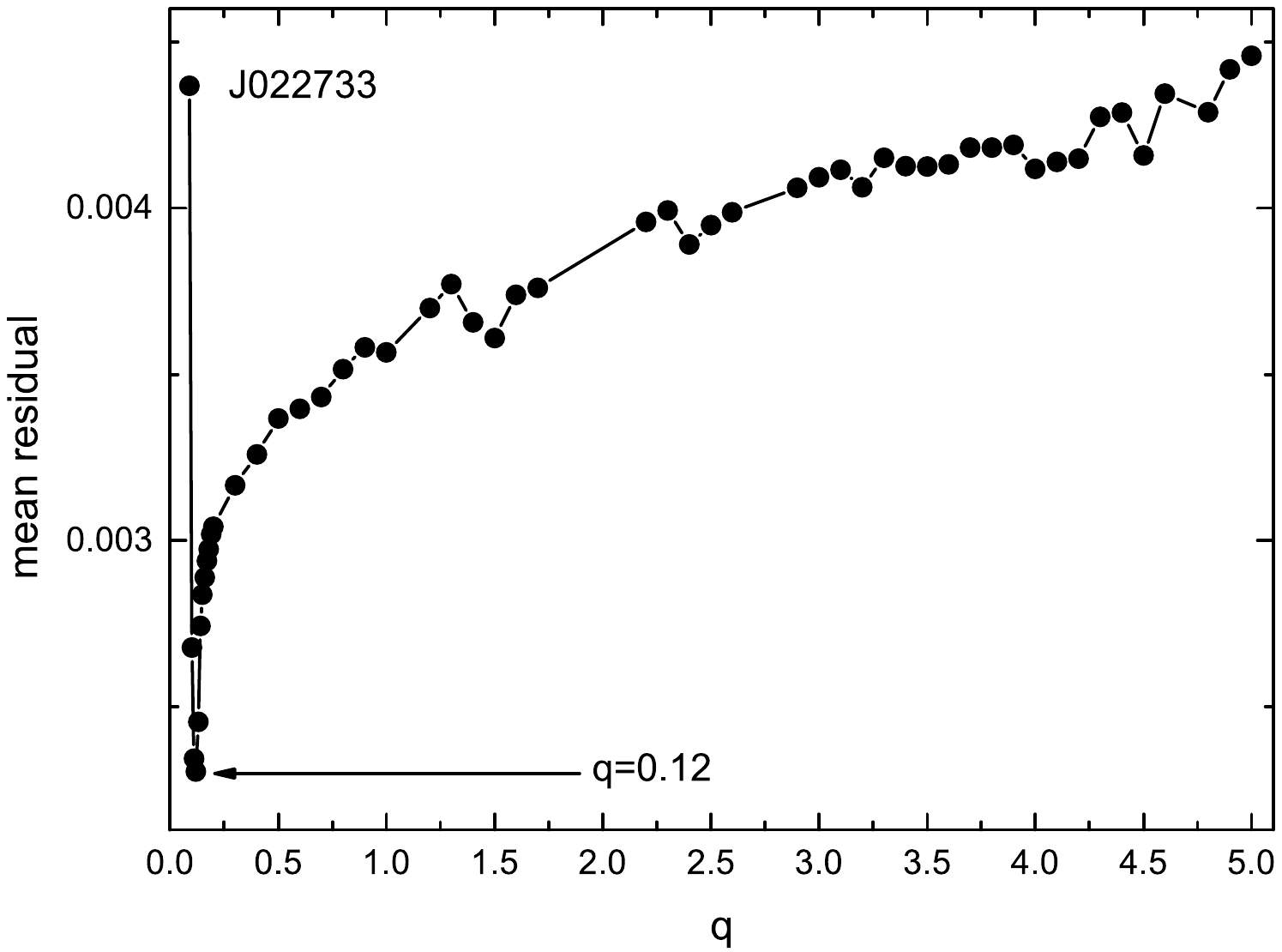}\hspace{-2.3cm}
\plotone{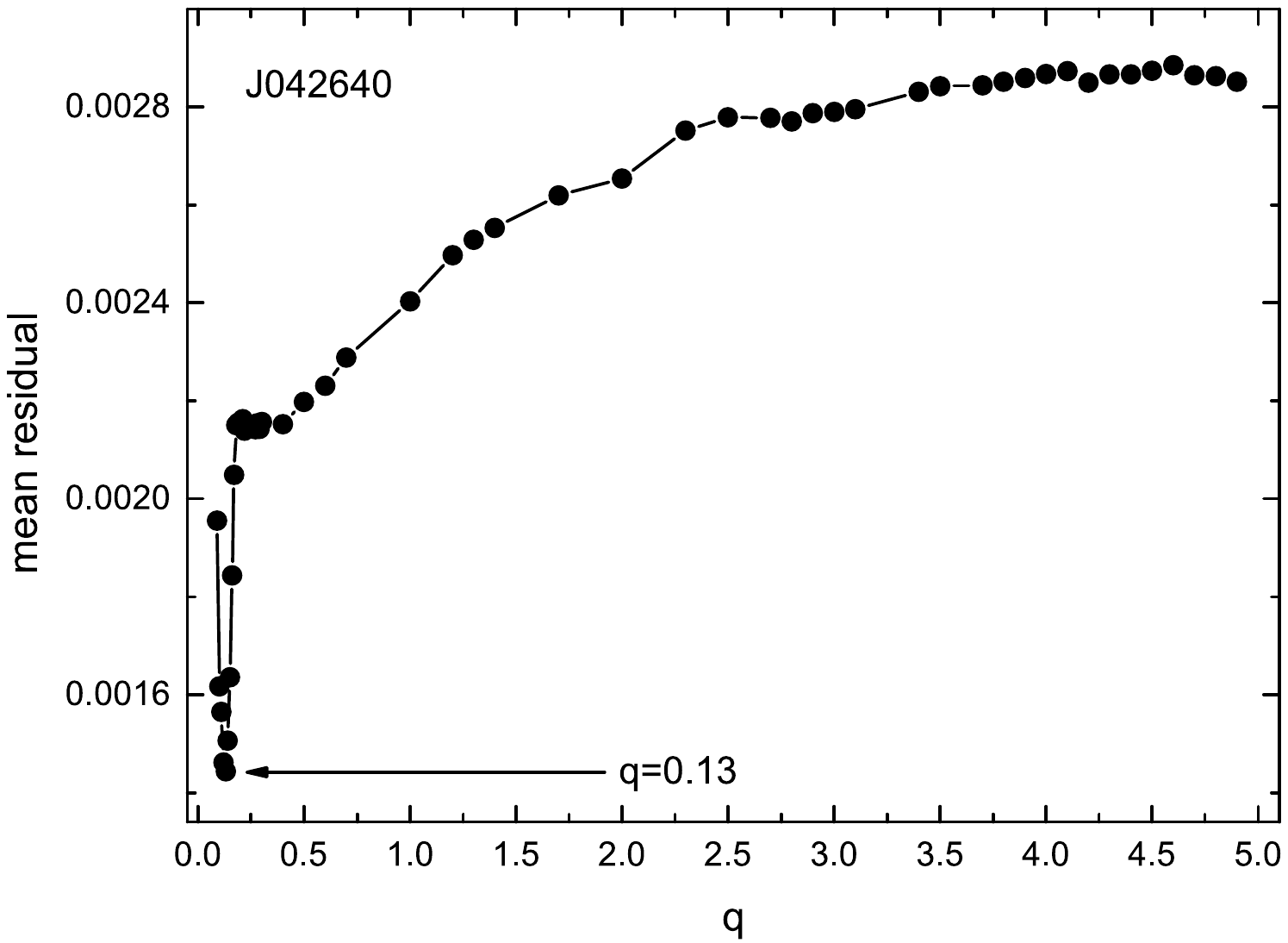}\\\vspace{-5cm}
\plotone{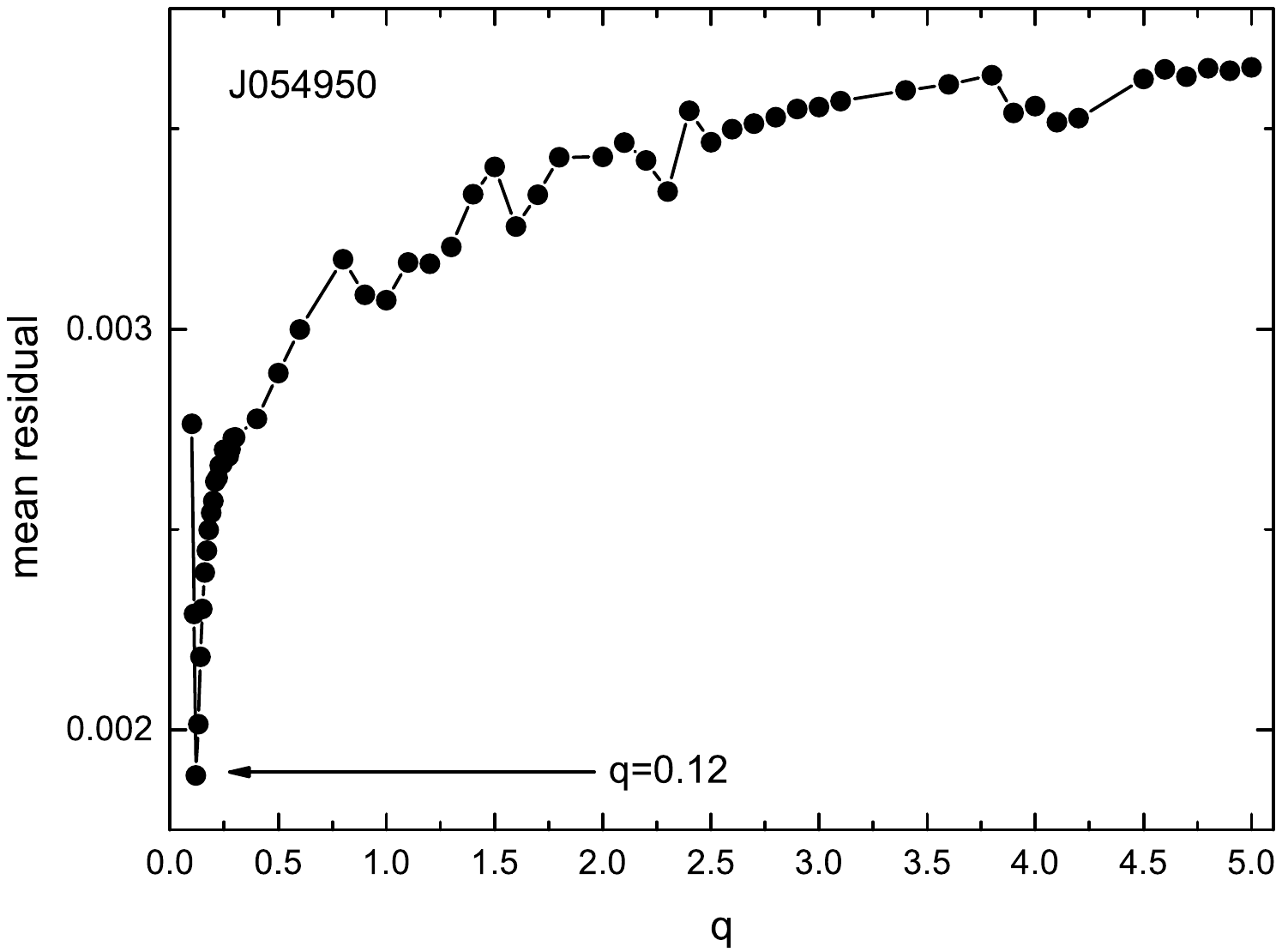}\hspace{-2.3cm}
\plotone{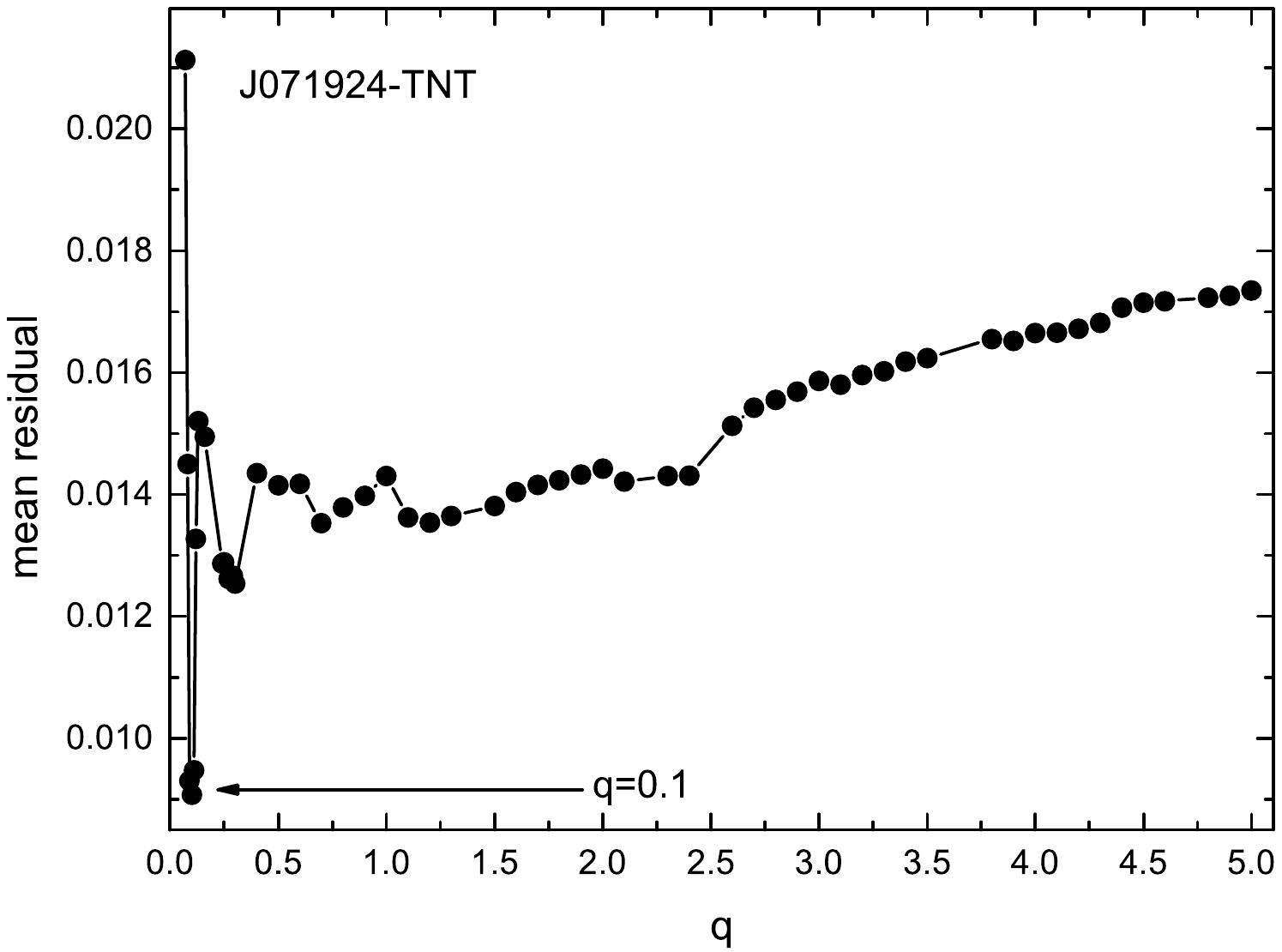}\hspace{-2.3cm}
\plotone{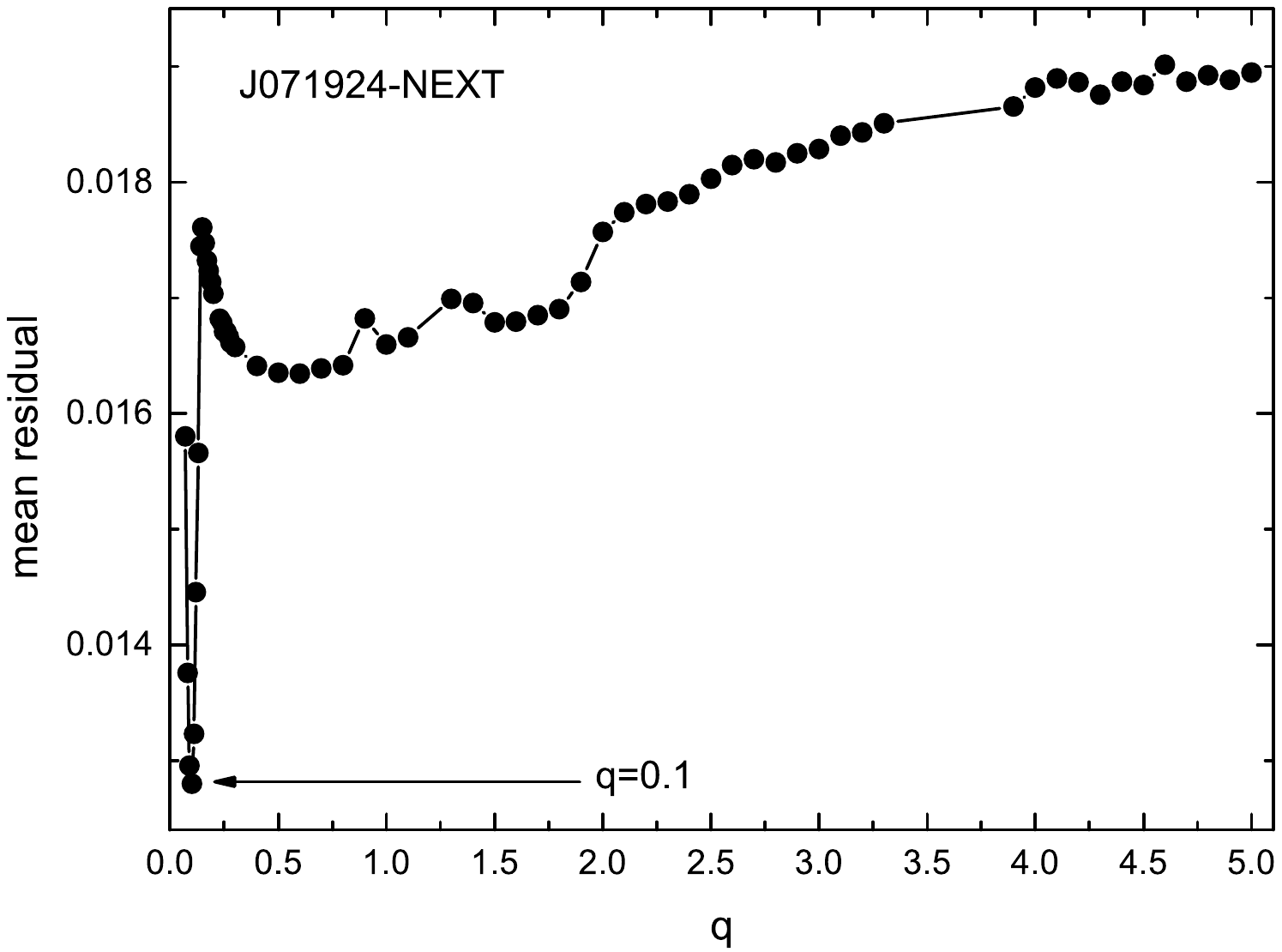}\\\vspace{-5cm}
\plotone{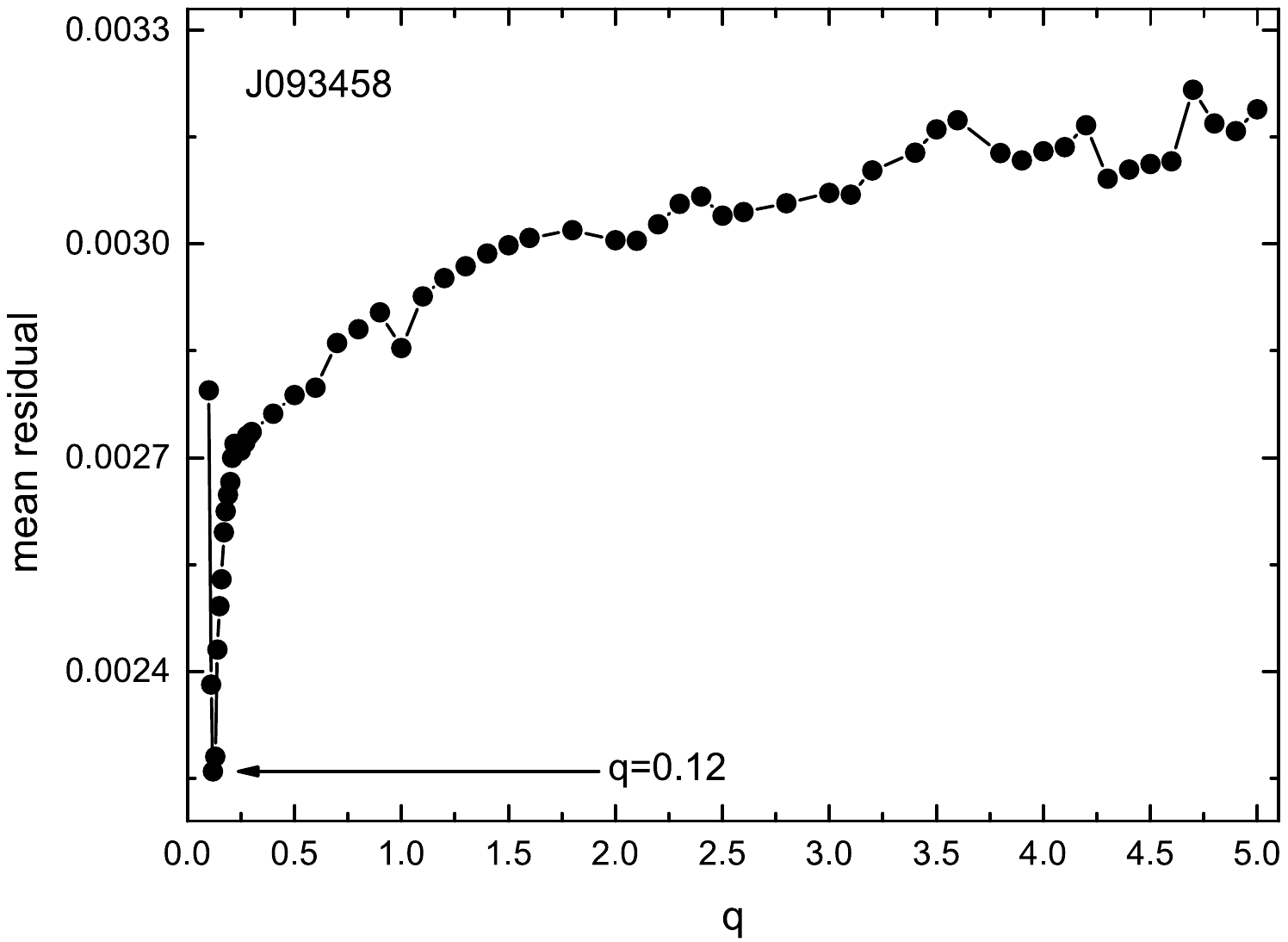}\hspace{-2.3cm}
\plotone{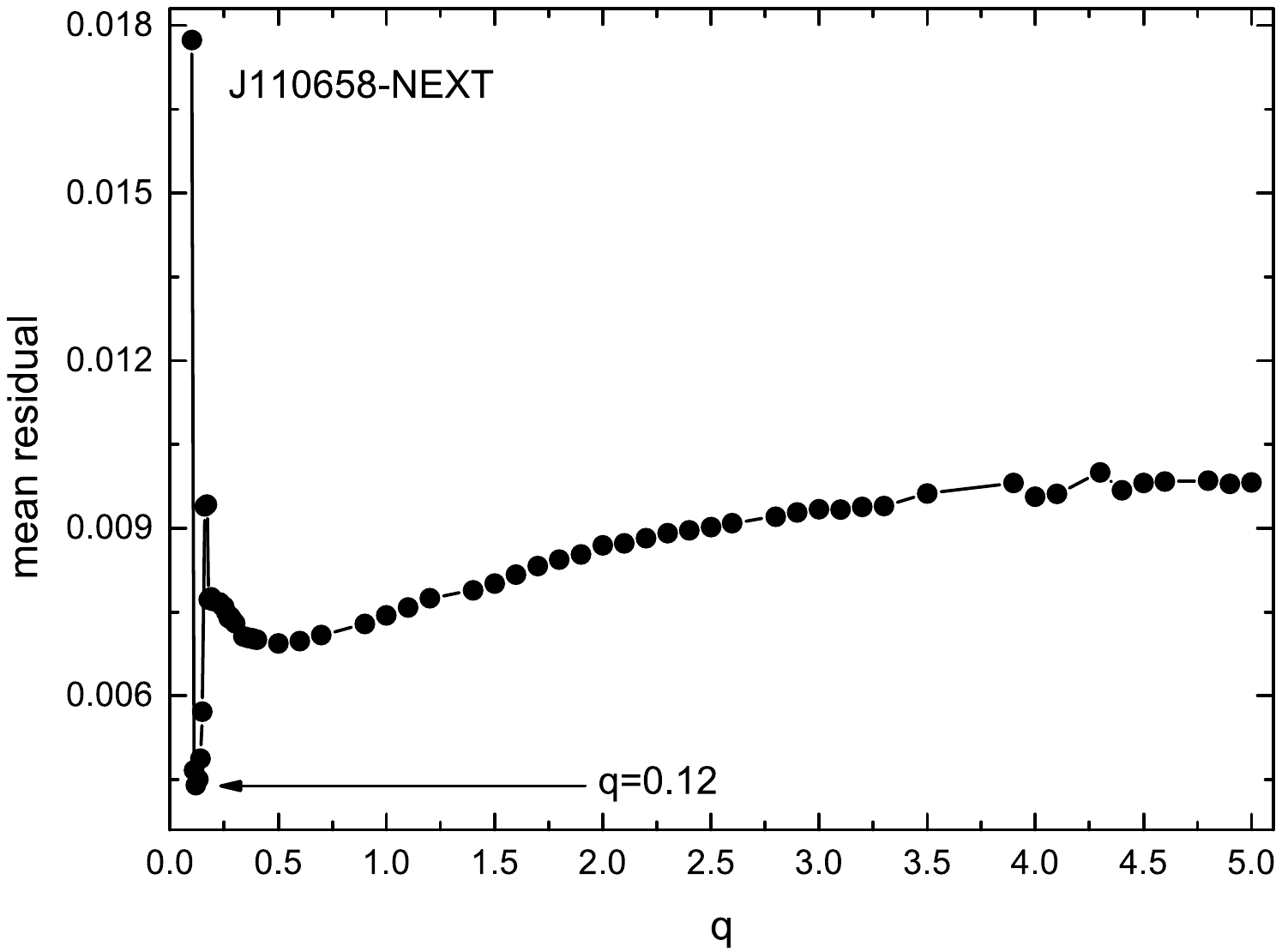}\hspace{-2.3cm}
\plotone{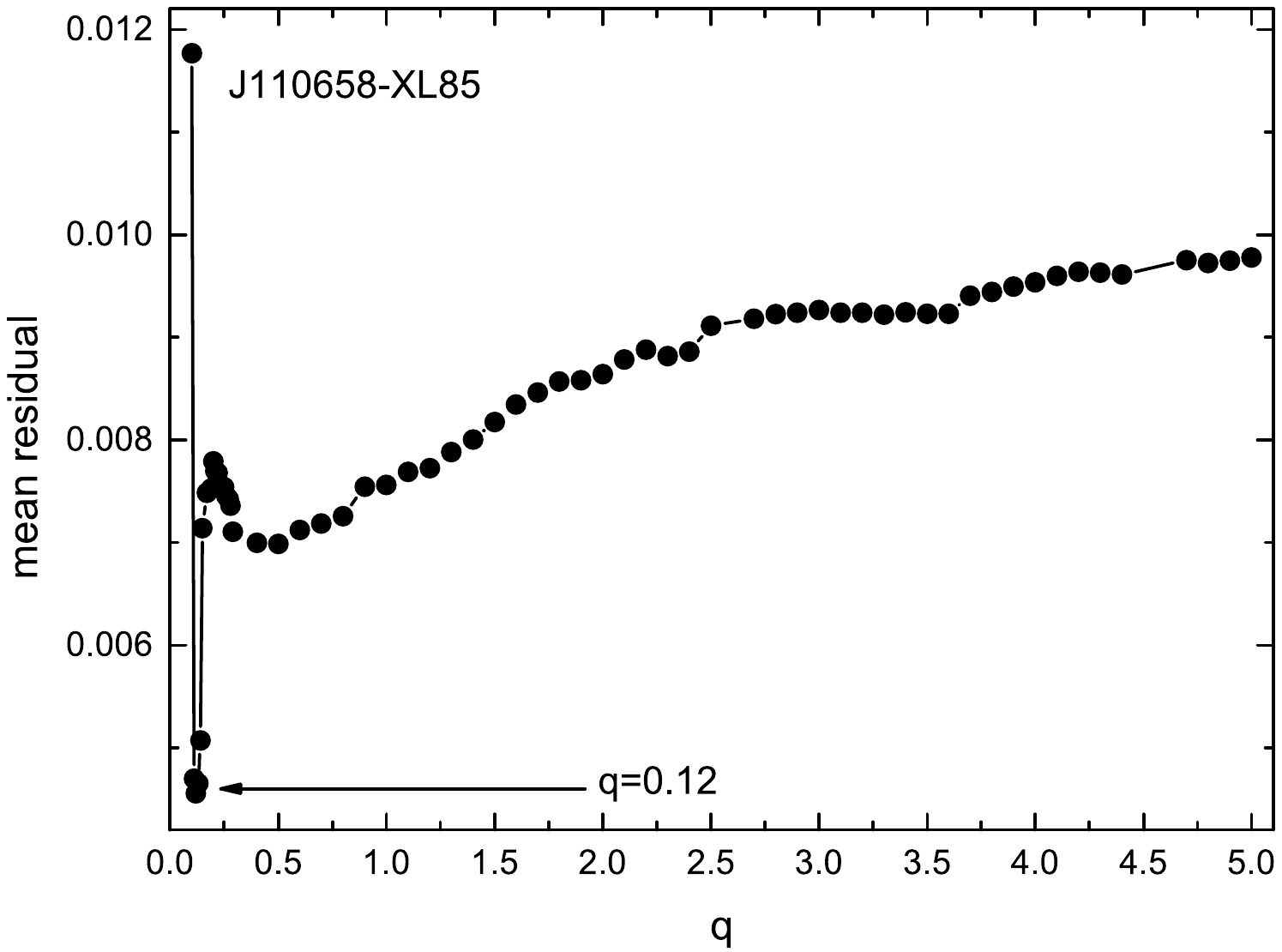}\\\vspace{-5cm}
\plotone{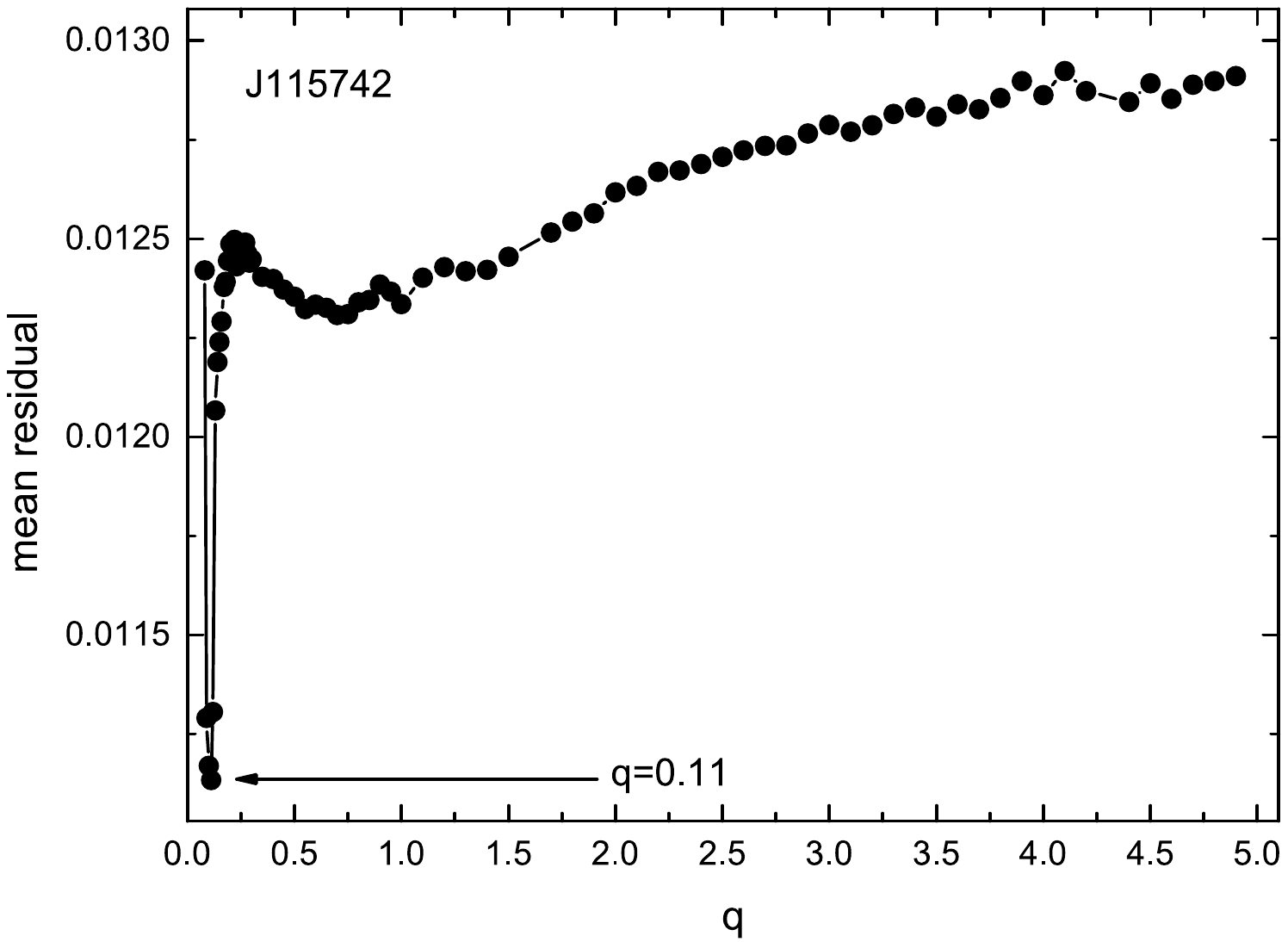}\hspace{-2.3cm}
\plotone{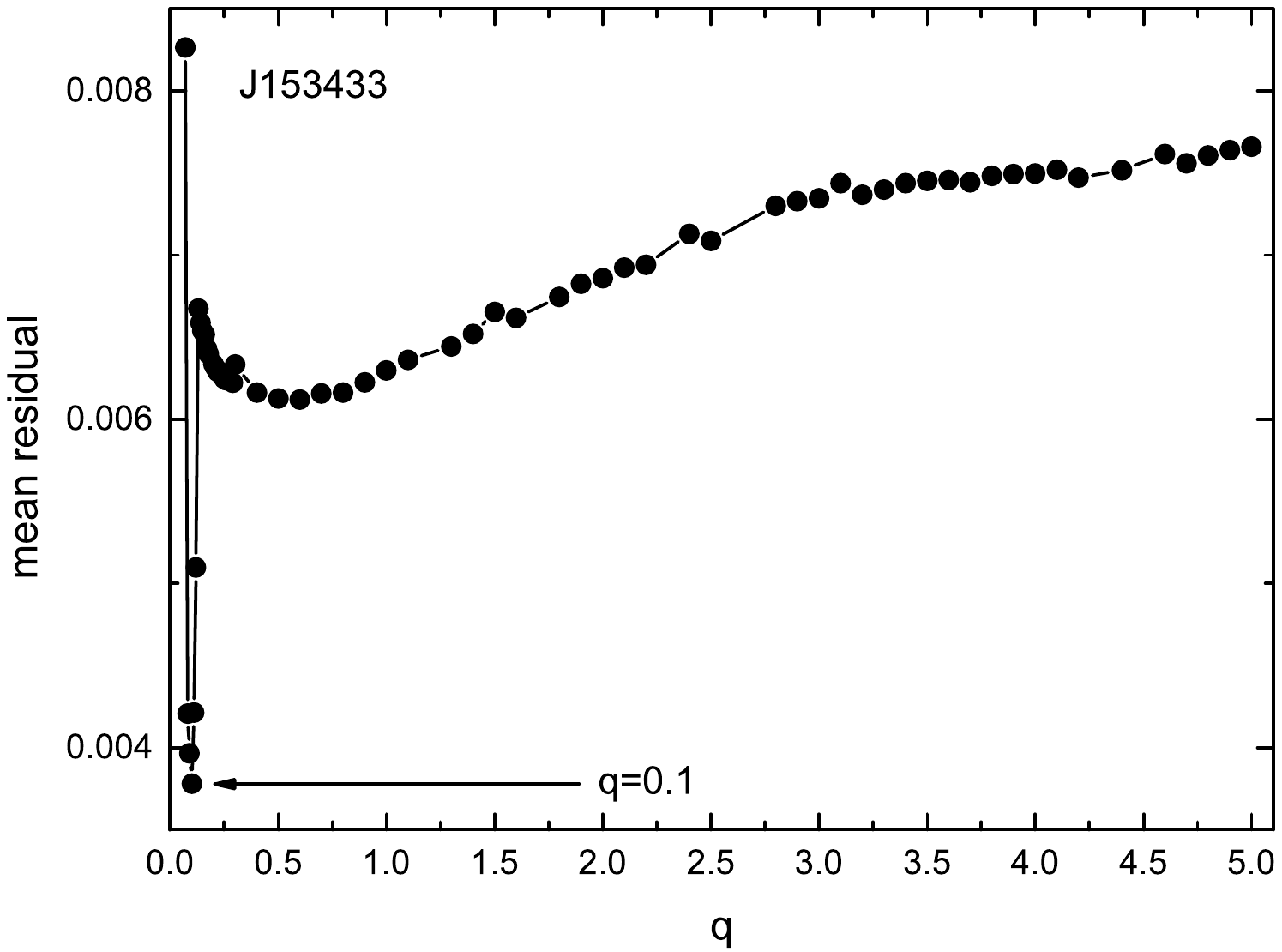}\hspace{-2.3cm}
\plotone{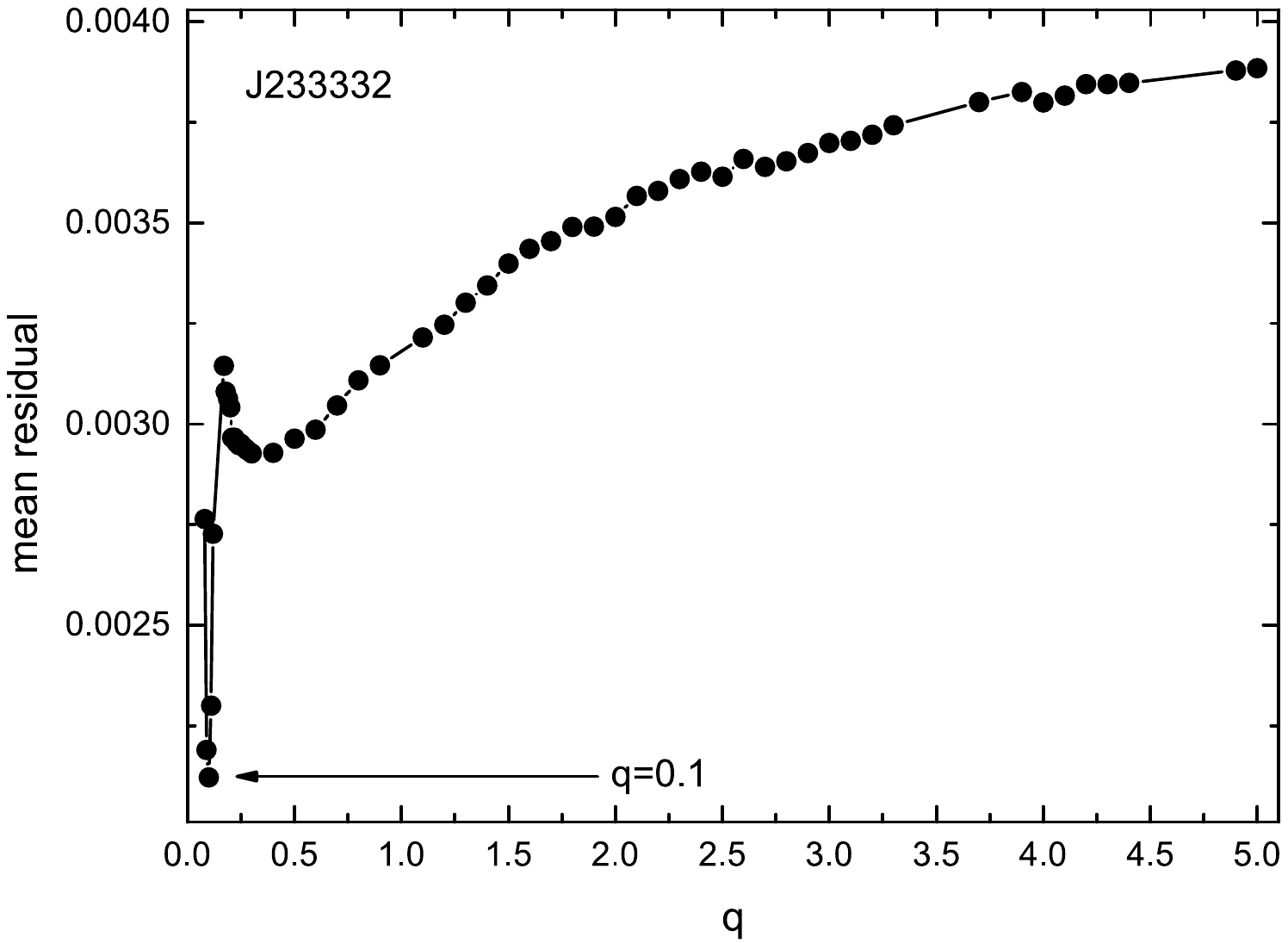}
\caption{The relationship between the mean residual and mass ratio $q$ for the ten targets.  \label{fig:q-search}}
\end{figure}

\begin{figure}
\epsscale{0.35}
\plotone{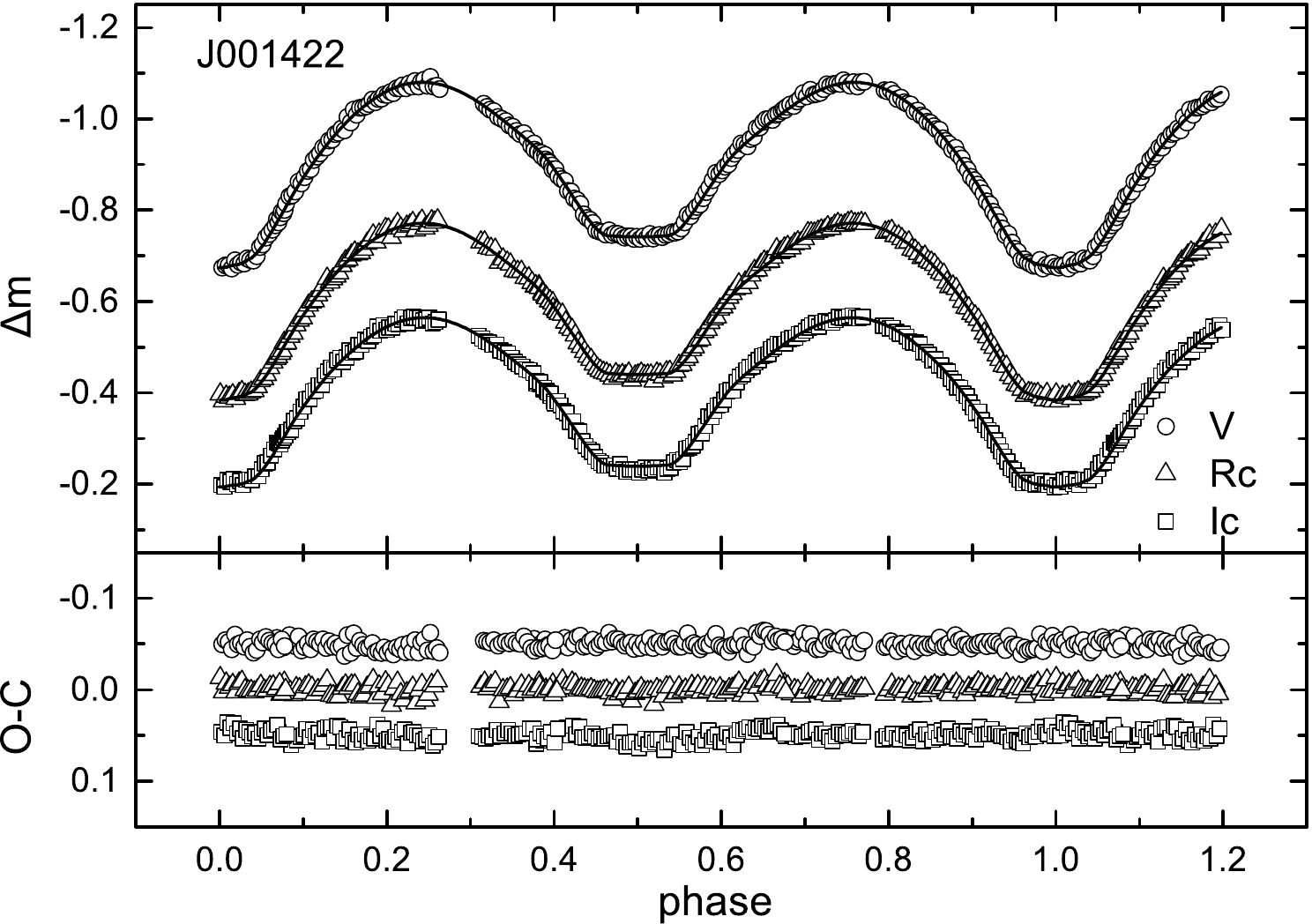}
\plotone{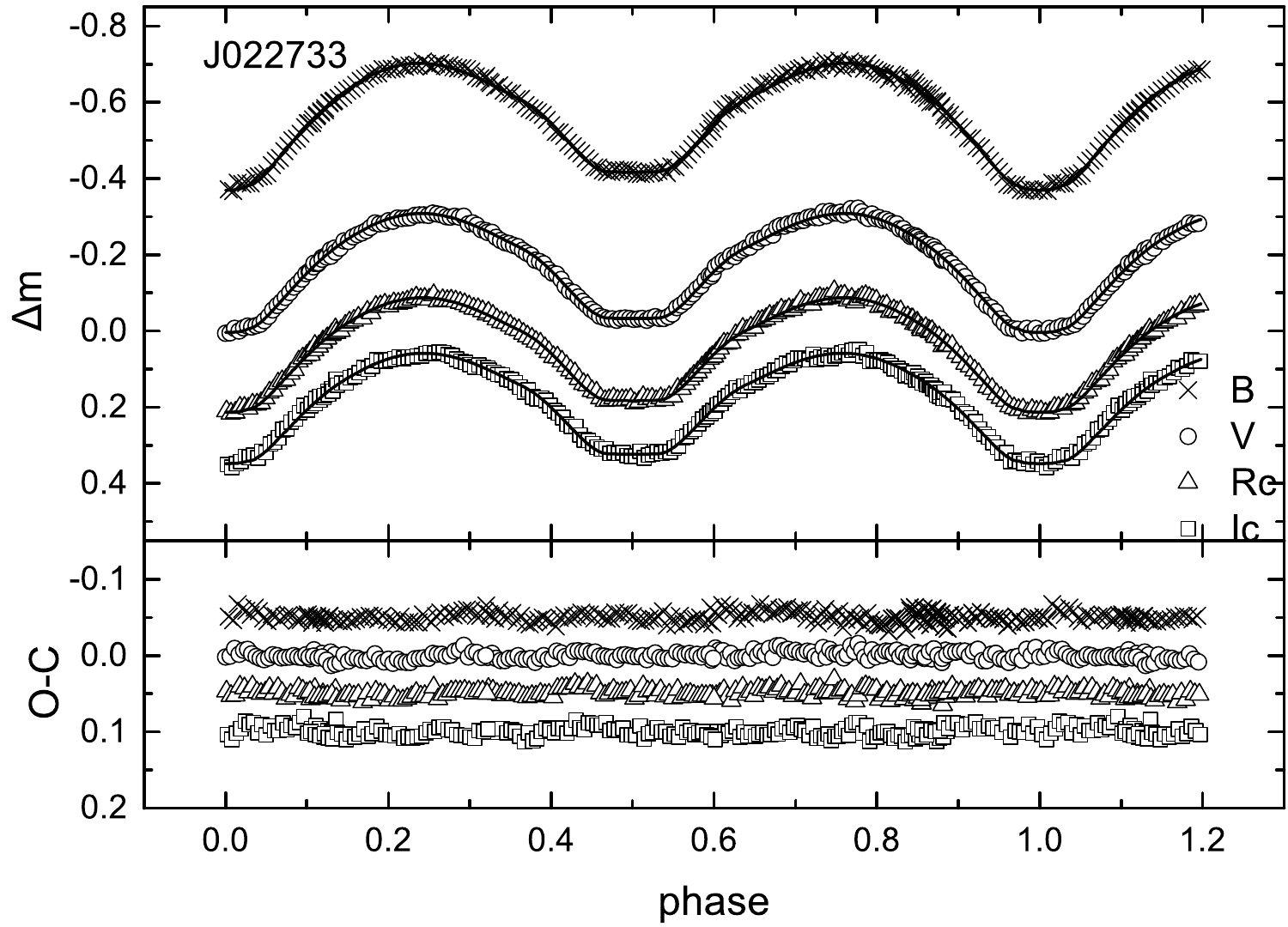}
\plotone{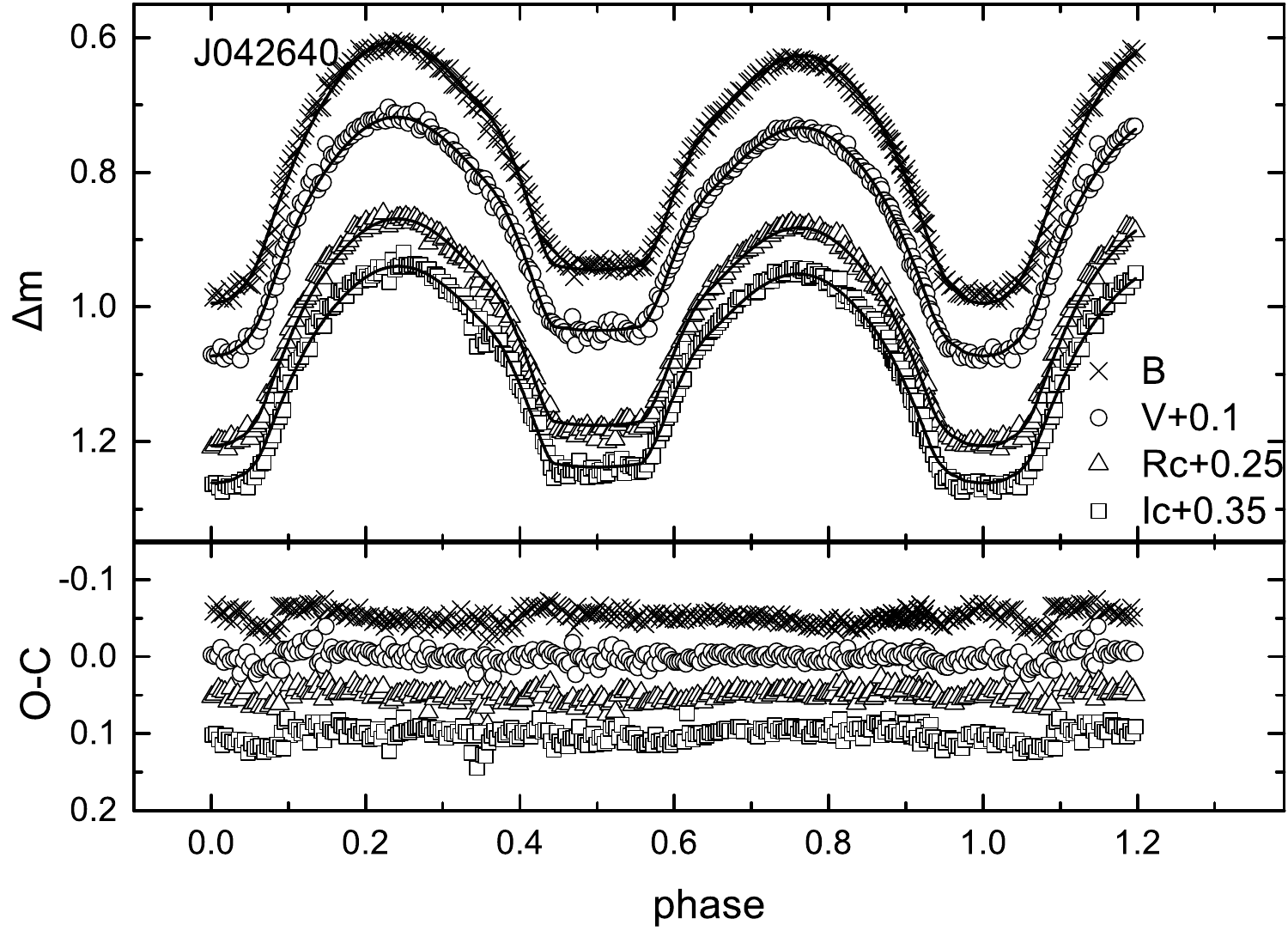}\\
\plotone{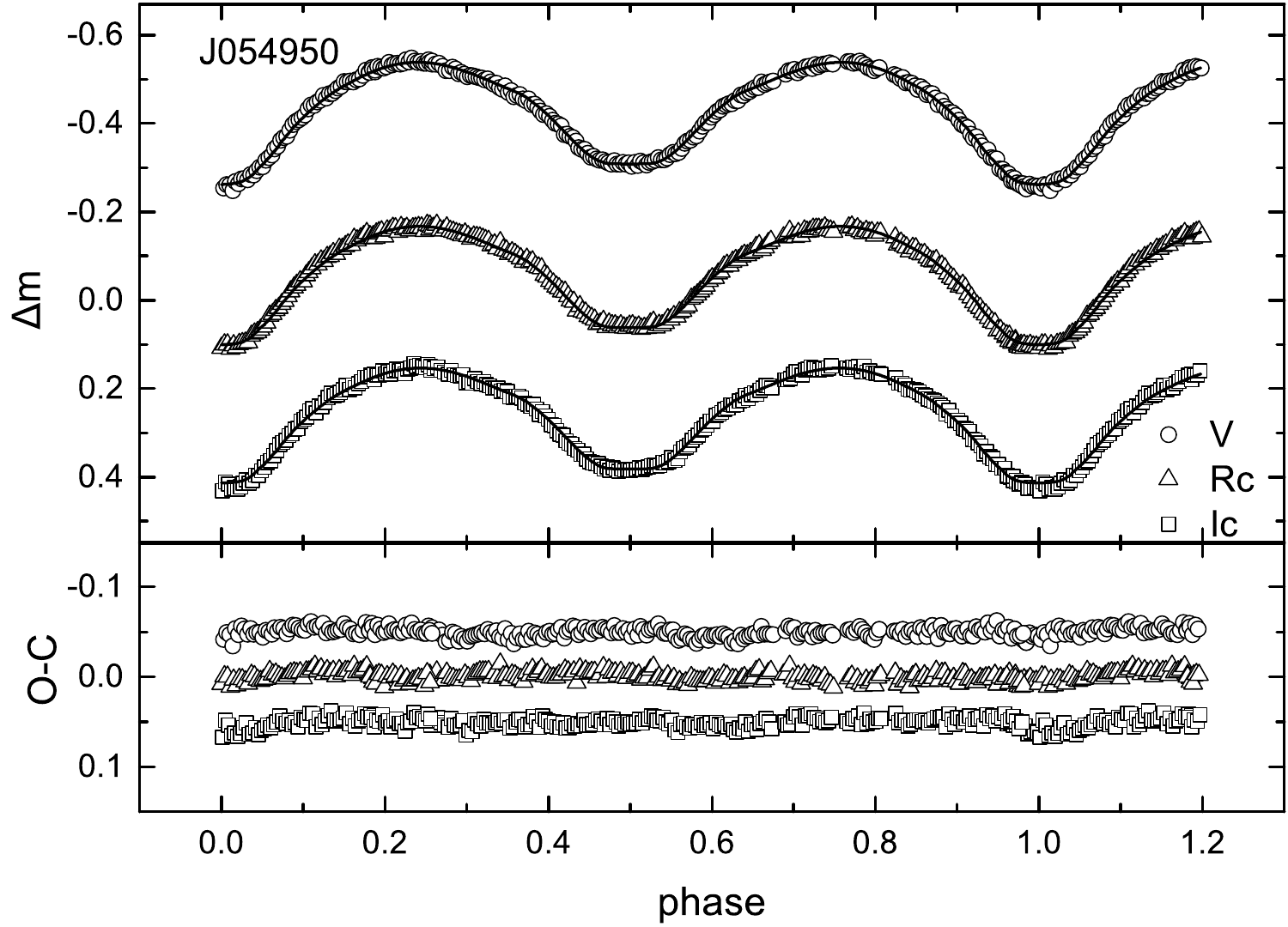}
\plotone{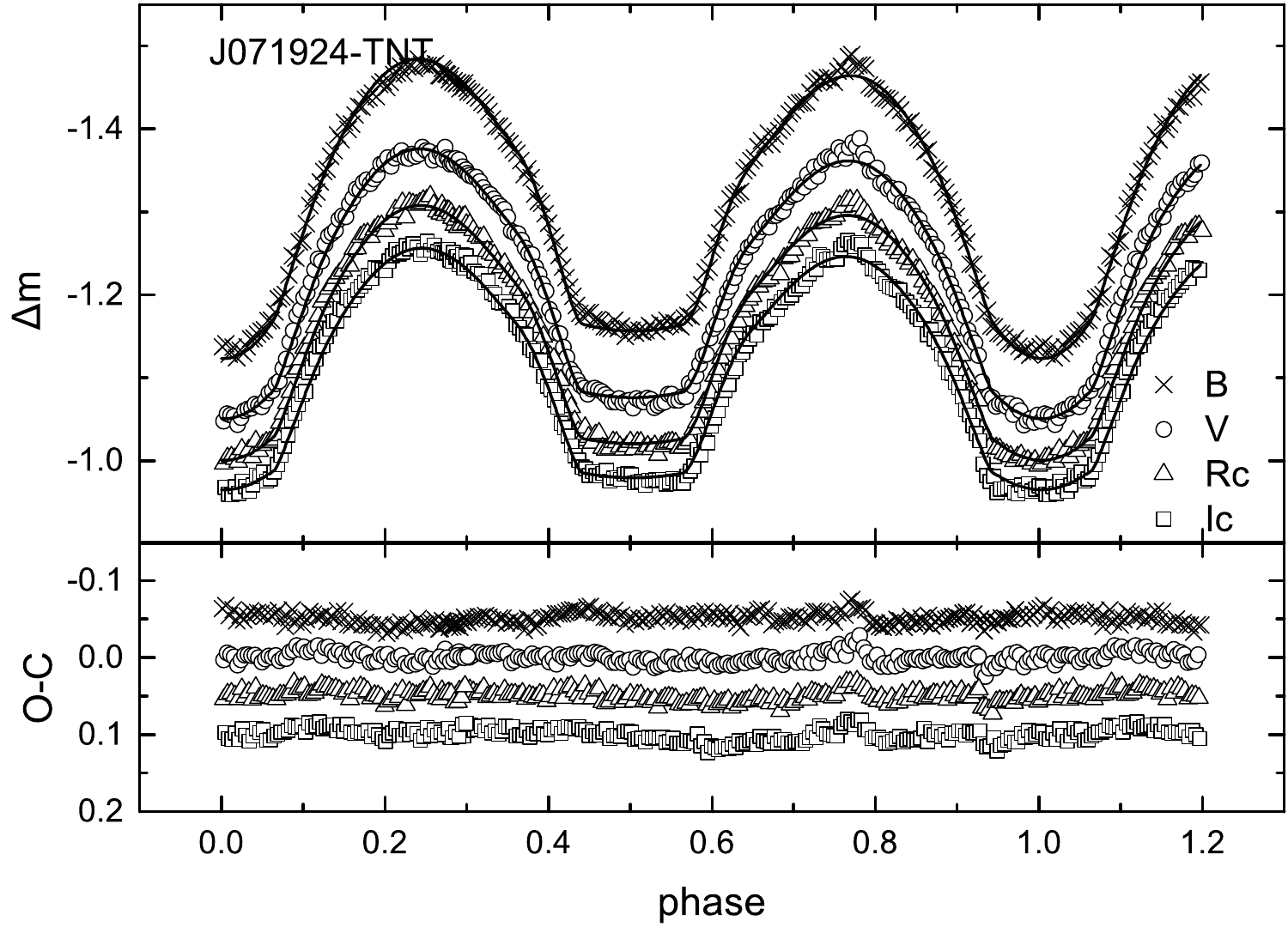}
\plotone{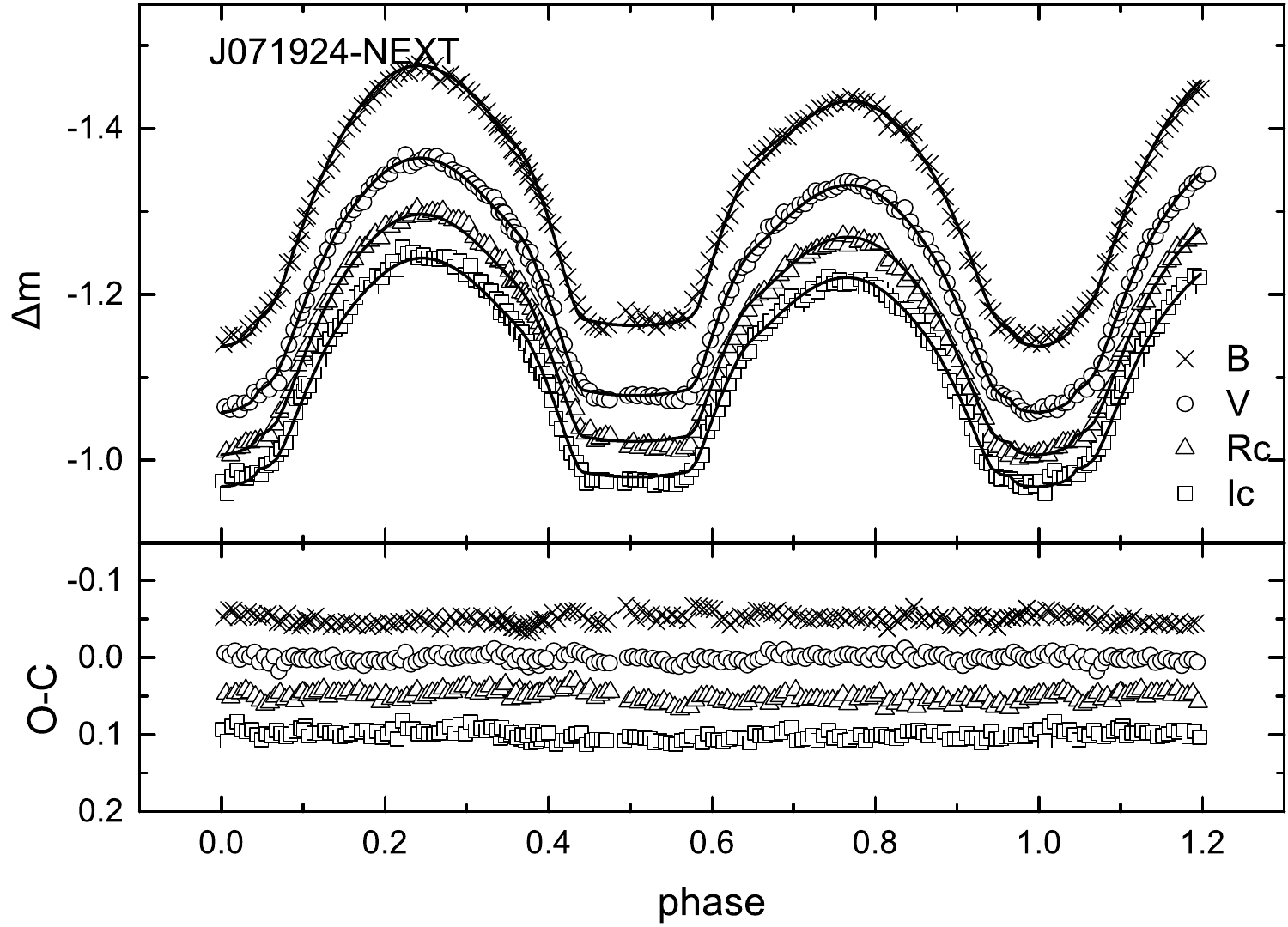}\\
\plotone{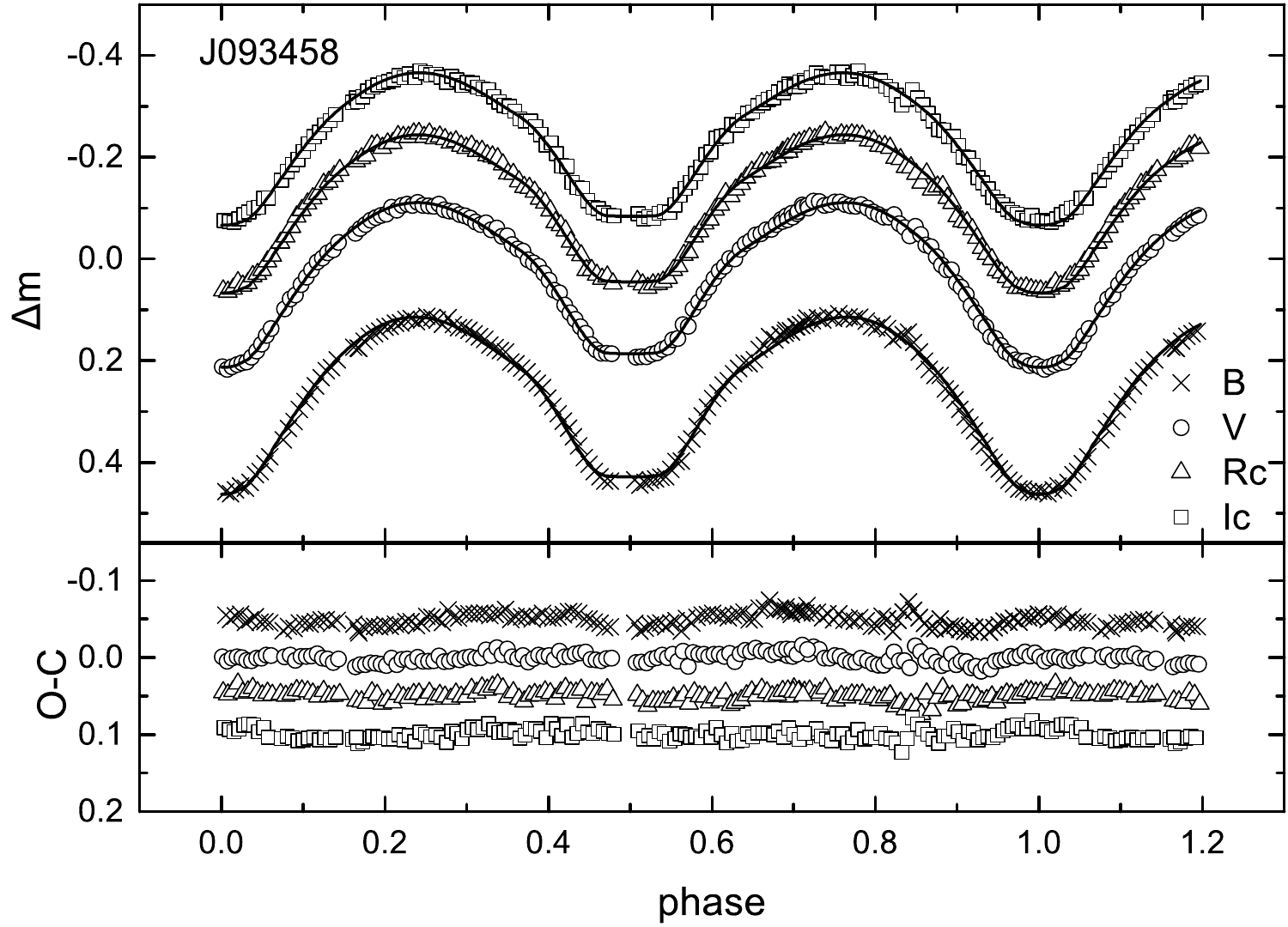}
\plotone{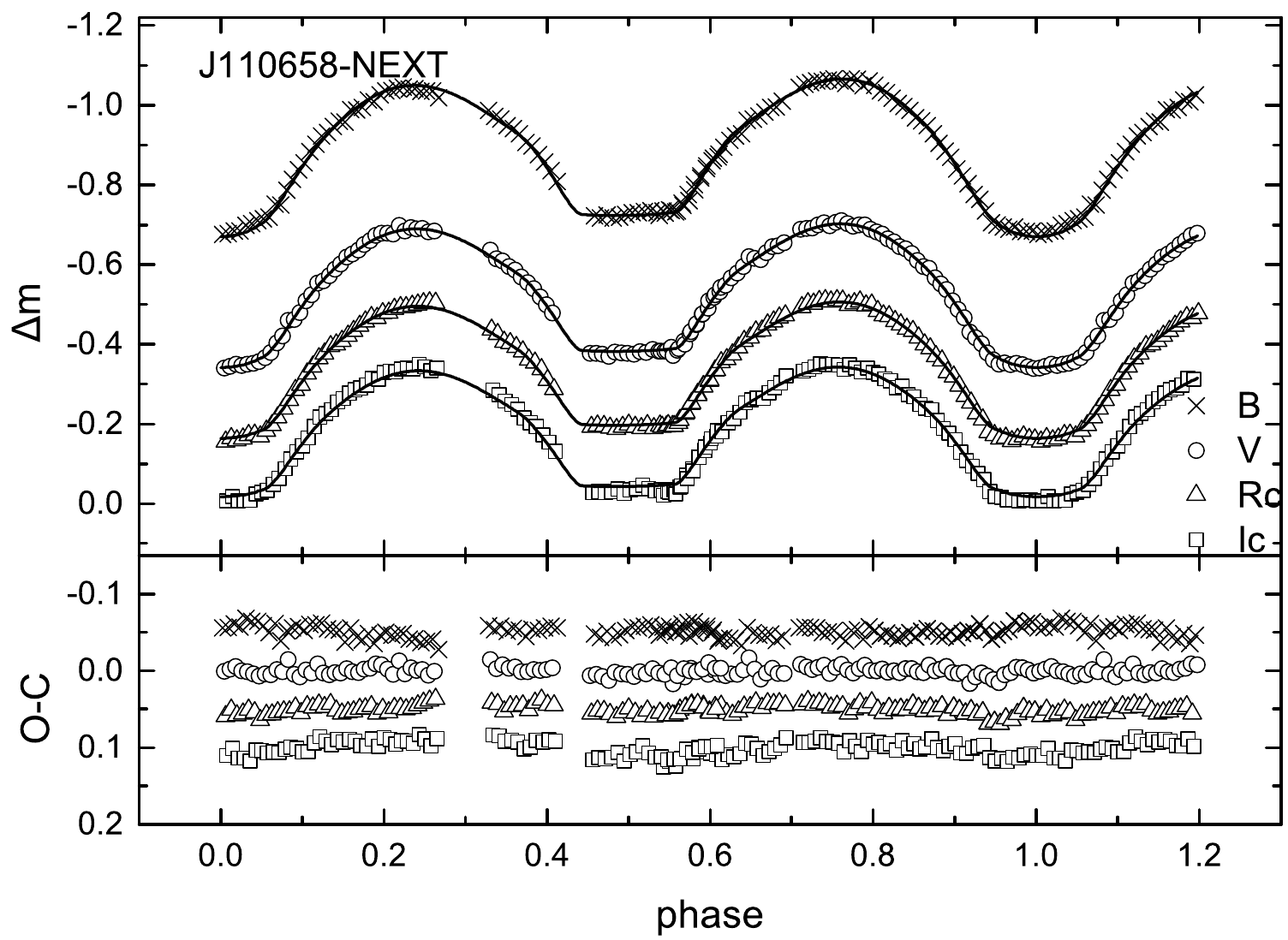}
\plotone{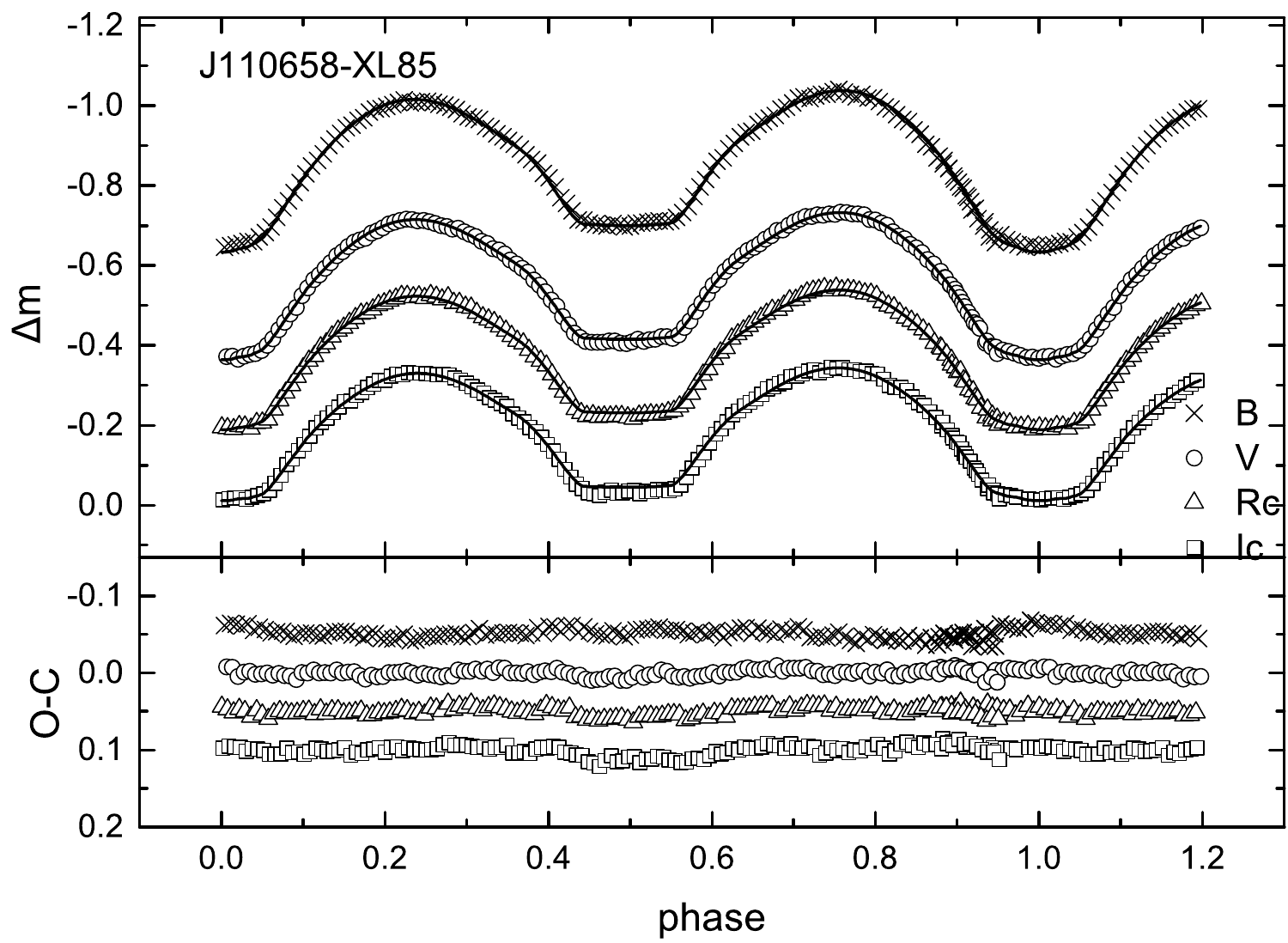}\\
\plotone{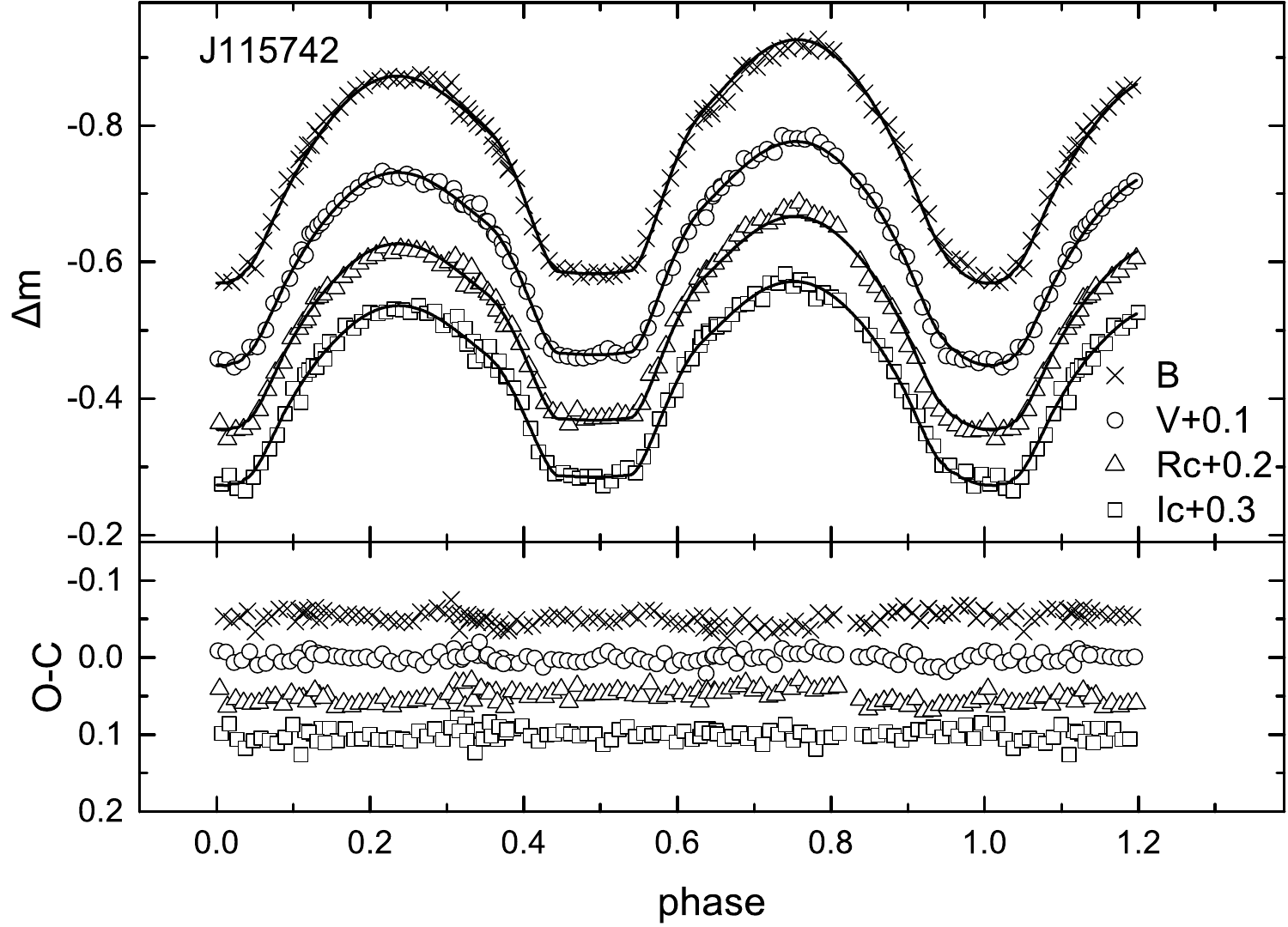}
\plotone{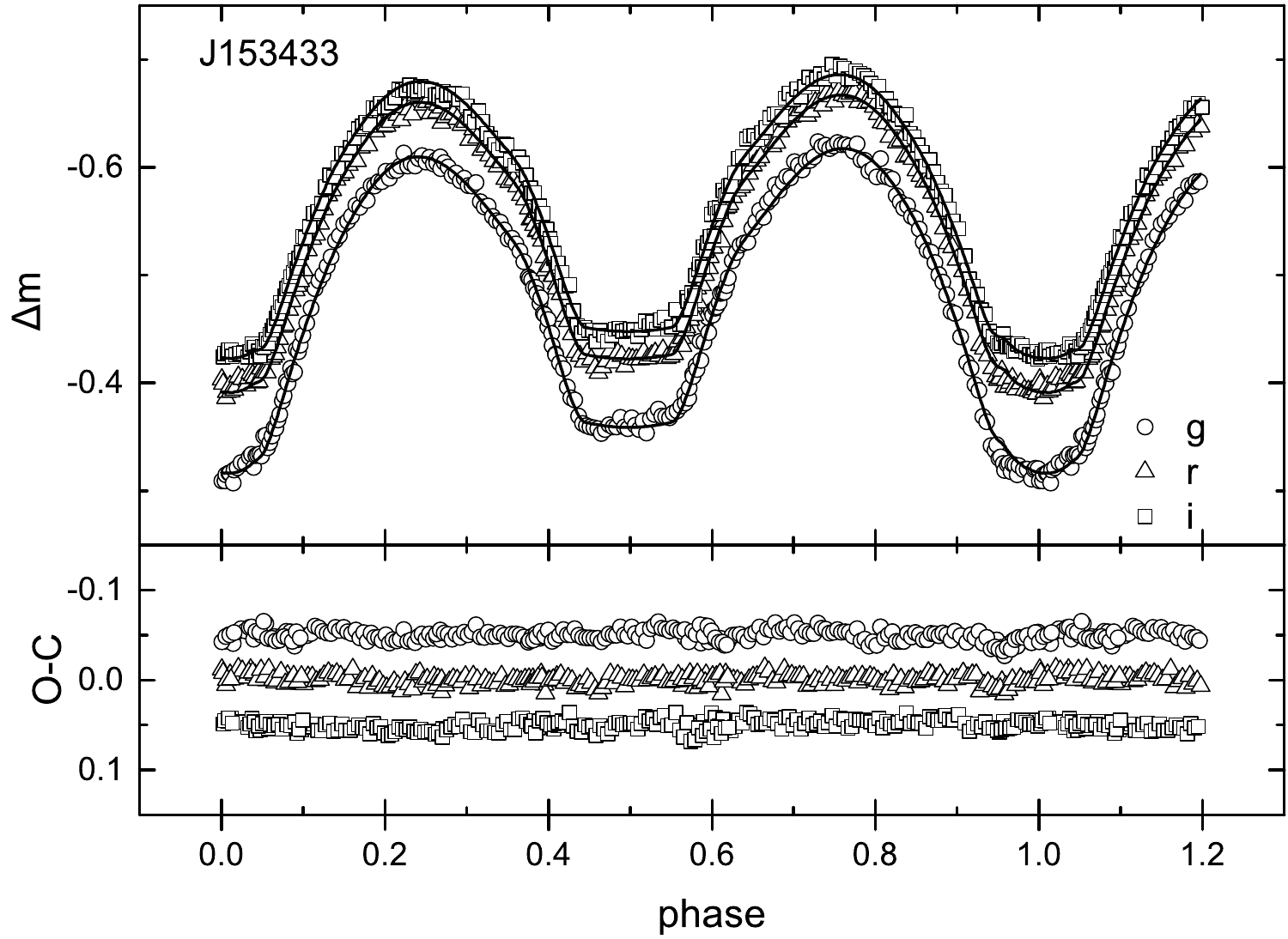}
\plotone{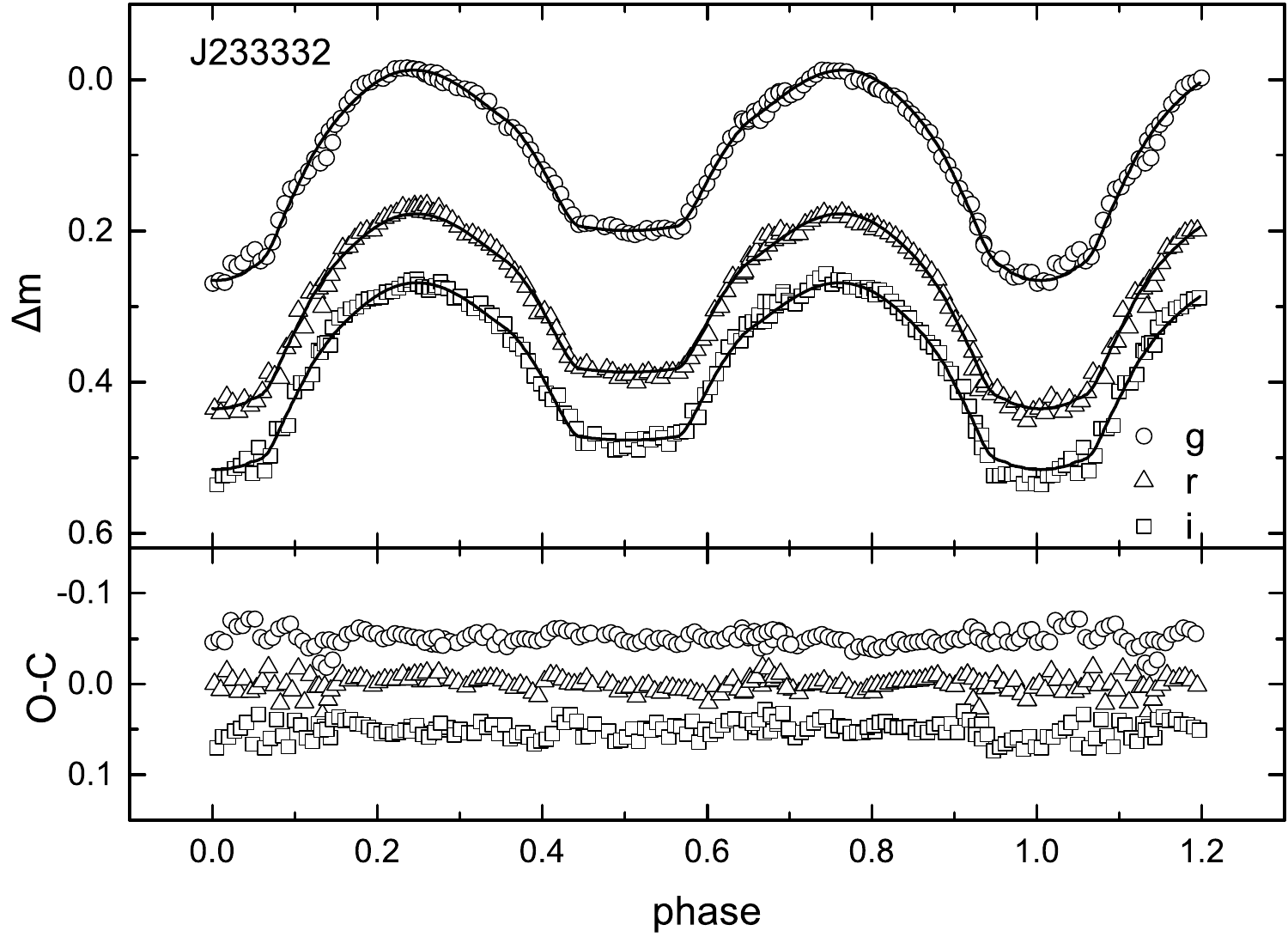}
\caption{The comparison between the synthetic and observed light curves of the ten targets. The lower panels show the O-C residuals. \label{fig:lcfit}}
\end{figure}

\begin{figure}
\epsscale{0.35}
\plotone{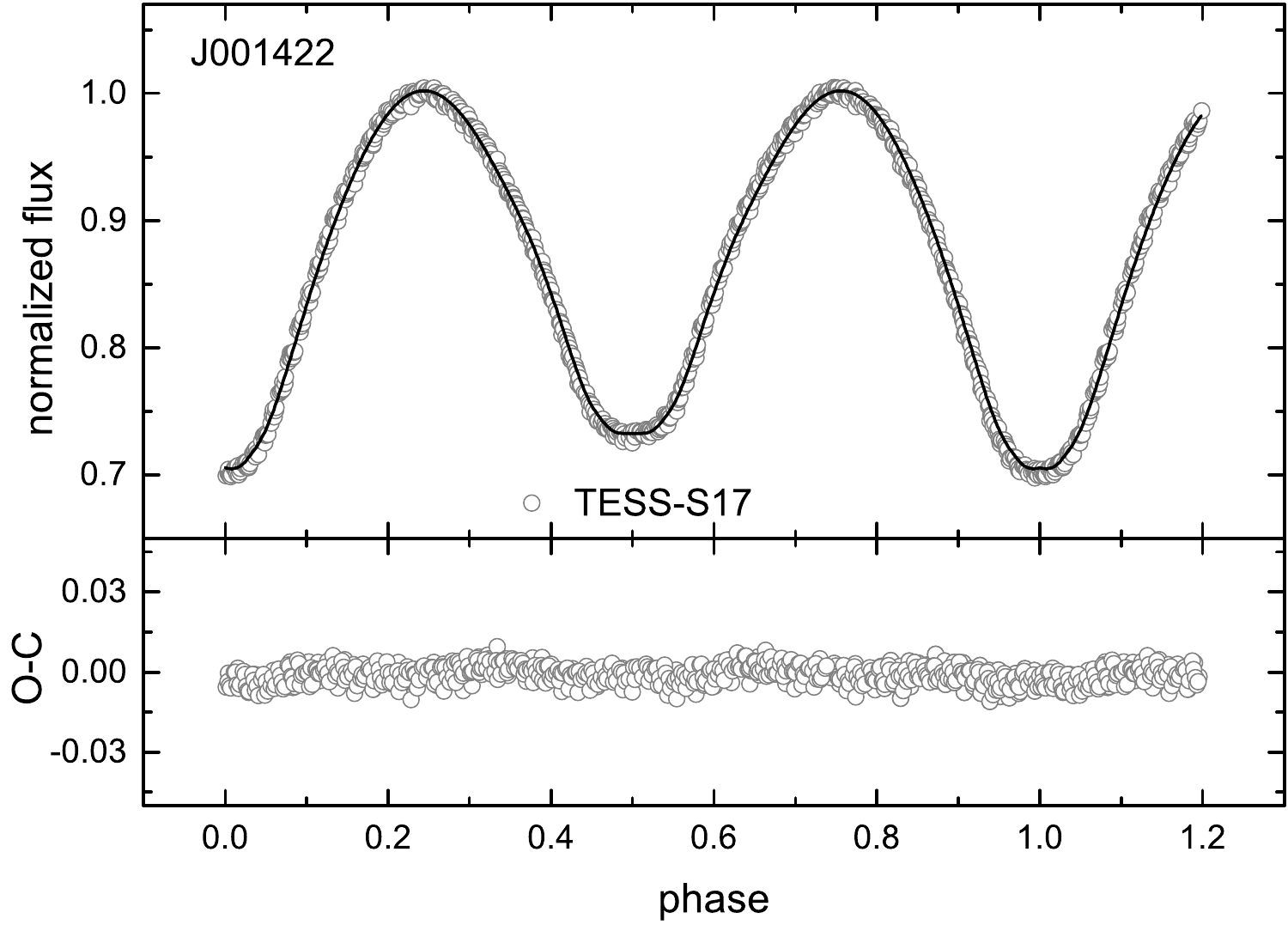}
\plotone{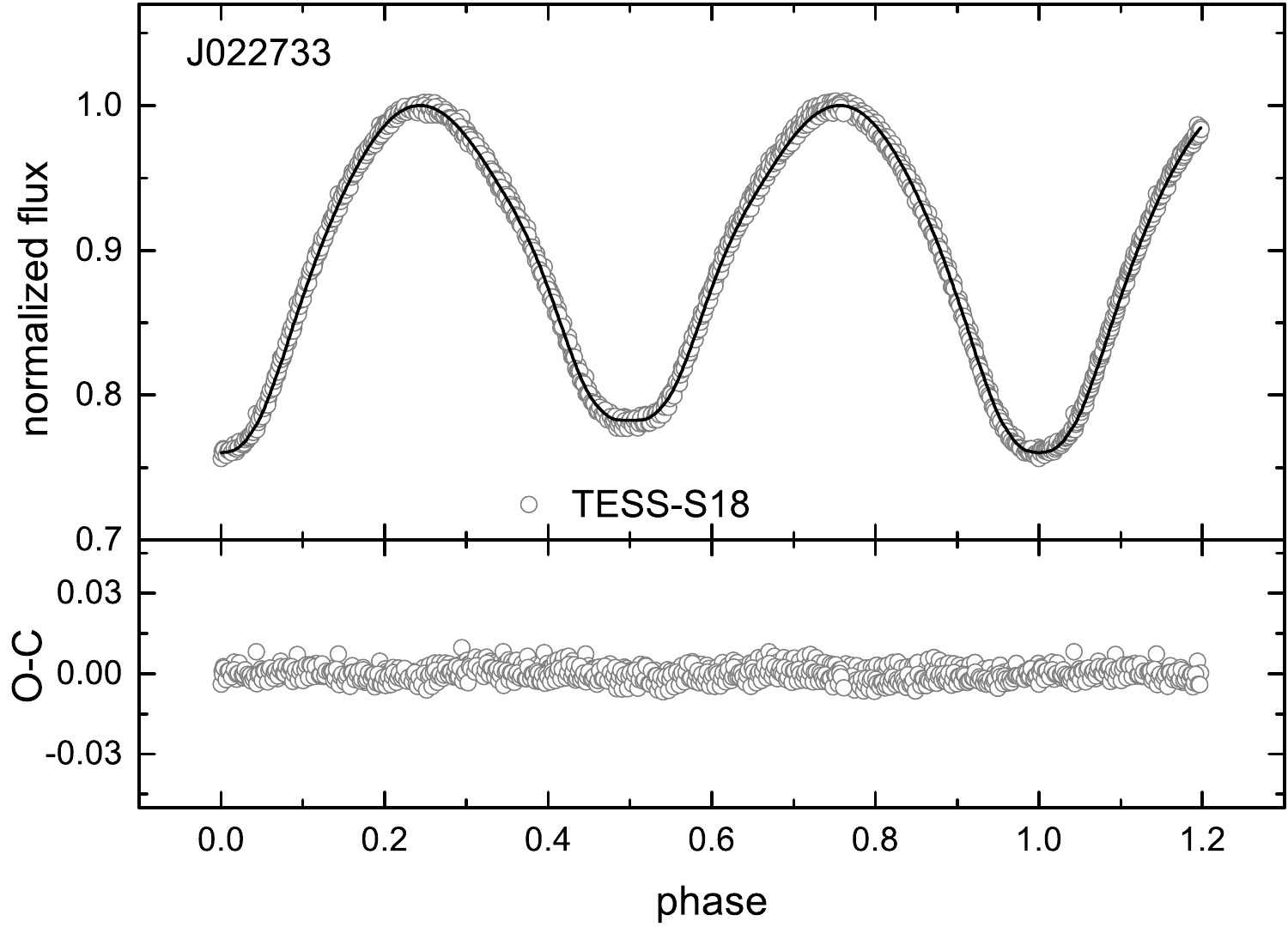}
\plotone{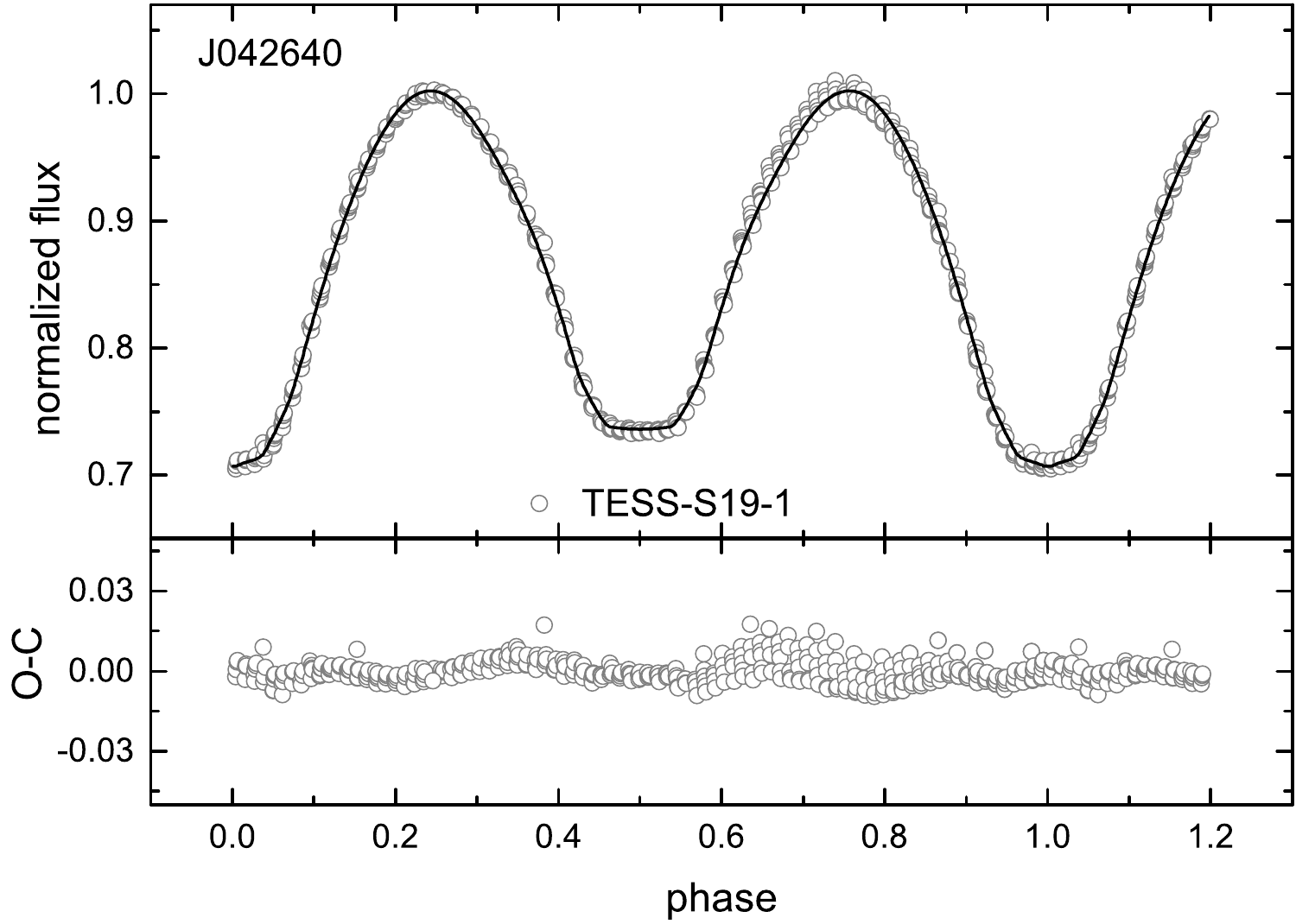}\\
\plotone{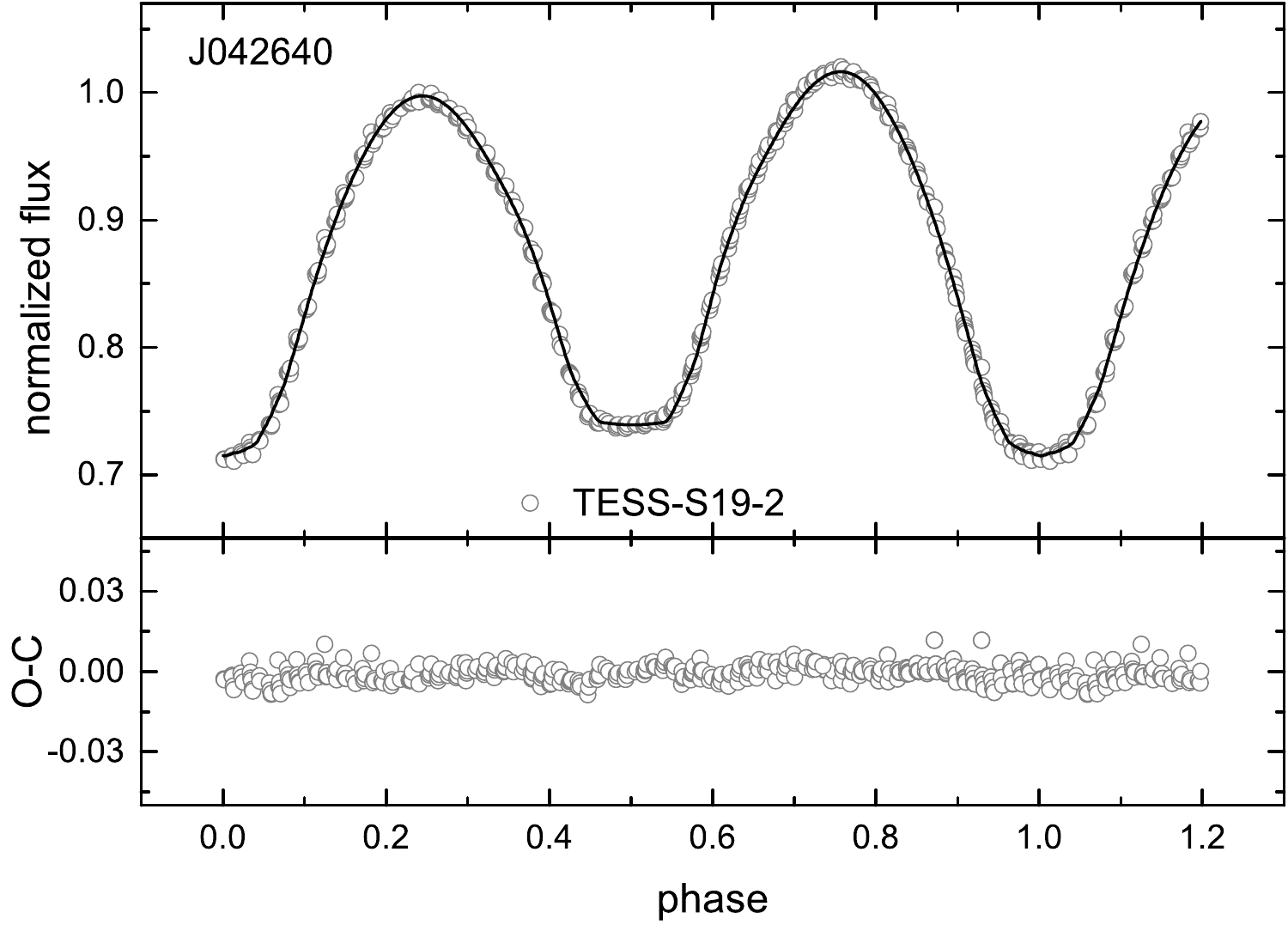}
\plotone{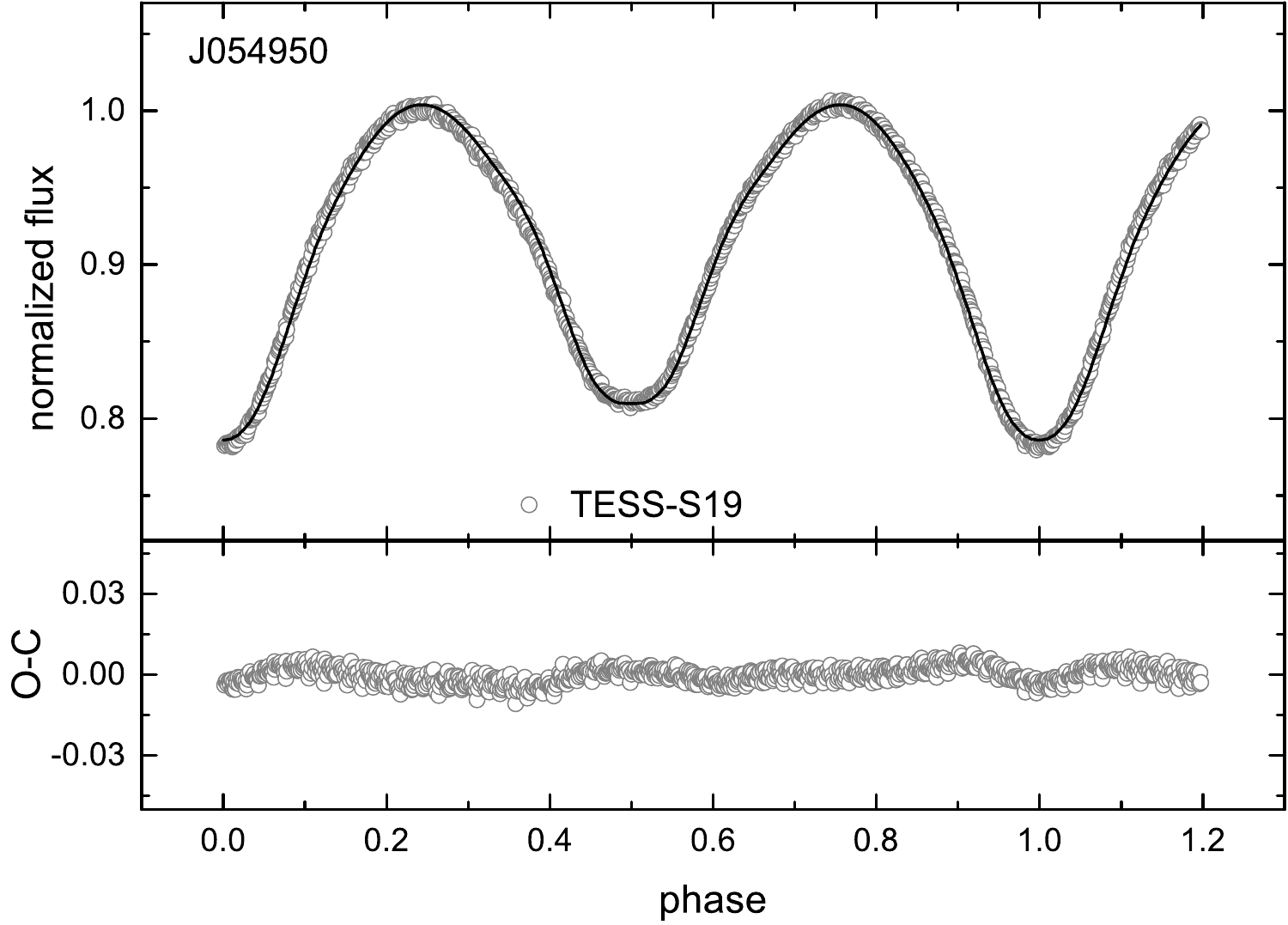}
\plotone{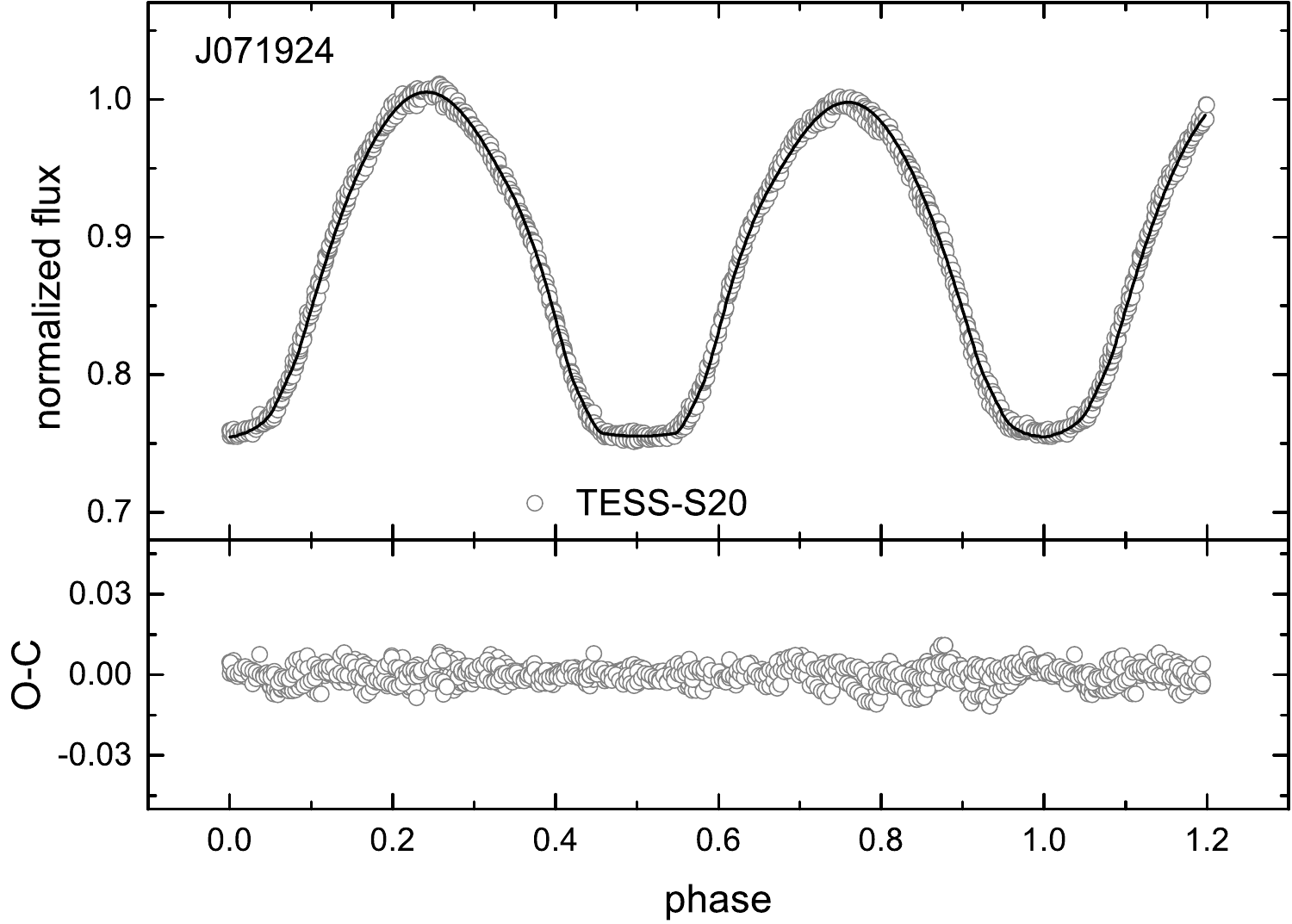}\\
\plotone{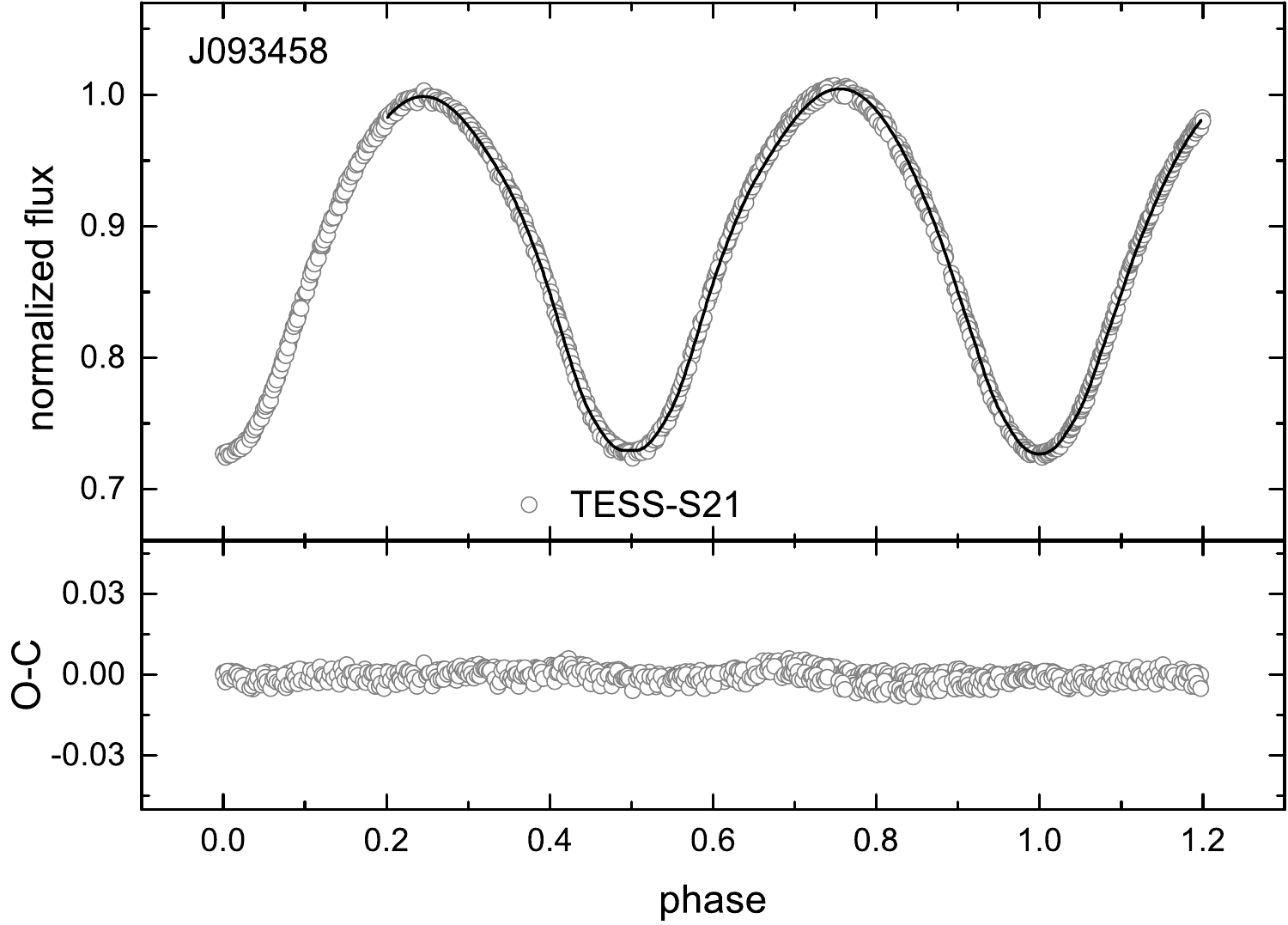}
\plotone{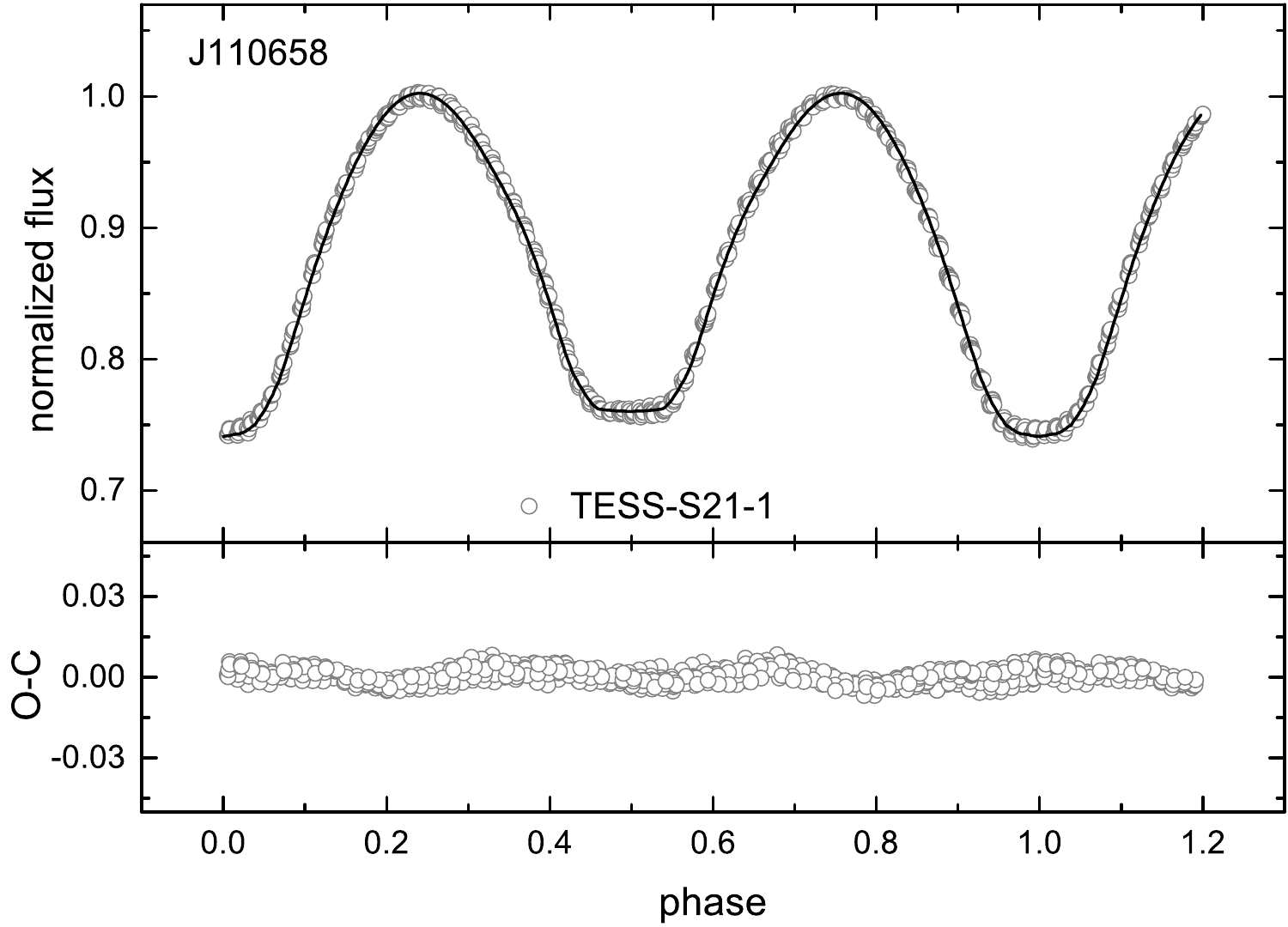}
\plotone{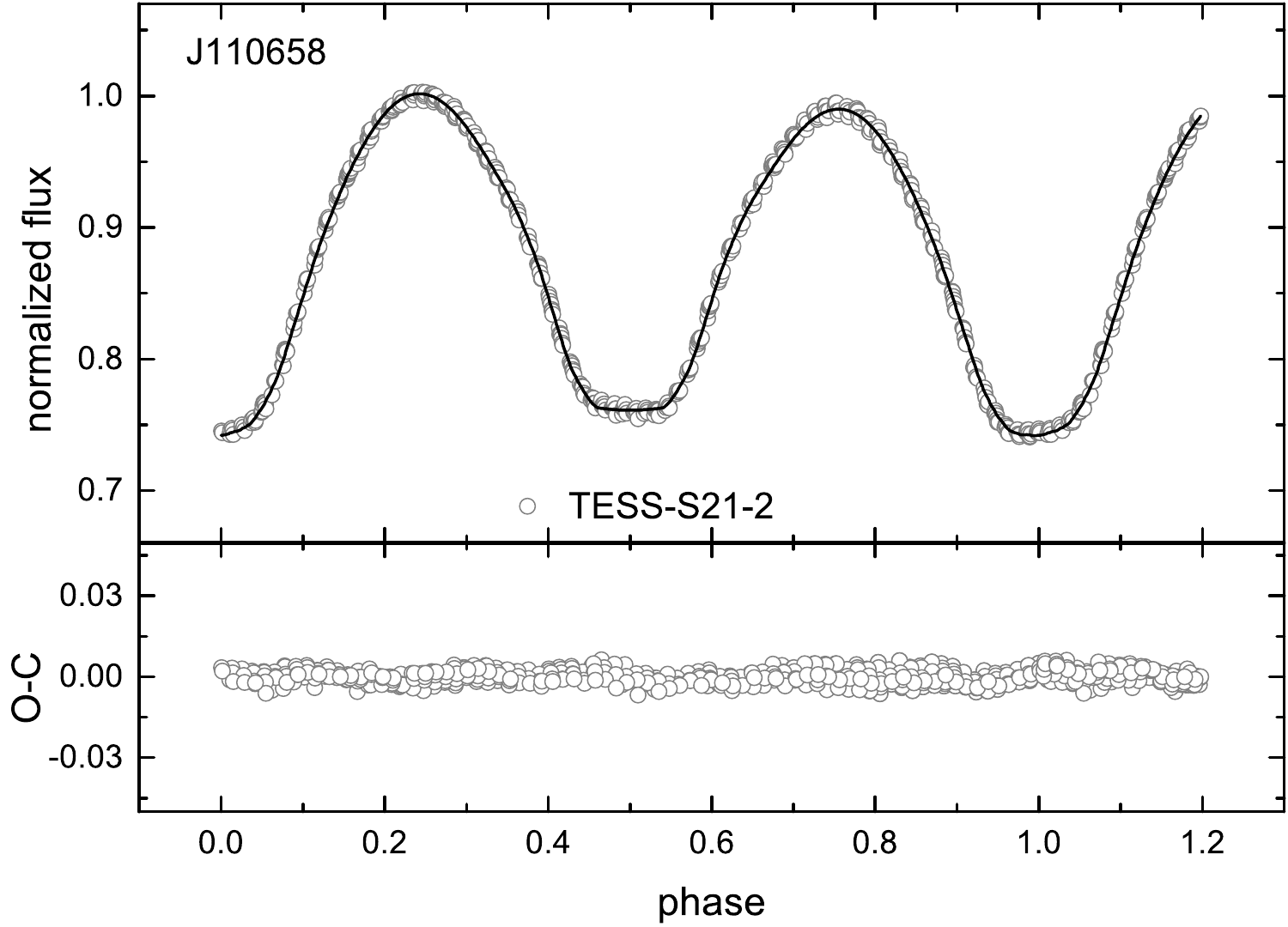}\\
\caption{The comparison between the synthetic and observed TESS light curves. The lower panels show the O-C residuals. \label{fig:tessfit}}
\end{figure}

All the ten systems were analyzed for the first time, no physical parameters have been obtained, especially the mass ratio. The generally used $q-$search method was performed for them. A series of photometric solutions with fixed $q$ from 0.05 to 5.0 (when $q<0.3$, the step size is 0.01, while 0.1 for $q>0.3$) were carried out. Because all the ten systems display typical W UMa type eclipsing binary light curves, mode 3 (contact mode) was adopted. During the process of $q-$search, the adjustable parameters are the orbital inclination $i$, the effective temperature of the secondary, $T_2$, the monochromatic luminosity of the primary, $L_1$, and the dimensionless potential of the primary ($\Omega_1=\Omega_2$). The relationship between the mean residual and mass ratio $q$ is shown in Figure \ref{fig:q-search}. It is clearly seen that all the $q-$search curves have very deep and sharp minima, indicating very precise mass ratio can be obtained (e.g., \citealt{2017MNRAS.466.1118Z,2019MNRAS.485.4588L,2021AJ....162...13L,2022Galax..10....8T}). The mass ratio corresponding to the minimum of the mean residual was chosen as the initial value and an adjustable parameter, new solutions were started to obtain a convergent result, then the errors of mass ratio can be determined. Some of the systems show very obvious O'Connell effect \cite{1951PRCO....2...85O}, meaning two unequal height maxima. Spot model was applied to these systems. We should declare that the latitude of the spot was fixed at $\pi/2$. After some iterations, the convergent solutions were obtained. Because the LAMOST spectrum is a mixture of the primary and secondary components, the temperature determined by the spectrum is a integral temperature of the two components. To solve this issue, the blackbody radiation was assumed for  the binaries, and the following equations were used to derive the individual temperatures \citep{2003A&A...404..333Z,2013AJ....146..157C},
\begin{eqnarray}
T_1&=&(((1+k^2)T_{eff}^4)/(1+k^2(T_2/T_1)^4))^{0.25},  \\\nonumber
T_2&=&T_1(T_2/T_1),
\end{eqnarray}
where $T_{eff}$ is the LAMOST temperature, $T_2/T_1$ and $k$ are the temperature and the radius ratio, respectively. Then, the derived temperatures were set as initial values, and the primary temperature was set as a fixed parameter while the secondary one was set as an adjustable parameter. Then, new solutions were carried out until convergent one was obtained. The results are listed in Table \ref{t:Parameters}, and the comparison between the synthetic and observed light curves are displayed in Figure \ref{fig:lcfit}. The TNT light curves of J071924 show a flare-like event near phase 0.8.
For J071924, the photometric elements determined by the TNT light curves were selected as the final results, and those derived by the XL85 light curves were adopted as the final results for J110658. The Transiting Exoplanet Survey Satellite (TESS) \citep{2015JATIS...1a4003R} has observed some of our targets. Sector 17 has observed J001422; Sector 18 has observed J022733; Sector 19 has observed J042640 and J054950; Sector 20 has observed J071924; Sector 21 has observed J093458 and J110658. All these data have very high precision and can be used for the light-curve investigation. Though the seven targets all have only one sector observation, some of them exhibit light curve variations. If the light curves were unchanged, the TESS data of one sector were analyzed together, while the light curves showing variations were divided into two parts for analyzing. During the analysis, the physical parameters determined by our observations were adopted as the initial parameters and adjusted. We should specify that although 2013 version of the W-D code does not include TESS filter, we still use the TESS band (No. 95) in the W-D analysis by downloading the TESS band limb-darkening coefficients through the personal website of Prof. Van Hamme\footnote{\url{https://faculty.fiu.edu/~vanhamme/lcdc2015/}}. Because the observational cadence of the TESS light curves is 30 minutes, the phase smearing effect should be considered during the investigations \citep{2021AJ....162...13L}. The derived photometric results were listed in Table \ref{tab:TessParameters}, and the corresponding theoretical light curves are shown in Figure \ref{fig:tessfit}. From Table \ref{tab:TessParameters}, we can find that the physical parameters determined by the TESS light curves are similar to those determined by our observed light curves. We should declare that the errors of these parameters listed in Tables \ref{t:Parameters} and \ref{tab:TessParameters} are determined by mathematical fitting by the W-D code. They are very small and surely artificial, espically for the mass ratio.

\renewcommand\arraystretch{1.3}
\begin{table*}
\centering
\tiny
\caption{The photometric results of the ten targets}
\label{t:Parameters}
\setlength{\tabcolsep}{0.6mm}{
\begin{tabular}{lcccccccccccc}
\hline\hline
Star & J001422& J022733& J042640 & J054950&\multicolumn2c{J071924}&  J093458 &\multicolumn2c{J110658} &  J115742&  J153433&  J233332\\
	 & & & & & TNT& NEXT&& NEXT& XL85 & & & \\\hline
	
$T_1(K)$            &6590$\pm13$      &6494$\pm39$      &6029$\pm93$      &7037$\pm37$      &5894$\pm32$      &5892$\pm35$      &5822$\pm96$      &6224$\pm40$      &6231$\pm40$      &5963$\pm65$      &6079$\pm370$     &6770$\pm39$      \\
$q(M_2/M_1) $       &0.142$\pm0.001$  &0.120$\pm0.001$  &0.128$\pm0.001$  &0.121$\pm0.001$  &0.099$\pm0.001$  &0.099$\pm0.001$  &0.128$\pm0.001$  &0.119$\pm0.001$  &0.117$\pm0.001$  &0.110$\pm0.001$  &0.096$\pm0.001$  &0.096$\pm0.001$  \\
$T_2(K)$            &6463$\pm24$      &6374$\pm52$      &6071$\pm116$     &6642$\pm55$      &6059$\pm48$      &6075$\pm56$      &5810$\pm132$     &6291$\pm60$      &6249$\pm60$      &6202$\pm107$     &6060$\pm430$     &6379$\pm72$      \\
$i(deg)$            &77.8$\pm0.1$     &74.6$\pm0.2$     &87.5$\pm0.2$     &73.1$\pm0.2$     &85.5$\pm0.2$     &87.4$\pm0.1$     &85.5$\pm0.2$     &86.4$\pm0.2$     &82.2$\pm0.2$     &78.0$\pm0.4$     &78.8$\pm0.2$     &82.5$\pm0.3$     \\
$\Omega_1=\Omega_2$ &1.994$\pm0.001$  &1.973$\pm0.003$  &1.994$\pm0.001$  &2.006$\pm0.002$  &1.905$\pm0.001$  &1.921$\pm0.001$  &2.005$\pm0.003$  &1.952$\pm0.001$  &1.941$\pm0.001$  &1.945$\pm0.004$  &1.912$\pm0.001$  &1.925$\pm0.002$  \\
$L_{2B}/L_{1B}$     &$-$              &0.146$\pm0.001$  &0.179$\pm0.001$  &$-$              &0.171$\pm0.001$  &0.166$\pm0.001$  &0.167$\pm0.001$  &0.180$\pm0.001$  &0.175$\pm0.001$  &0.193$\pm0.001$  &$-$              &$-$              \\
$L_{2V}/L_{1V}$     &0.188$\pm0.001$  &0.150$\pm0.001$  &0.177$\pm0.001$  &0.120$\pm0.001$  &0.163$\pm0.001$  &0.158$\pm0.001$  &0.167$\pm0.001$  &0.176$\pm0.001$  &0.173$\pm0.001$  &0.181$\pm0.001$  &$-$              &$-$              \\
$L_{2R_c}/L_{1R_c}$ &0.190$\pm0.001$  &0.151$\pm0.001$  &0.176$\pm0.001$  &0.125$\pm0.001$  &0.160$\pm0.001$  &0.154$\pm0.001$  &0.168$\pm0.001$  &0.175$\pm0.001$  &0.173$\pm0.001$  &0.176$\pm0.001$  &$-$              &$-$              \\
$L_{2I_c}/L_{1I_c}$ &0.192$\pm0.001$  &0.153$\pm0.001$  &0.175$\pm0.001$  &0.130$\pm0.001$  &0.157$\pm0.001$  &0.151$\pm0.001$  &0.168$\pm0.001$  &0.173$\pm0.001$  &0.172$\pm0.001$  &0.171$\pm0.001$  &$-$              &$-$              \\
$L_{2g}/L_{1g}$     & $-$             & $-$             & $-$             & $-$             & $-$             & $-$             & $-$             & $-$             & $-$             & $-$             & 0.131$\pm0.001$ & 0.096$\pm0.001$ \\
$L_{2r}/L_{1r}$     & $-$             & $-$             & $-$             & $-$             & $-$             & $-$             & $-$             & $-$             & $-$             & $-$             & 0.132$\pm0.001$ & 0.104$\pm0.001$ \\
$L_{2i}/L_{1i}$     & $-$             & $-$             & $-$             & $-$             & $-$             & $-$             & $-$             & $-$             & $-$             & $-$             & 0.132$\pm0.001$ & 0.108$\pm0.001$ \\
$r_1$               & 0.588$\pm0.001$ &0.585$\pm0.001$  &0.580$\pm0.001$  &0.572$\pm0.001$  &0.604$\pm0.001$  &0.597$\pm0.001$  &0.576$\pm0.001$  &0.593$\pm0.001$  &0.598$\pm0.001$  &0.592$\pm0.001$  &0.599$\pm0.001$  &0.594$\pm0.001$  \\
$r_2$               & 0.272$\pm0.002$ &0.240$\pm0.006$  &0.244$\pm0.001$  &0.225$\pm0.003$  &0.234$\pm0.003$  &0.225$\pm0.003$  &0.239$\pm0.006$  &0.249$\pm0.002$  &0.253$\pm0.002$  &0.234$\pm0.009$  &0.223$\pm0.003$  &0.216$\pm0.007$  \\
$f$                 & 96.3$\pm0.4$\%  &58.3$\pm3.7$\%   &57.3$\pm0.8$\%   &20.3$\pm2.4$\%   &80.3$\pm0.6$\%   &55.3$\pm2.1$\%   &45.2$\pm4.0$\%   &81.2$\pm0.5$\%   &92.6$\pm0.5$\%   &61.0$\pm6.0$\%   &56.1$\pm1.8$\%   &37.4$\pm3.6$\%   \\
Spot                & $-$             &$-$              & star 2          &$-$              & star 2          & star 2          &$-$              & star 1          & star 1          & star 1          & star 1          &$-$              \\
$\theta(deg)$       & $-$             &$-$              &90               &$-$              &90               &90               &$-$              &90               &90               &90               &90               &$-$              \\
$\lambda(deg)$      & $-$             &$-$              &$290\pm35$       &$-$              &$282\pm9$        &$261\pm20$       &$-$              &$282\pm5$        &$278\pm3$        &$287\pm12$       &$356\pm15$       &$-$              \\
$r_s(deg)$          & $-$             &$-$              &21$\pm3$         &$-$              &23$\pm1$         &32$\pm2$         &$-$              &8$\pm1$          &9$\pm1$          &14$\pm1$         &12$\pm1$         &$-$              \\
$T_s$               & $-$             &$-$              &0.820$\pm0.009$  &$-$              &0.857$\pm0.007$  &0.807$\pm0.011$  &$-$              &0.851$\pm0.005$  &0.787$\pm0.004$  &0.775$\pm0.008$  &0.820$\pm0.009$  &$-$              \\
\hline
\end{tabular}}
\end{table*}

\setlength{\tabcolsep}{0.6mm}{
\begin{deluxetable*}{lccccccccc}
\tablecaption{The photometric results by analyzing the TESS light curves\label{tab:TessParameters}}
\tablewidth{0pt}
\tablehead{
\colhead{Star} &\colhead{J001422}&\colhead{J022733}&\multicolumn2c{J042640}&\colhead{J054950}&\colhead{J071924}& \colhead{J093458} &\multicolumn2c{J110658} \\
\nocolhead{Star} &\colhead{S17}&\colhead{S18}&\colhead{S19-1}&\colhead{S19-2}&\colhead{S19}&\colhead{S20}&\colhead{S21} &\colhead{S21-1}&\colhead{S21-2} \\
}
\startdata
$T_1(K)$            &6590             &6494             &6029             &6029             &7037             &5894             &5822             &6231             &6231             \\
$q(M_2/M_1) $       &0.146$\pm0.001$  &0.119$\pm0.001$  &0.136$\pm0.001$  &0.135$\pm0.001$  &0.123$\pm0.001$  &0.099$\pm0.001$  &0.138$\pm0.001$  &0.111$\pm0.001$  &0.115$\pm0.001$  \\
$T_2(K)$            &6532$\pm6$       &6328$\pm7$       &6051$\pm8$       &6060$\pm6$       &6661$\pm6$       &6363$\pm6$       &6197$\pm7$       &6330$\pm5$       &6357$\pm6$       \\
$i(deg)$            &77.2$\pm0.2$     &74.2$\pm0.1$     &85.9$\pm0.2$     &87.6$\pm0.1$     &73.3$\pm0.1$     &85.6$\pm0.1$     &75.4$\pm0.1$     &81.4$\pm0.1$     &82.1$\pm0.1$     \\
$\Omega_1=\Omega_2$ &1.999$\pm0.001$  &1.963$\pm0.003$  &1.978$\pm0.001$  &1.987$\pm0.001$  &2.005$\pm0.002$  &1.900$\pm0.001$  &2.017$\pm0.002$  &1.922$\pm0.001$  &1.941$\pm0.001$  \\
$L_{2T}/L_{1T}$     &0.205$\pm0.001$  &0.152$\pm0.001$  &0.198$\pm0.001$  &0.192$\pm0.001$  &0.133$\pm0.001$  &0.188$\pm0.001$  &0.209$\pm0.001$  &0.174$\pm0.001$  &0.175$\pm0.001$  \\
$r_1$               & $0.588\pm0.001$ &$0.589\pm0.001$  &$0.591\pm0.001$  &$0.587\pm0.001$  &$0.576\pm0.001$  &$0.606\pm0.001$  &$0.578\pm0.001$  &$0.603\pm0.001$  &$0.598\pm0.001$  \\
$r_2$               & $0.276\pm0.002$ &$0.244\pm0.004$  &$0.268\pm0.001$  &$0.261\pm0.001$  &$0.230\pm0.001$  &$0.237\pm0.001$  &$0.264\pm0.001$  &$0.252\pm0.001$  &$0.253\pm0.001$  \\
$f$                 & 99.7$\pm0.4$\%  &69.4$\pm2.0$\%   &97.3$\pm0.6$\%   &84.6$\pm0.4$\%   &26.8$\pm1.3$\%   &87.5$\pm0.4$\%   &59.4$\pm1.7$\%   &98.7$\pm0.4$\%   &83.9$\pm0.3$\%   \\
Spot                & $-$             &$-$              &$-$              & star 2          &$-$              & star 2          &star 2         &$-$              & star 1          \\
$\theta(deg)$       & $-$             &$-$              &$-$              &90               &$-$              &90               &$90$             &$-$              &90               \\
$\lambda(deg)$      & $-$             &$-$              &$-$              &$294\pm2$        &$-$              &$205\pm2$        &$169\pm1$        &$-$              &$229\pm4$        \\
$r_s(deg)$          & $-$             &$-$              &$-$              &25$\pm1$         &$-$              &21$\pm1$         &22$\pm1$         &$-$              &24$\pm1$         \\
$T_s$               & $-$             &$-$              &$-$              &1.170$\pm0.003$  &$-$              &0.844$\pm0.004$  &0.711$\pm0.010$  &$-$              &0.831$\pm0.005$  \\
\enddata
\end{deluxetable*}}

\section{O-C analysis} \label{subsec:O-C}
\begin{figure}
\epsscale{0.4}
\plotone{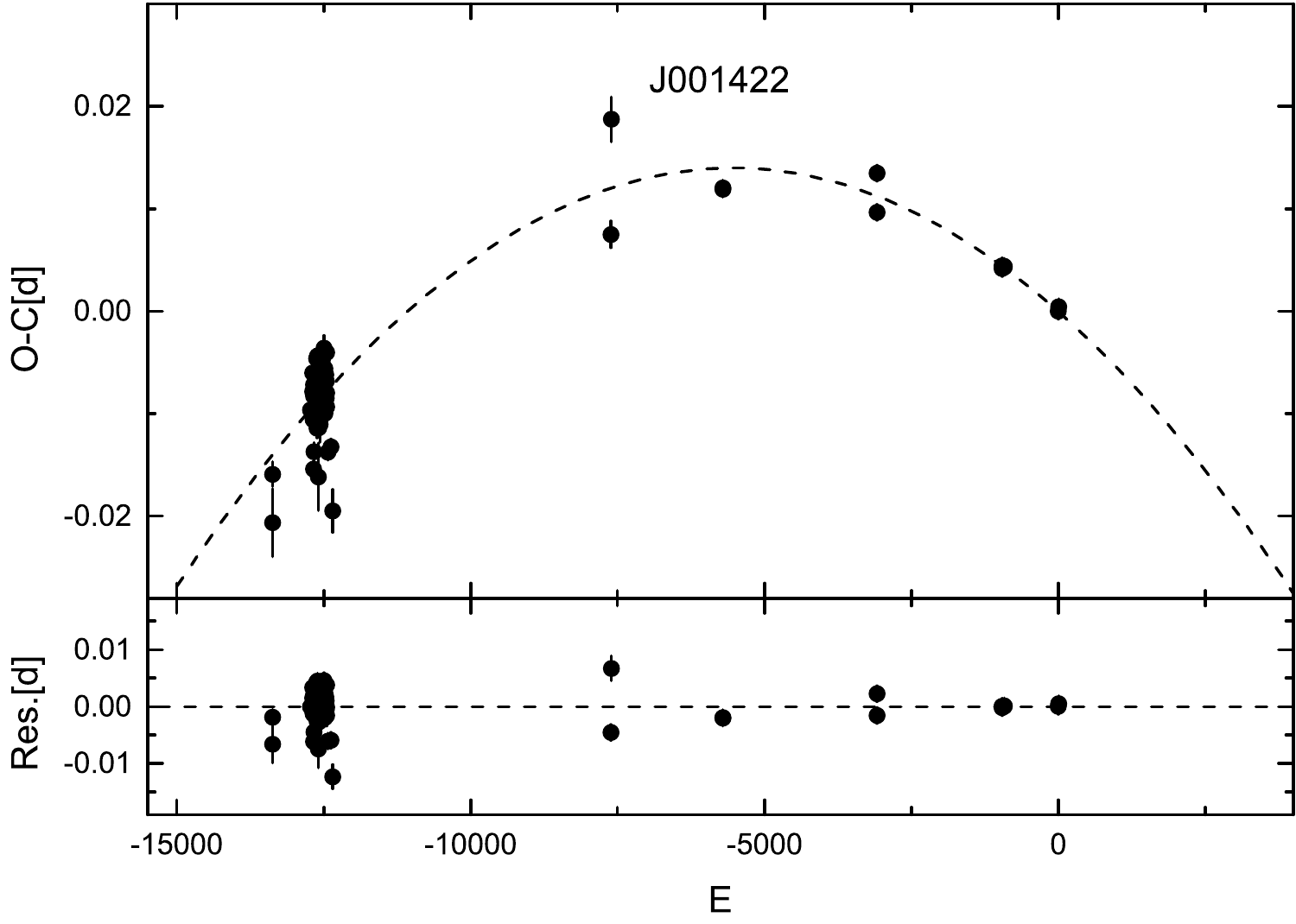}
\plotone{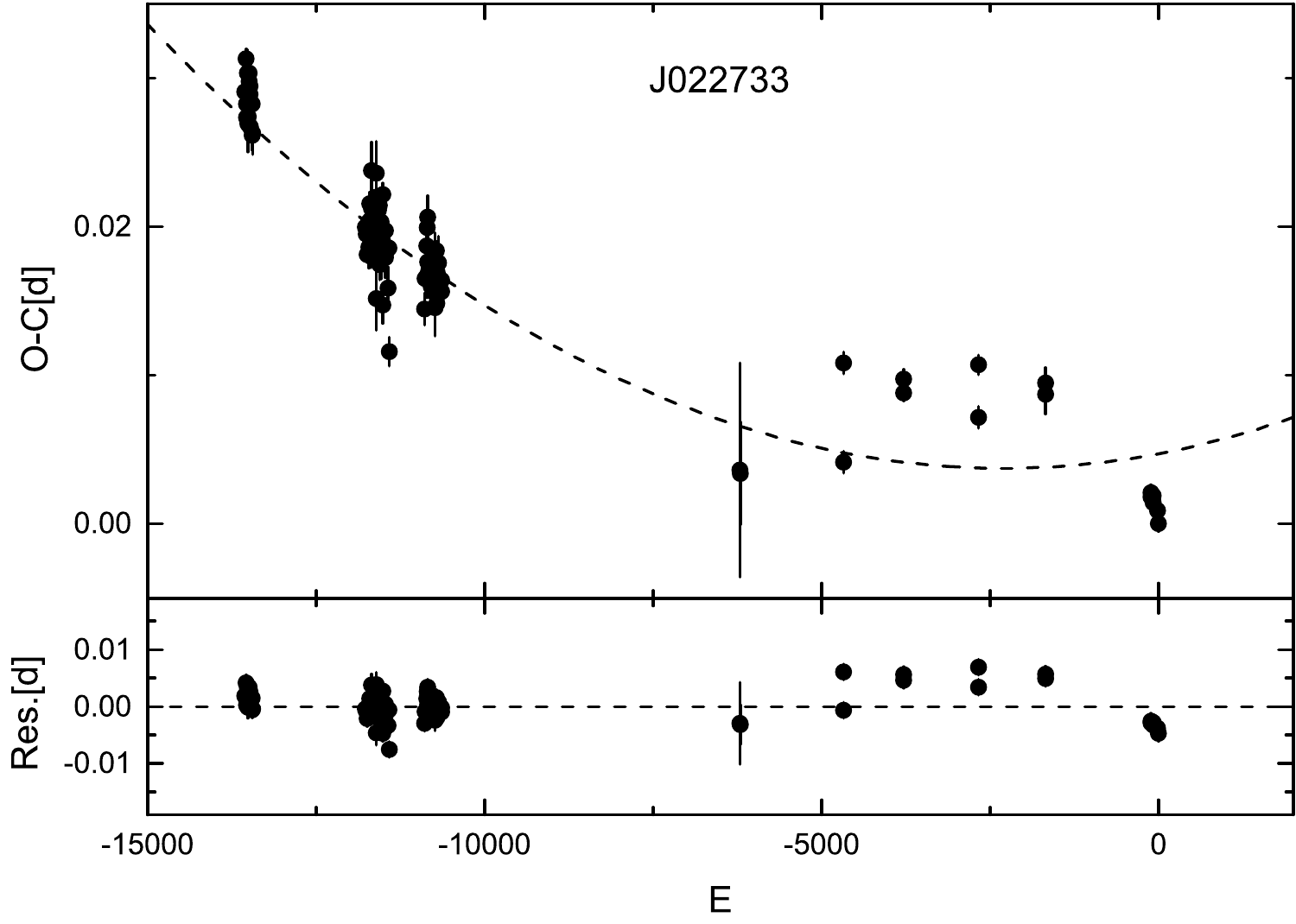}\\
\plotone{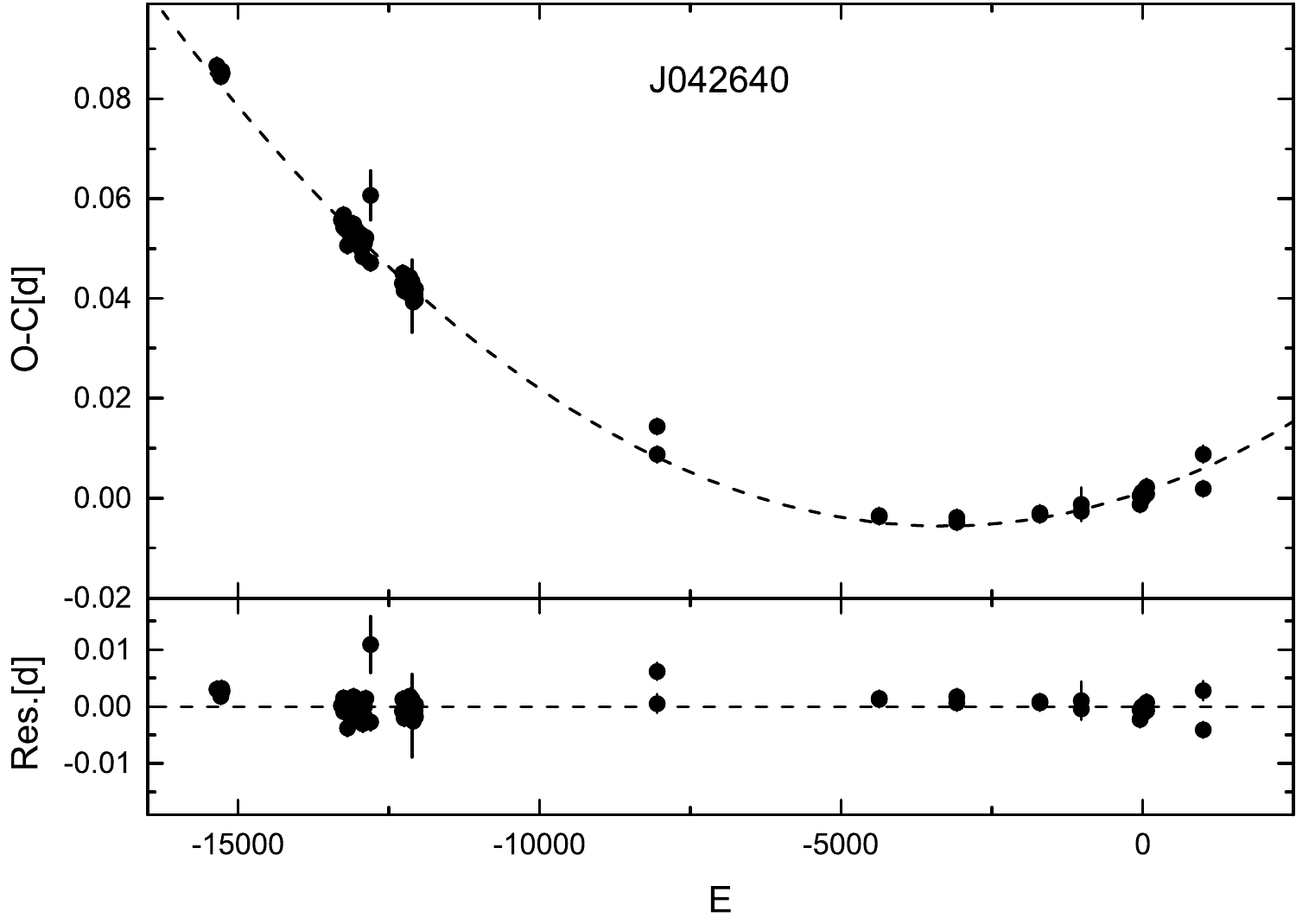}
\plotone{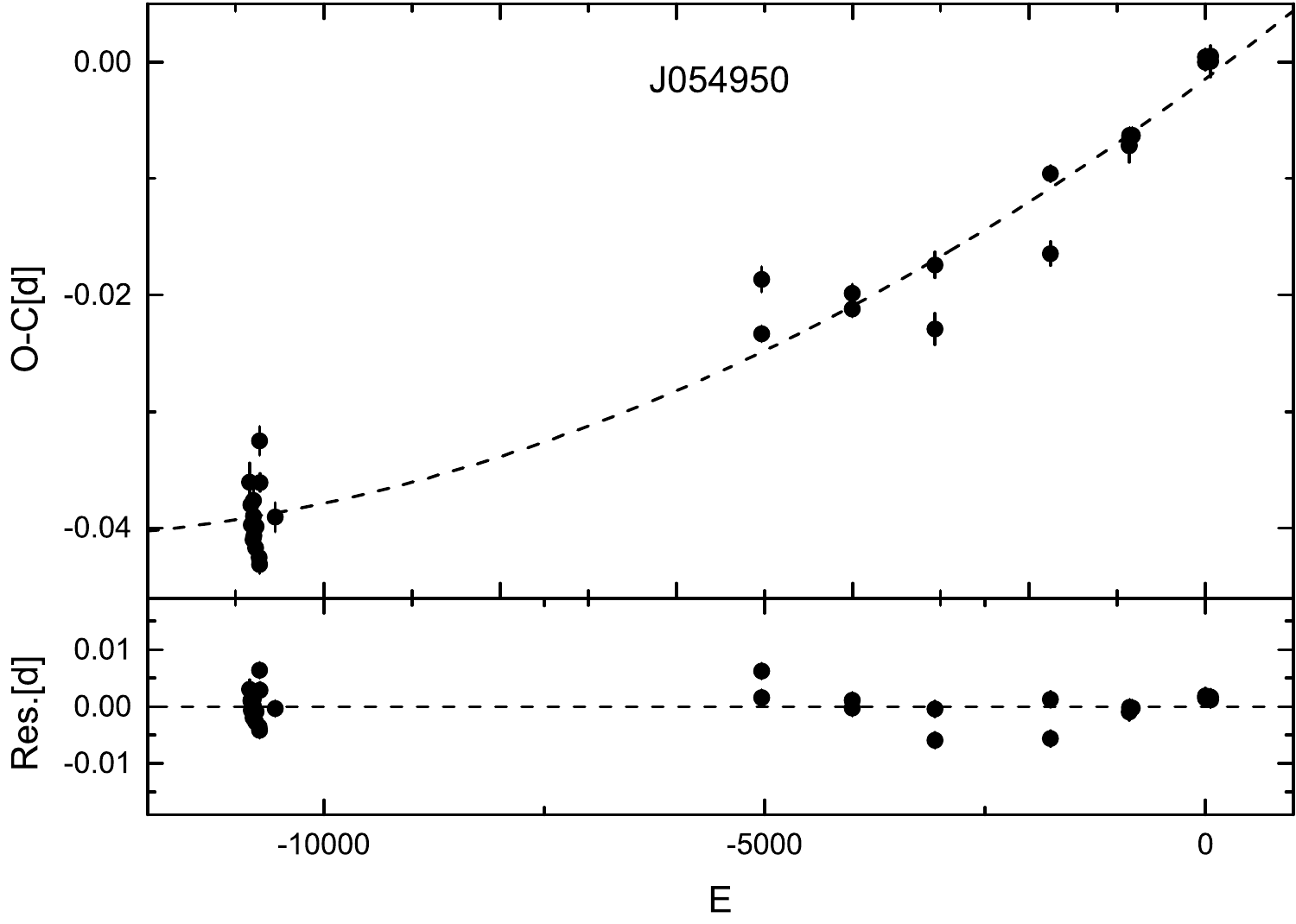}\\
\plotone{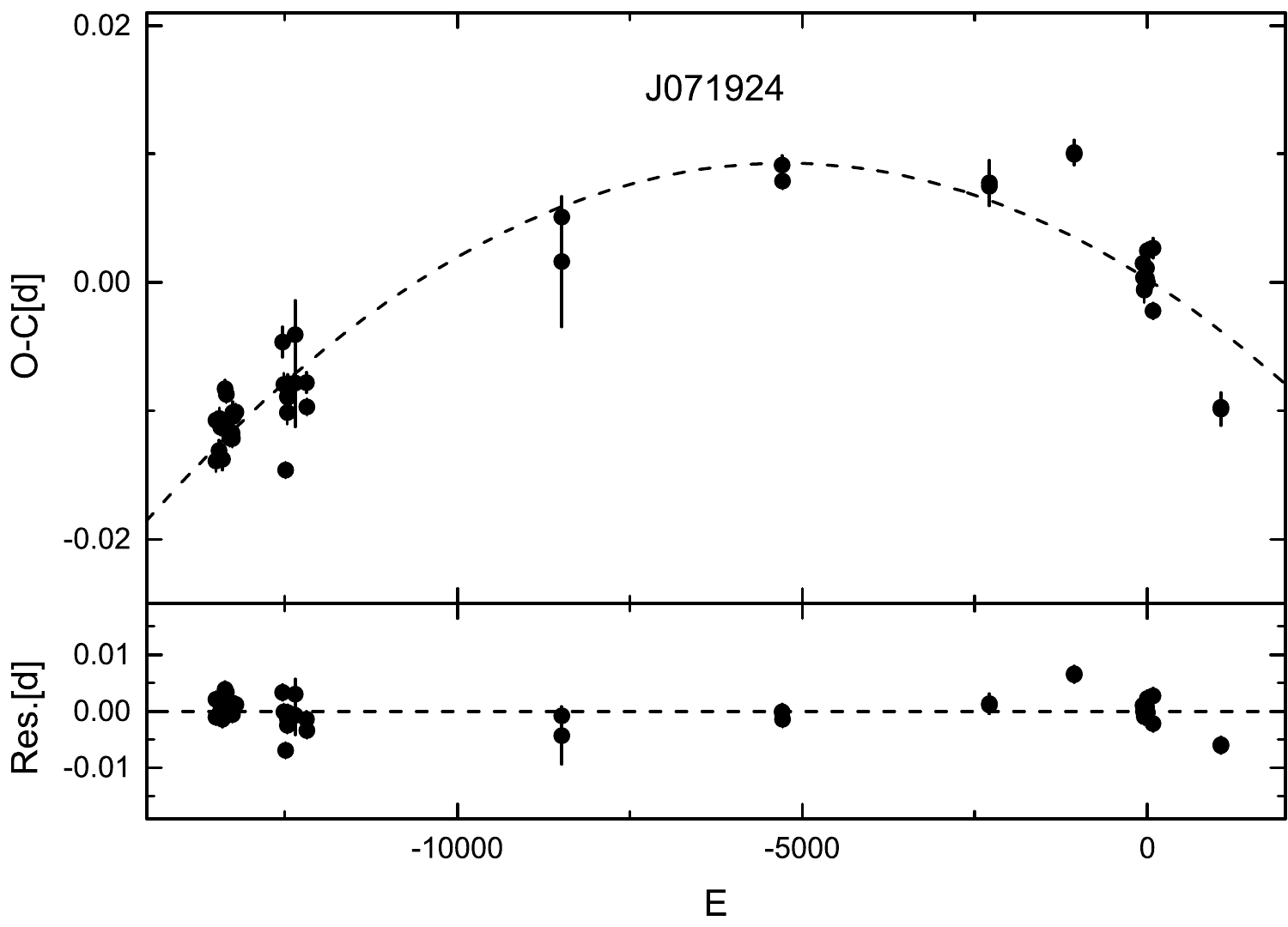}
\plotone{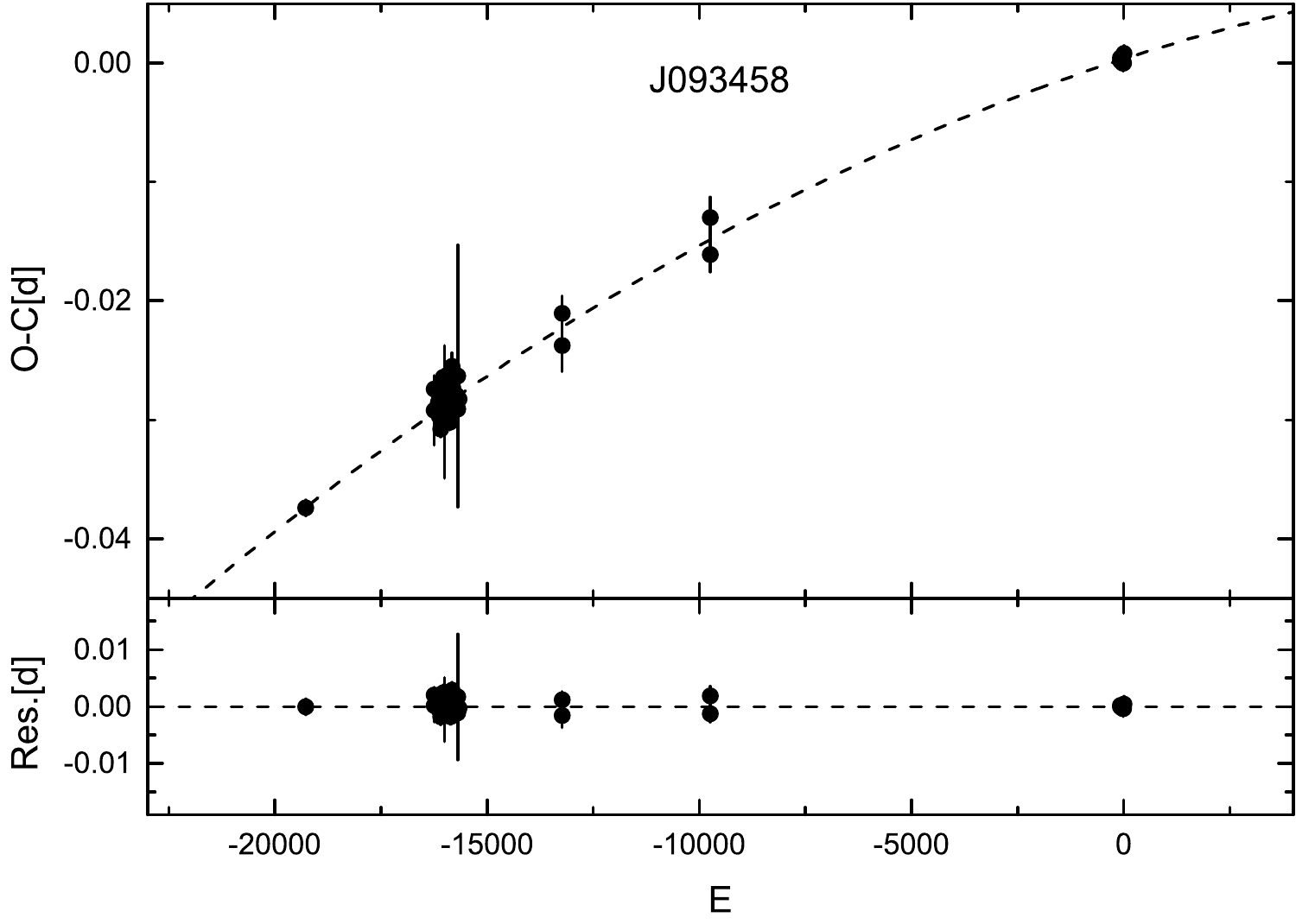}\\
\plotone{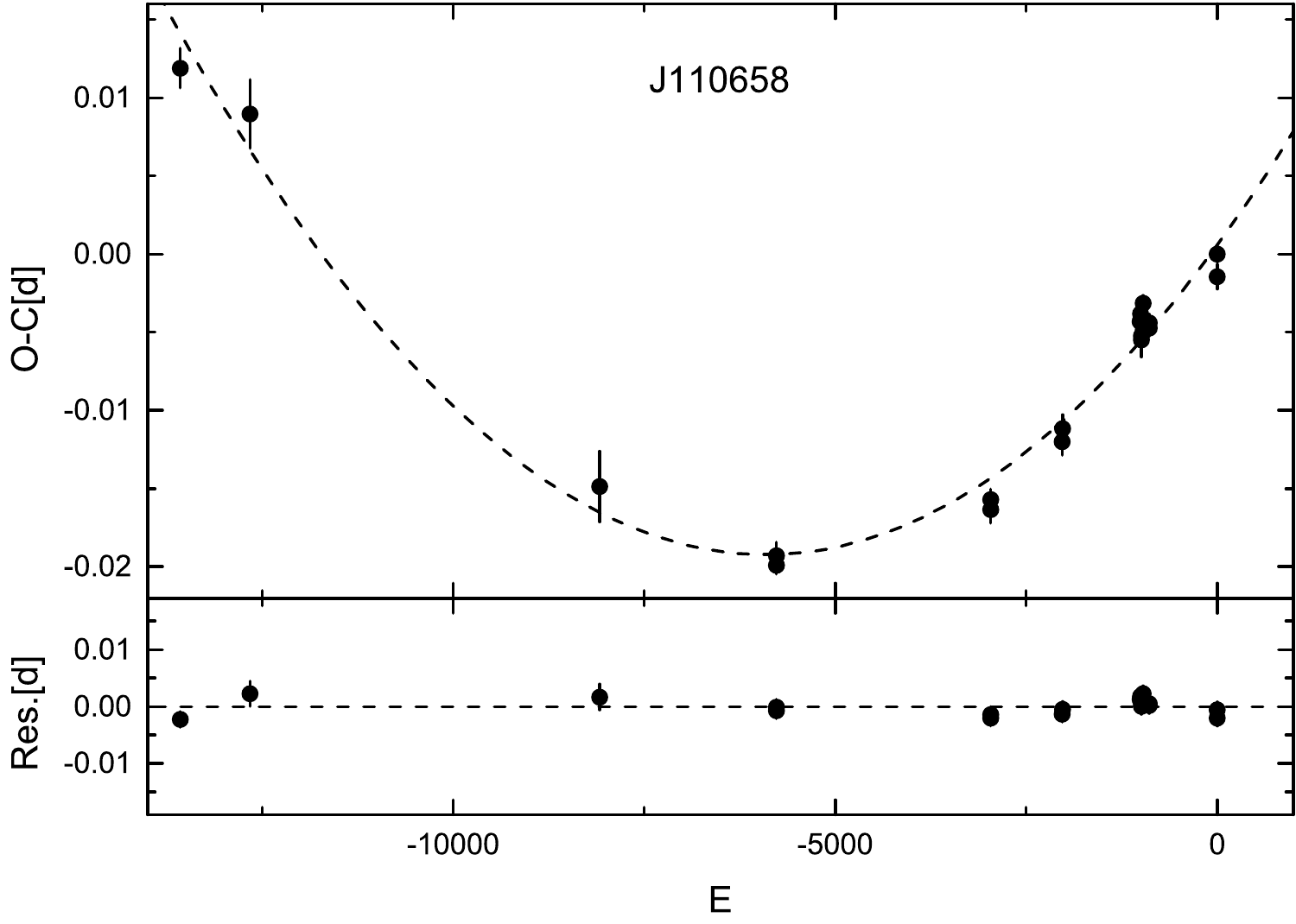}
\plotone{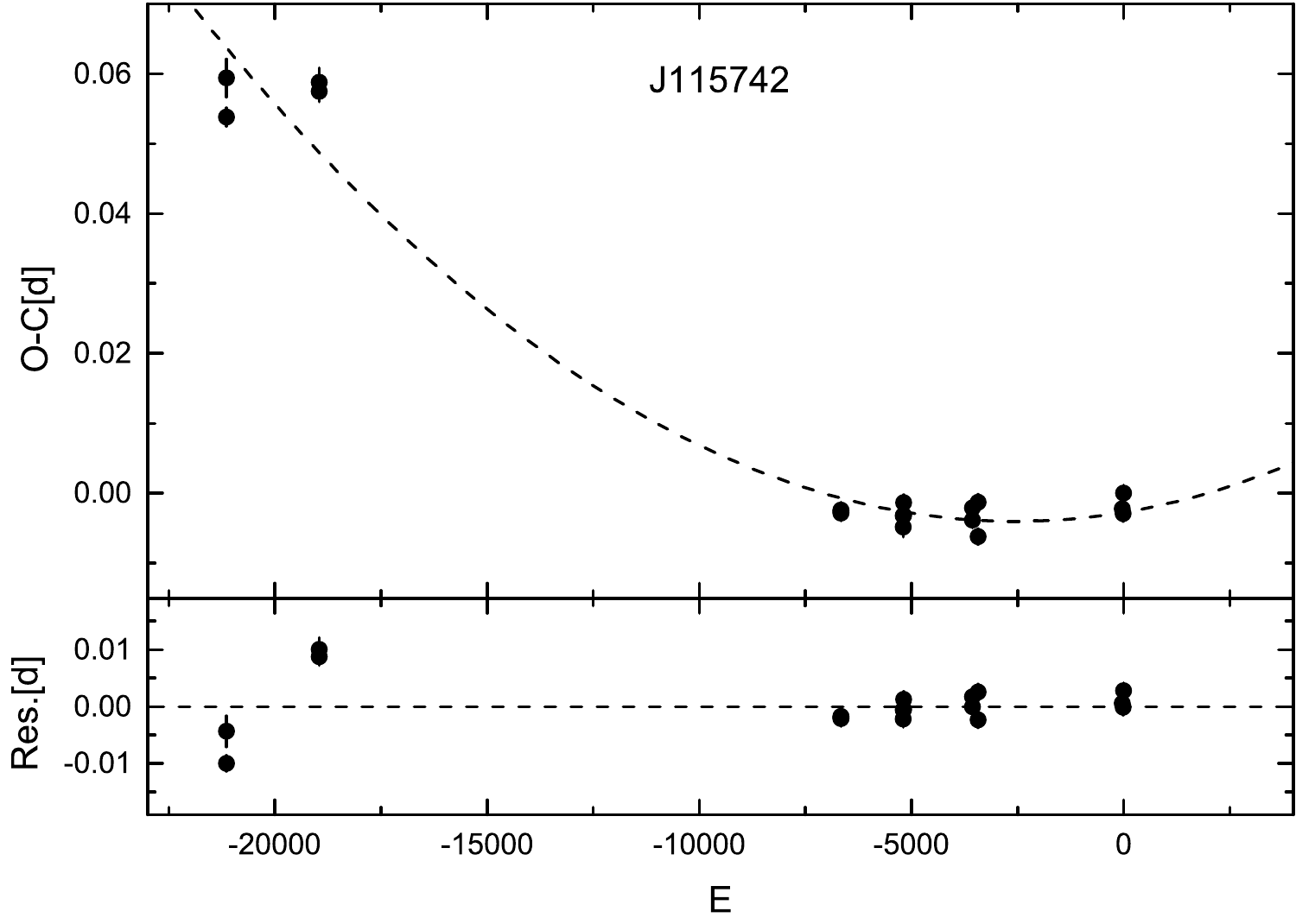}\\
\plotone{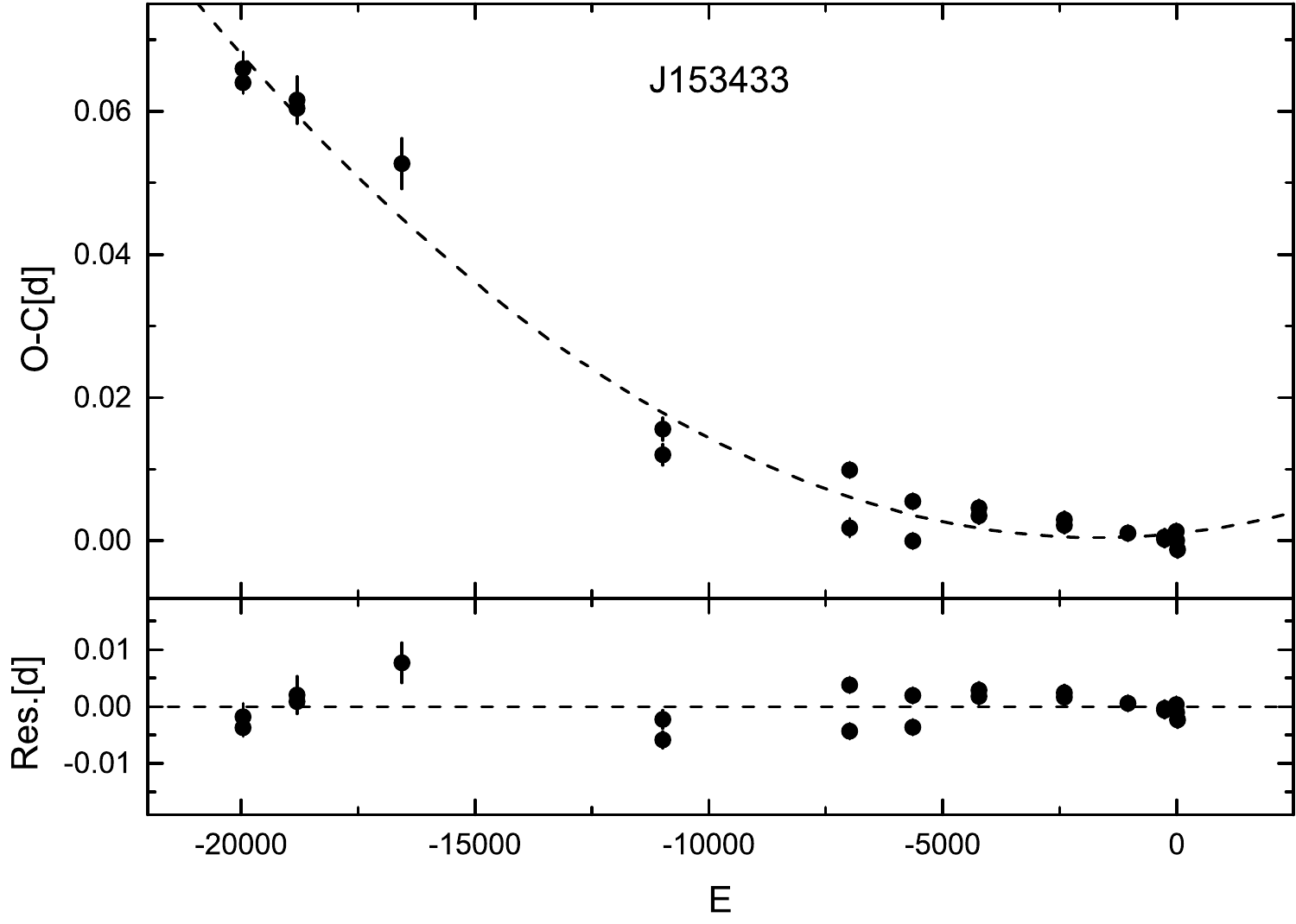}
\plotone{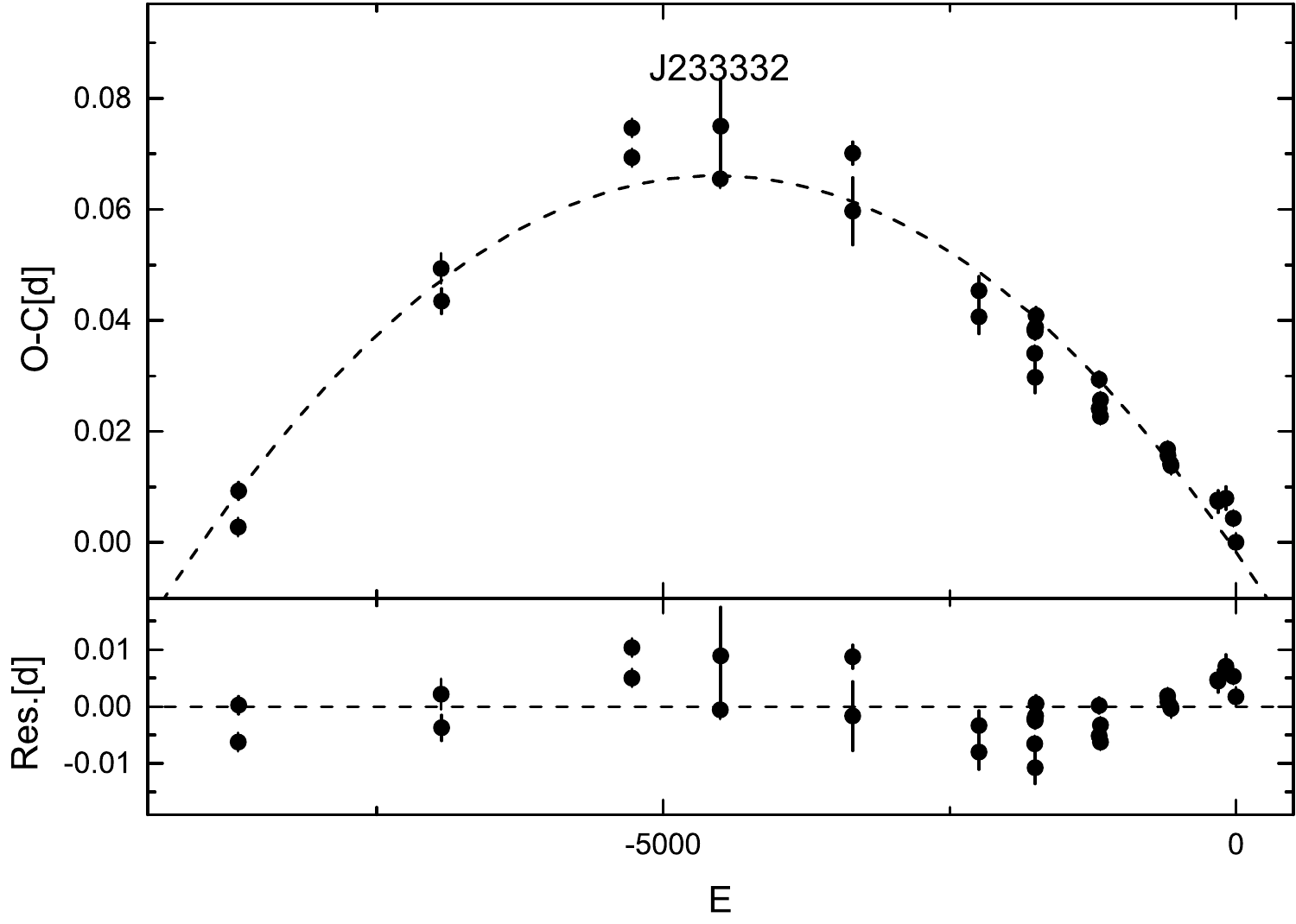}
\caption{The O-C diagrams of the ten binariess.  \label{fig:oc}}
\end{figure}

The O-C (observed eclipsing minimum minus calculated eclipsing minimum) analysis is a very powerful tool to study the evolution and search for potential additional companions of contact binaries. Therefore, we try to collect as many eclipsing times as possible to construct the O-C diagrams for the ten contact binaries. Super Wide Angle Search for Planets (SuperWASP, \citealt{2010A&A...520L..10B}) provides very high time resolution and continuous observations and has observed some of our targets, then we can calculate the eclipsing times directly from the SuperWASP data for these targets using the K-W method \citep{1956BAN....12..327K}. The All Sky Automated Survey (ASAS, \citealt{1997AcA....47..467P}), the Catalina Sky Surveys (CSS, \citealt{2009ApJ...696..870D}), the All-Sky Automated Survey for SuperNovae (ASAS-SN, \citealt{2014ApJ...788...48S,2017PASP..129j4502K}), and the Zwicky Transient Facility (ZTF) survey \citep{2019PASP..131a8002B,2019PASP..131a8003M} have observed some of our targets. However, these observations are very dispersed, one day one point or several days one point. The TESS light curves are consecutive observations, yet the observational cadence is 30 minutes. Therefore, we cannot calculate the eclipsing times directly for these surveys' data. Then, we used the phase shift method proposed by \cite{2020AJ....159..189L,2021ApJ...922..122L,2021AJ....162...13L} to compute the times of minimum light for these surveys' data. Because the observational time of TESS is BJD, we transferred all the other times of minimum light from HJD to BJD using the online procedure\footnote{\url{https://astroutils.astronomy.osu.edu/time/}}  provided by \cite{2010PASP..122..935E}. All the derived eclipsing times including those determined by our observations are listed in Table \ref{tab:minima}. Then, using the following linear ephemeris,
\begin{eqnarray}
T=T_0+P_0\times E,
\end{eqnarray}
we calculated the O-C values for the ten targets and tabulate them in Table \ref{tab:minima}. In this equation, $T$ is the eclipsing times, $T_0$ and $P_0$ are respectively the initial primary minimum and the orbital period which are listed in Table \ref{tab:information} ($HJD_0$ has been transferred to $BJD_0$ during the calculations), $E$ is the cycle number. The constructed O-C diagrams of the ten targets are displayed in Figure \ref{fig:oc}. As shown in this figure, very obvious long-term variation can be seen for all the ten stars. So, we used the following equation to fit their O-C diagrams,
\begin{eqnarray}
O-C= \Delta T_0 + \Delta P_0\times E+{\beta \over 2}\times E^2,
\end{eqnarray}
where $\Delta T_0$ is the correction for the initial epoch, $\Delta P_0$ is the correction for the orbital period, and $\beta$ is the long-term changing rate of the orbital period. These results are listed in Table \ref{tab:o-cresults} and the fitted curves are plotted in Figure \ref{fig:oc}. From Figure \ref{fig:oc}, we can see that the residuals are almost flat except for J022733, which may show more complex period changes, future observations are needed to confirm this. Four of the ten binaries exhibit long-term orbital period decrease, while the others show orbital period increase. The period changing rates for the ten binary systems are all in the order of 10$^{-7}$ d yr$^{-1}$. This result is consistent with other contact binaries (e.g., \citealt{2016AJ....151...67Z,2019AJ....157..207L,2022ApJ...927..183L}).

\setlength{\tabcolsep}{5mm}{
\begin{deluxetable*}{ccccccc}
\tablecaption{The eclipsing times of the ten targets\label{tab:minima}}
\tablewidth{0pt}
\tablehead{
\colhead{Star} & \colhead{BJD} & \colhead{Errors} & \colhead{E} & \colhead{O-C} & \colhead{Residuals} & \colhead{Source} \\
}
\startdata
J001422 & 2454059.7129 &  0.0033 &  -13376.5& -0.0206&  -0.0066   & (1)   \\
		& 2454059.9073 &  0.0012 &  -13376  & -0.0159&  -0.0019   & (1)   \\
		& 2454306.6465 &  0.0013 &  -12725.5& -0.0096&  -0.0001   & (2)   \\
		& 2454318.5962 &  0.0007 &  -12694  & -0.0079&  0.0015    & (2)   \\
		& 2454318.5980 &  0.0010 &  -12694  & -0.0060&  0.0034    & (2)   \\
		& 2454320.6800 &  0.0014 &  -12688.5& -0.0101&  -0.0008   & (2)   \\
		& 2454321.6278 &  0.0010 &  -12686  & -0.0106&  -0.0012   & (2)   \\
		& 2454322.5771 &  0.0011 &  -12683.5& -0.0096&  -0.0002   & (2)   \\
		& 2454324.6573 &  0.0006 &  -12678  & -0.0154&  -0.0062   & (2)   \\
		& 2454325.6104 &  0.0011 &  -12675.5& -0.0106&  -0.0013   & (2)   \\
		& 2454325.6138 &  0.0013 &  -12675.5& -0.0072&  0.0021    & (2)   \\
		& 2454326.5610 &  0.0010 &  -12673  & -0.0083&  0.0010    & (2)   \\
		& 2454327.6997 &  0.0008 &  -12670  & -0.0074&  0.0018    & (2)   \\
\enddata                                                             \tablecomments{(1) CSS; (2) SuperWASP; (3)  ASAS-SN; (4) TESS; (5) NEXT;
(6) ZTF; (7) XL85; (8) TNT; (9) ASAS; (10) WHOT. \\
(This table is available in its entirety in machine-readable form in the online version of this article.)}
\end{deluxetable*} }

\begin{deluxetable*}{crrrr}
\tablecaption{The orbital period change parameters of the ten binaries\label{tab:o-cresults}}
\tablewidth{0pt}
\tablehead{
\colhead{Star} & \colhead{$\Delta T_0$}&\colhead{ $\Delta P_0$}& \colhead{$\beta$}  & \colhead{dM$_1$/dt} \\
\nocolhead{Star} &\colhead{$\times10^{-4}$ d}&\colhead{$\times10^{-6}$ d}&\colhead{$\times10^{-7}$ d yr$^{-1}$}& \colhead{$\times10^{-7}$ $M_\odot$ yr$^{-1}$} \\
}
\startdata
J001422 & $ 0.40\pm0.01$ & $ 5.06\pm0.61$ & $-8.79\pm0.82$ & $-1.90\pm0.18$ \\
J022733 & $ 47.2\pm8.37$ & $ 0.85\pm0.27$ & $3.27 \pm0.34$ & $0.57 \pm0.06$ \\
J042640 & $ 12.4\pm5.01$ & $ 4.10\pm0.21$ & $12.4 \pm0.31$ & $2.48 \pm0.06$ \\
J054950 & $-14.7\pm10.4$ & $ 5.68\pm0.63$ & $3.35 \pm0.87$ & $0.57 \pm0.15$ \\
J071924 & $ 2.31\pm7.10$ & $-3.44\pm0.41$ & $-6.70\pm0.61$ & $-1.04\pm0.10$ \\
J093458 & $ 3.66\pm4.53$ & $ 1.15\pm0.16$ & $-1.02\pm0.23$ & $-0.23\pm0.05$ \\
J110658 & $ 5.51\pm6.04$ & $ 6.70\pm0.33$ & $10.3 \pm0.45$ & $1.78 \pm0.08$ \\
J115742 & $-27.6\pm24.6$ & $ 1.00\pm0.68$ & $4.99 \pm0.75$ & $0.99 \pm0.15$ \\
J153433 & $ 10.2\pm13.1$ & $ 0.68\pm0.40$ & $4.46 \pm0.43$ & $0.70 \pm0.07$ \\
J233332 & $ 17.6\pm18.4$ & $-29.8\pm1.33$ & $-3.82\pm1.84$ & $-4.37\pm0.21$ \\
\enddata
\end{deluxetable*}

\section{LAMOST spectra investigation} \label{subsec:LAMOST}
\begin{figure}
\epsscale{0.18}
\plotone{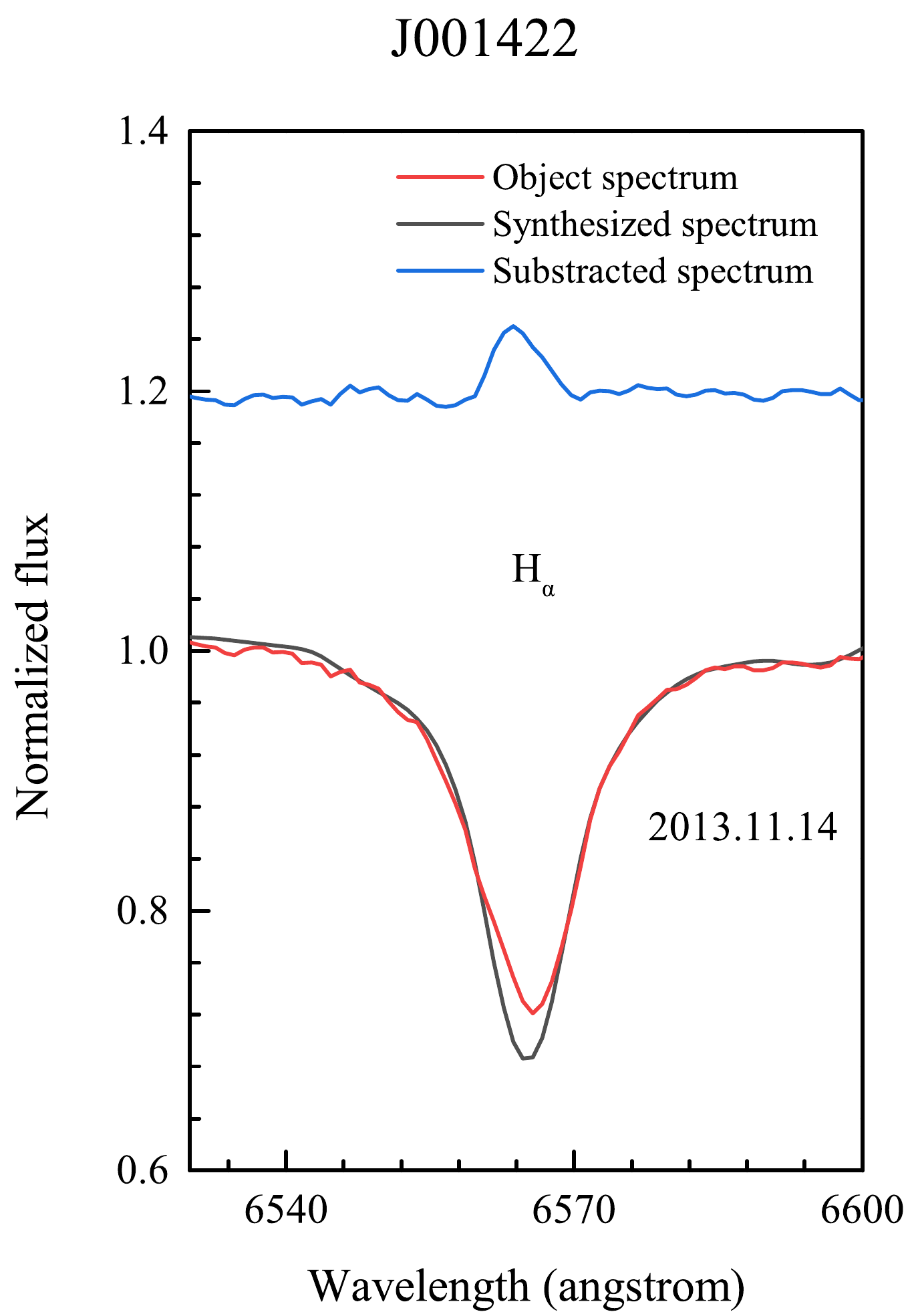}
\plotone{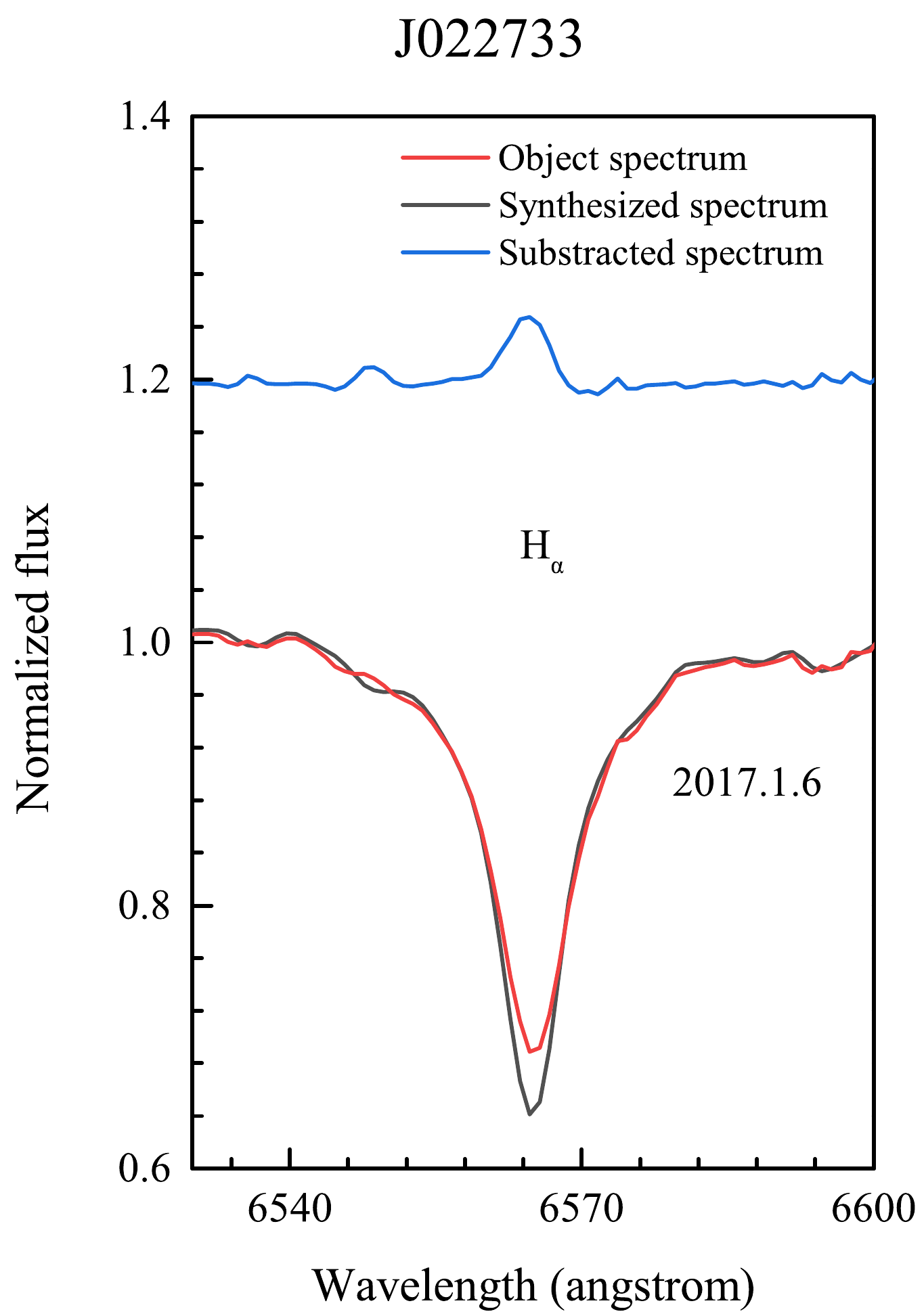}
\plotone{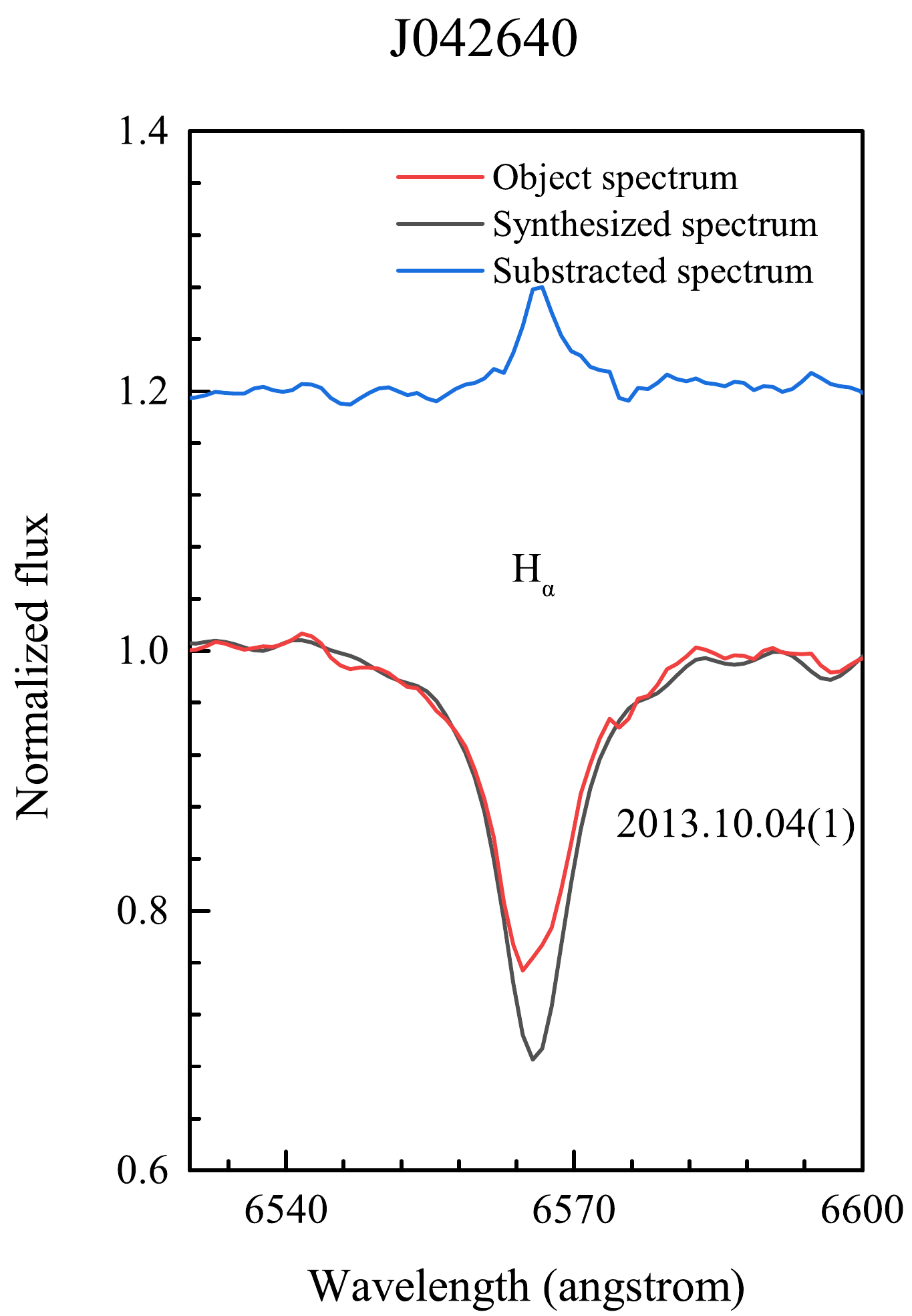}
\plotone{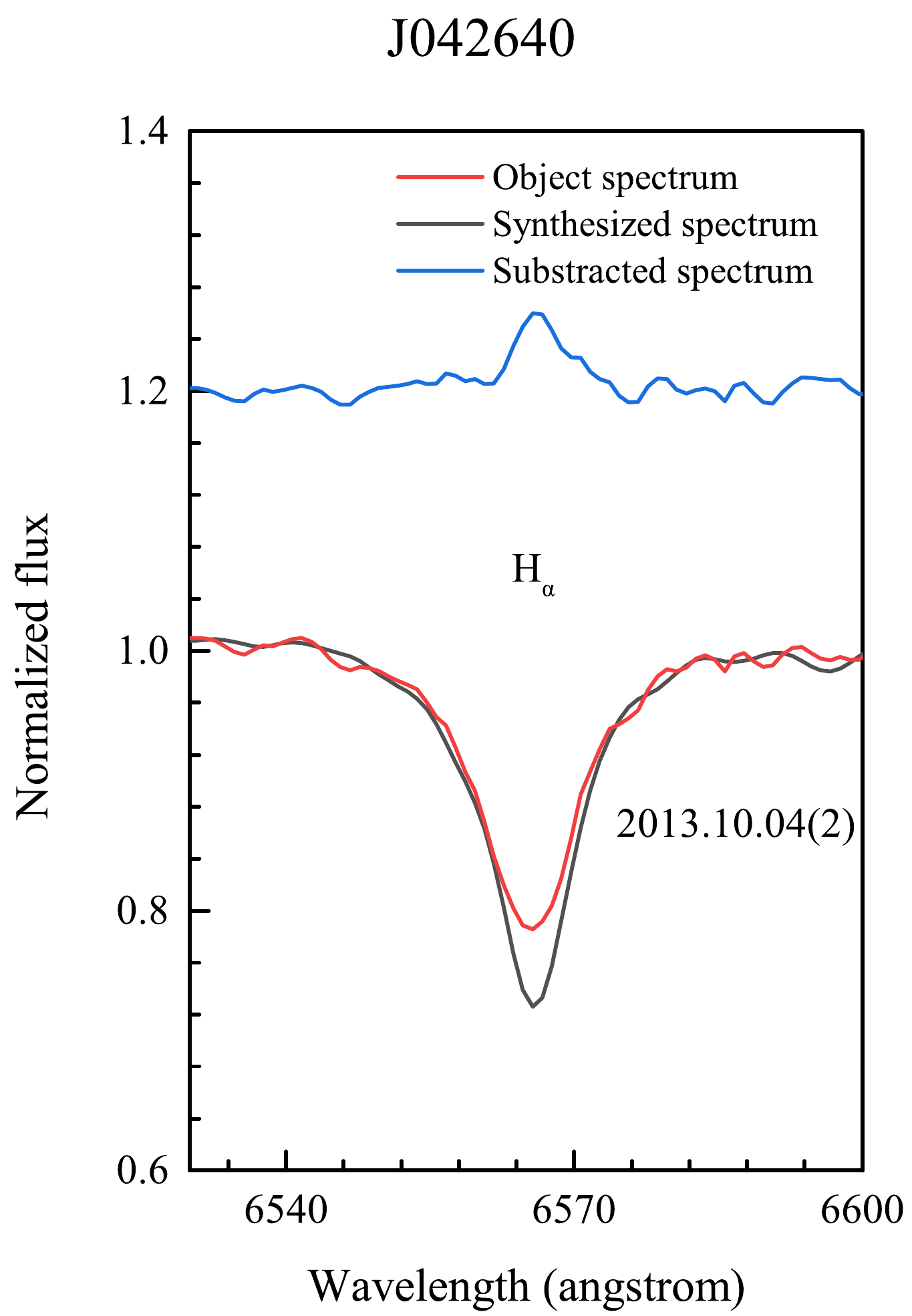}
\plotone{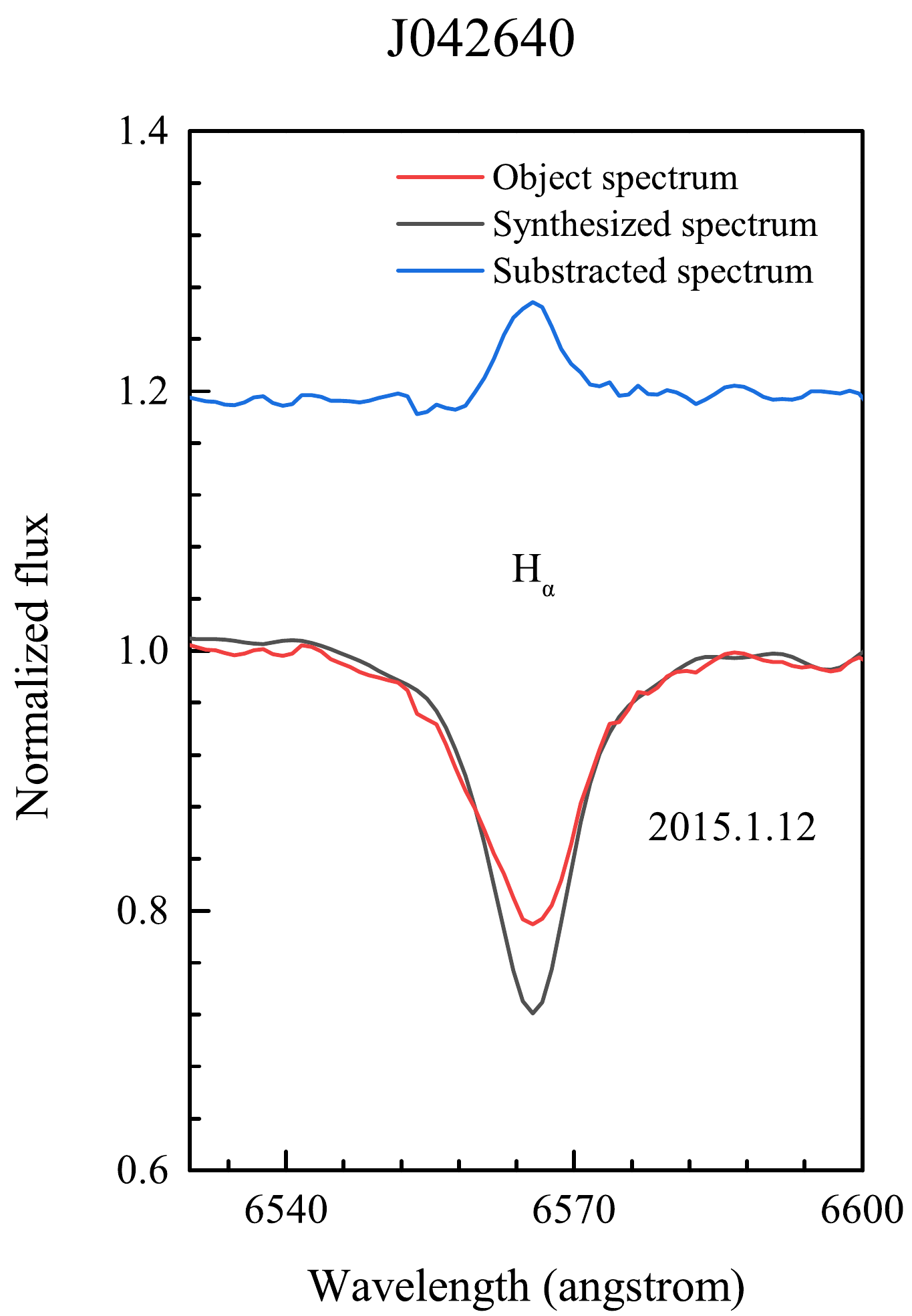}
\plotone{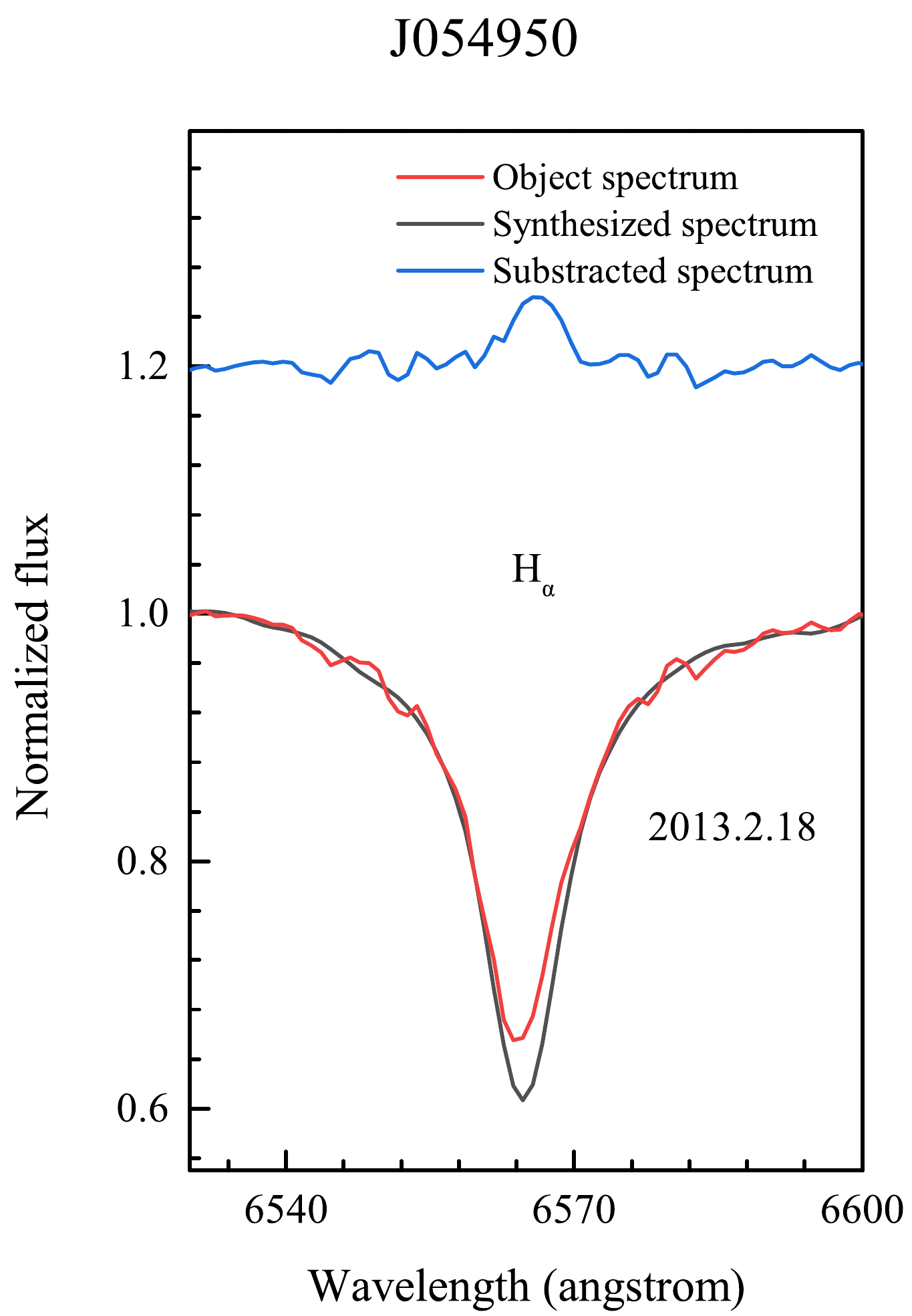}\\
\plotone{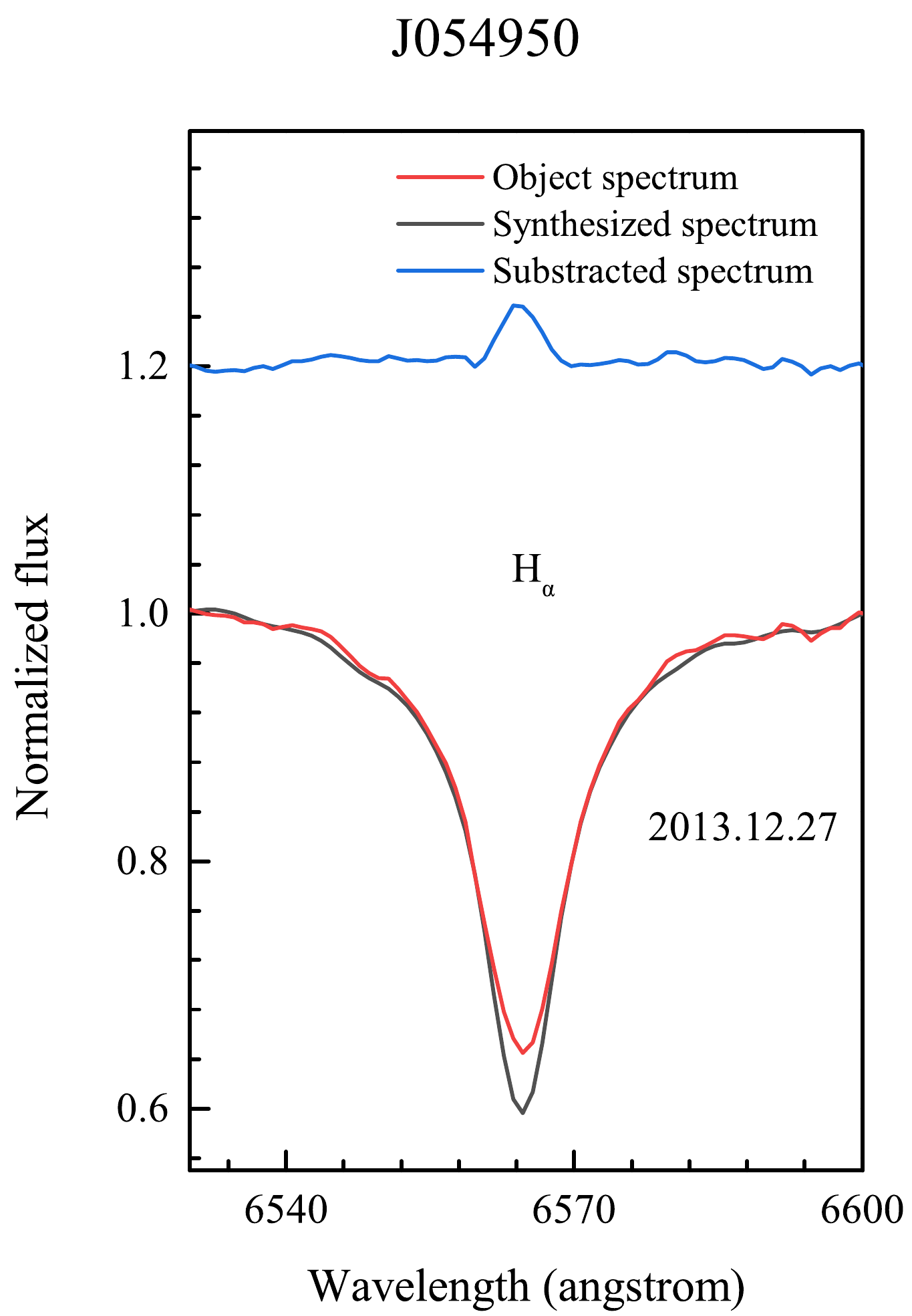}
\plotone{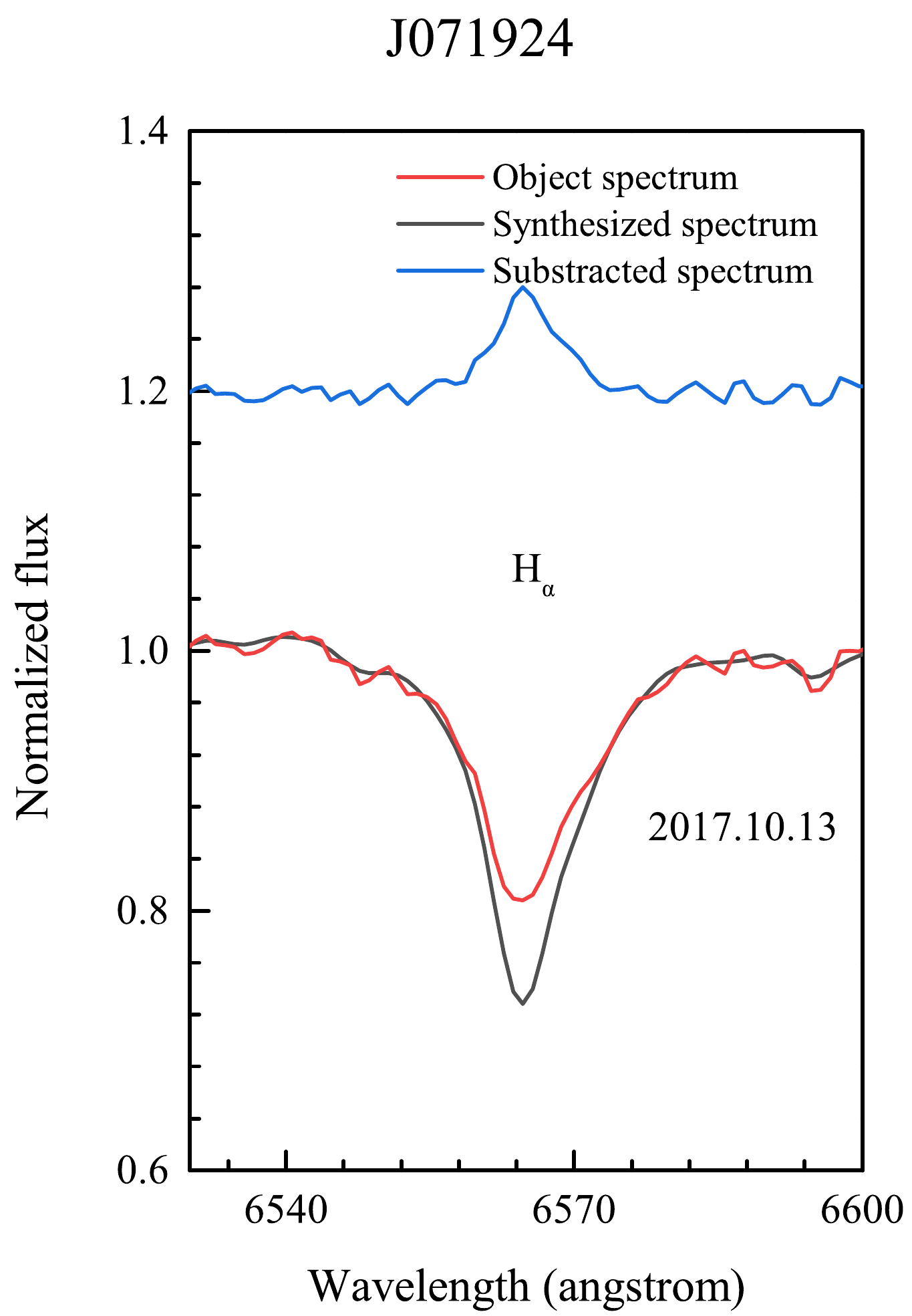}
\plotone{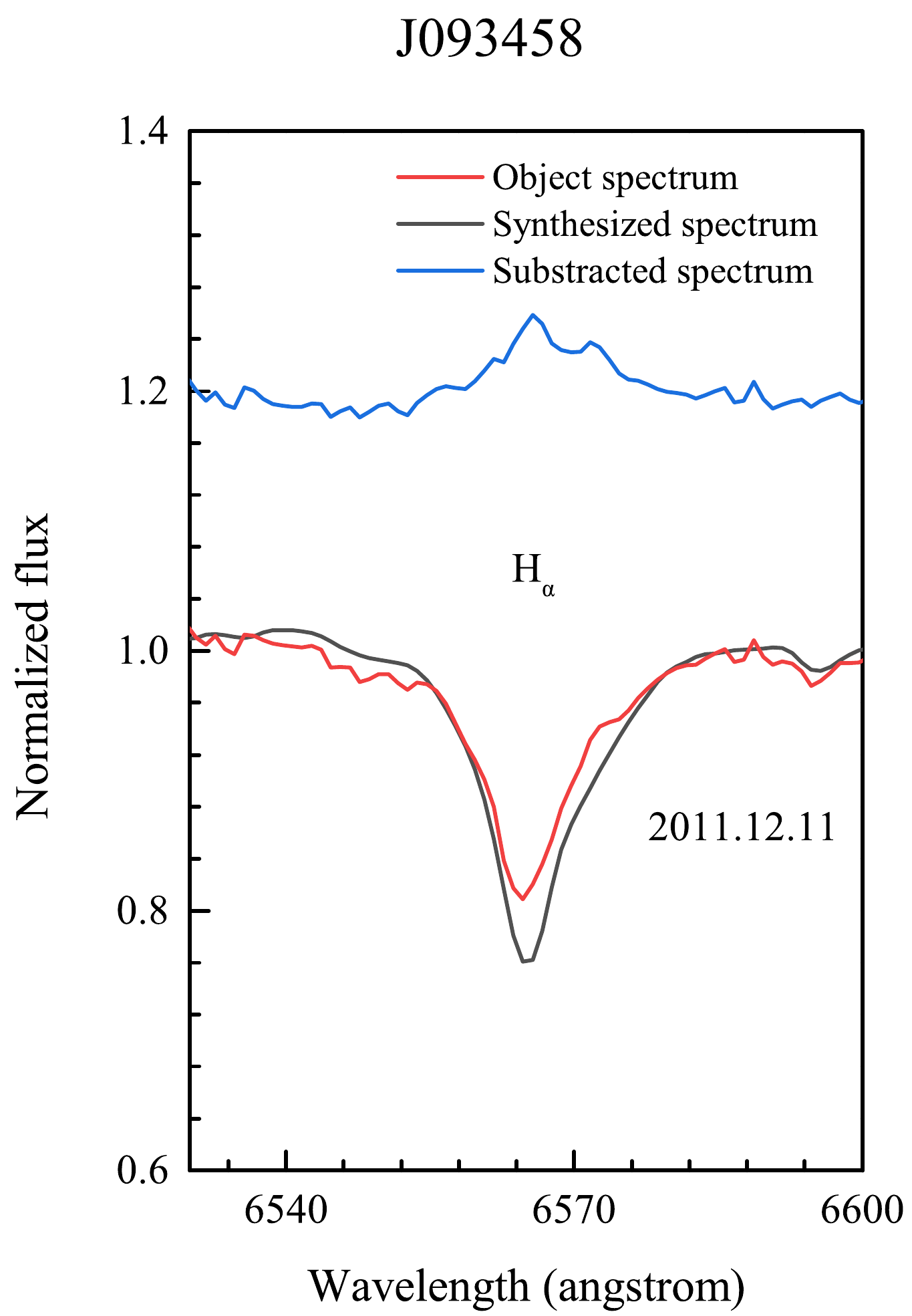}
\plotone{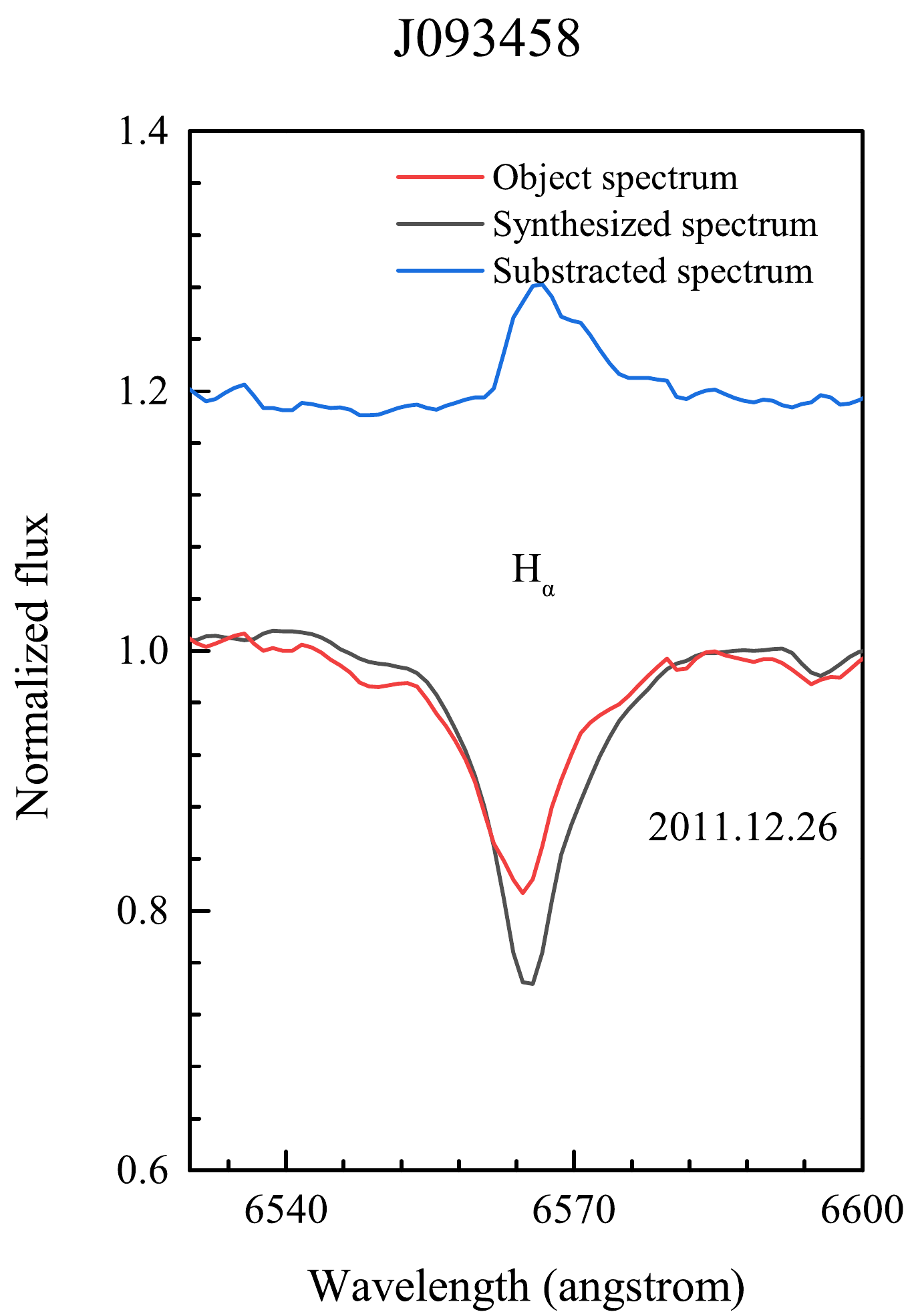}
\plotone{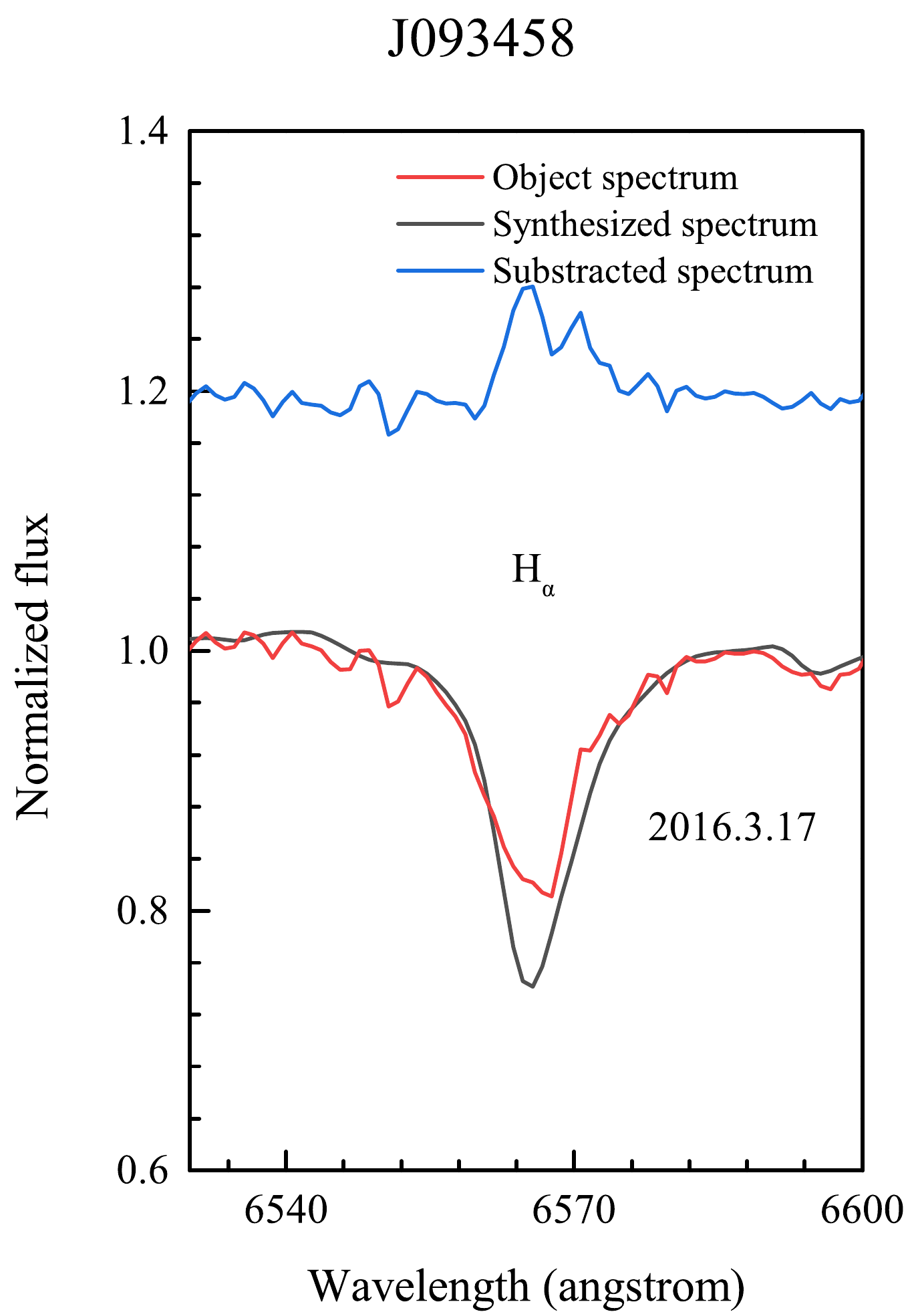}
\plotone{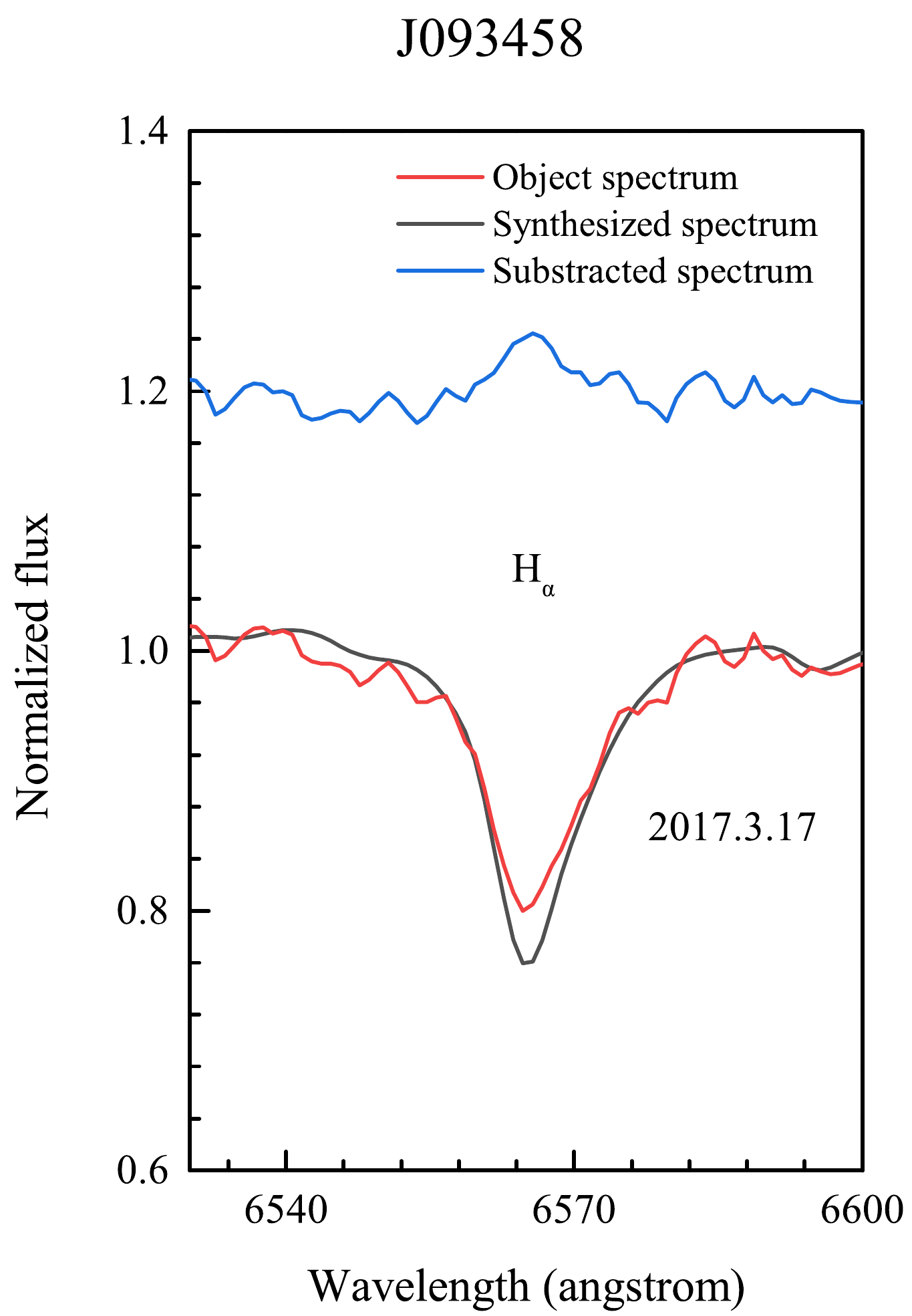}\\
\plotone{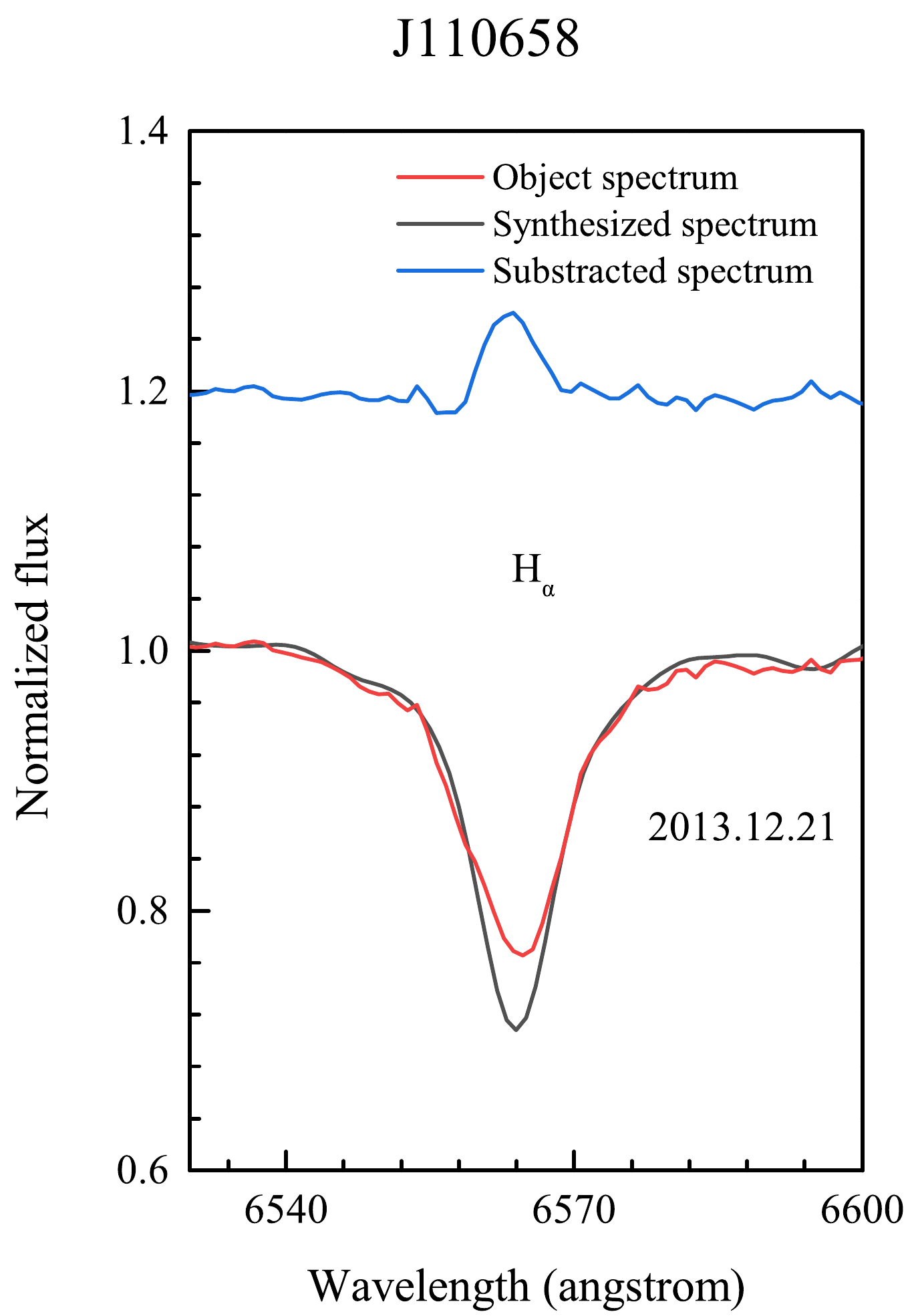}
\plotone{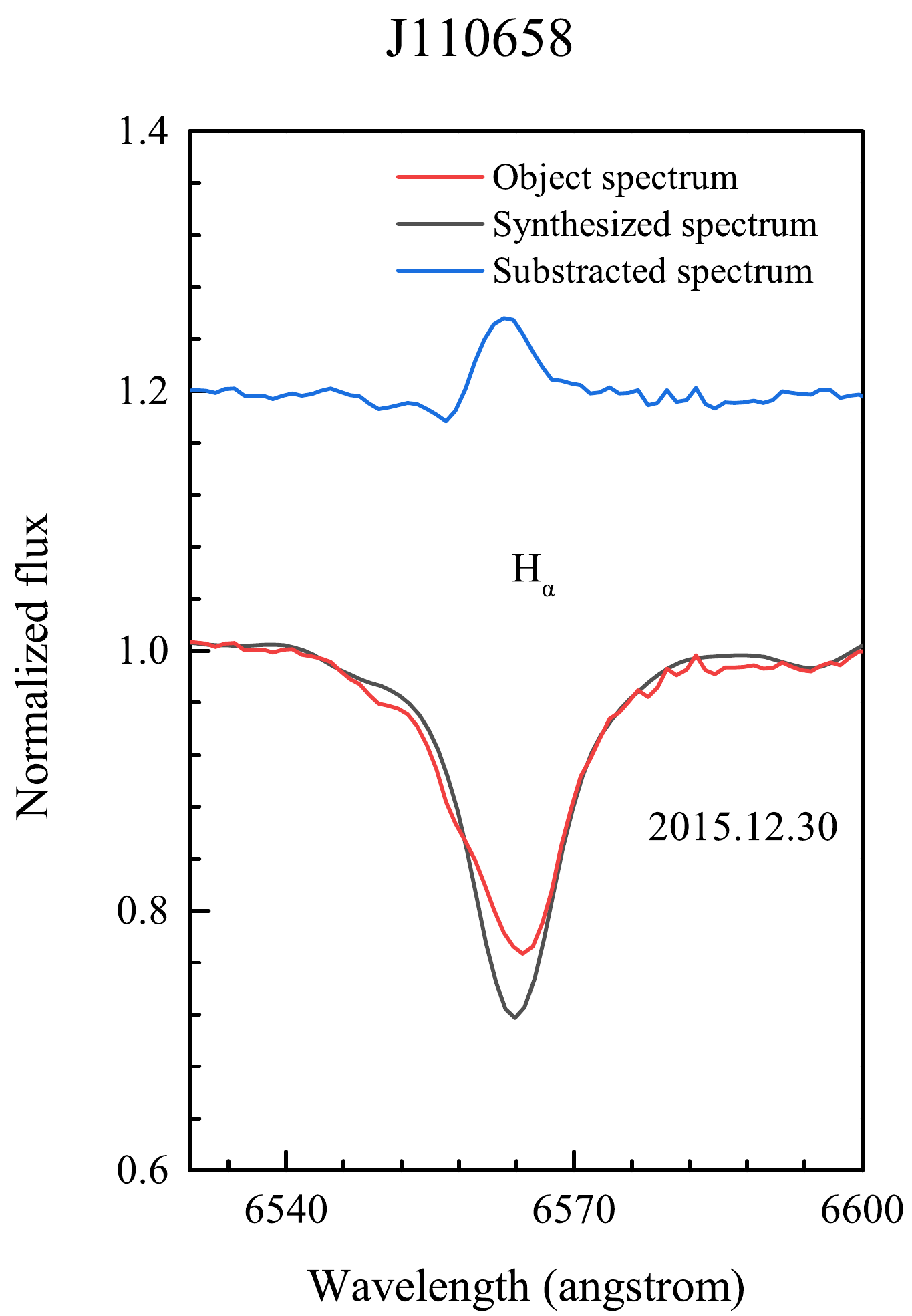}
\plotone{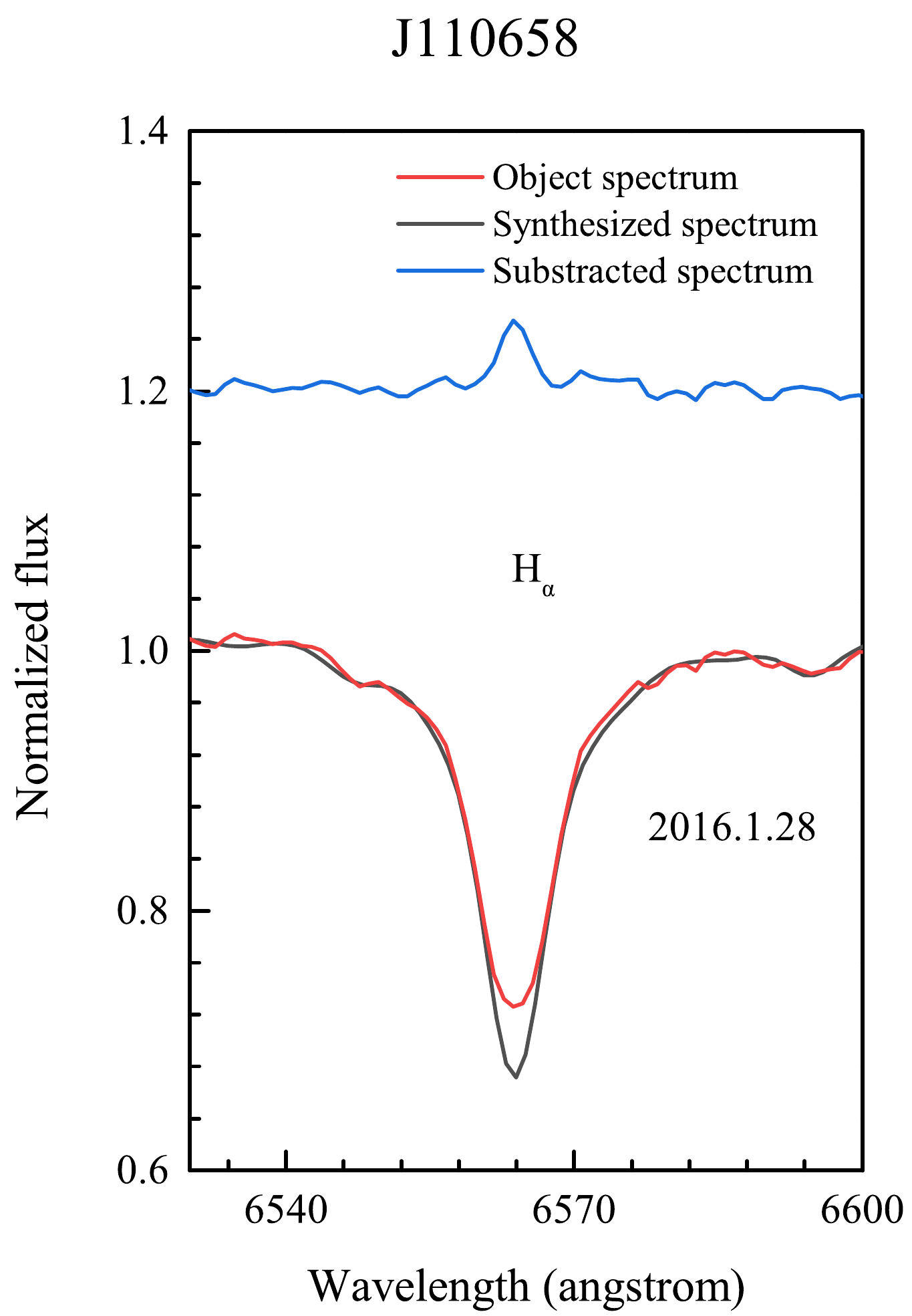}
\plotone{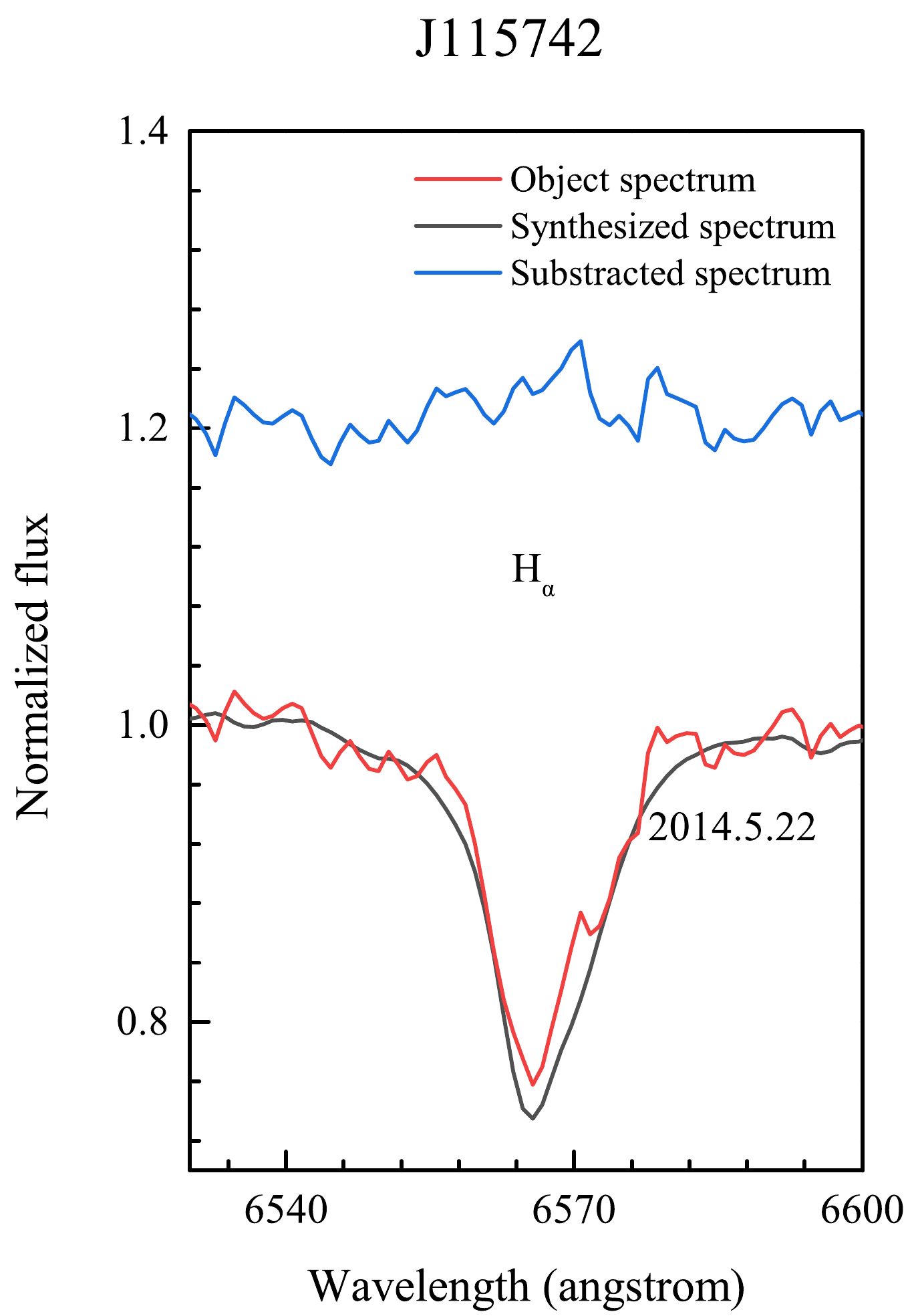}
\plotone{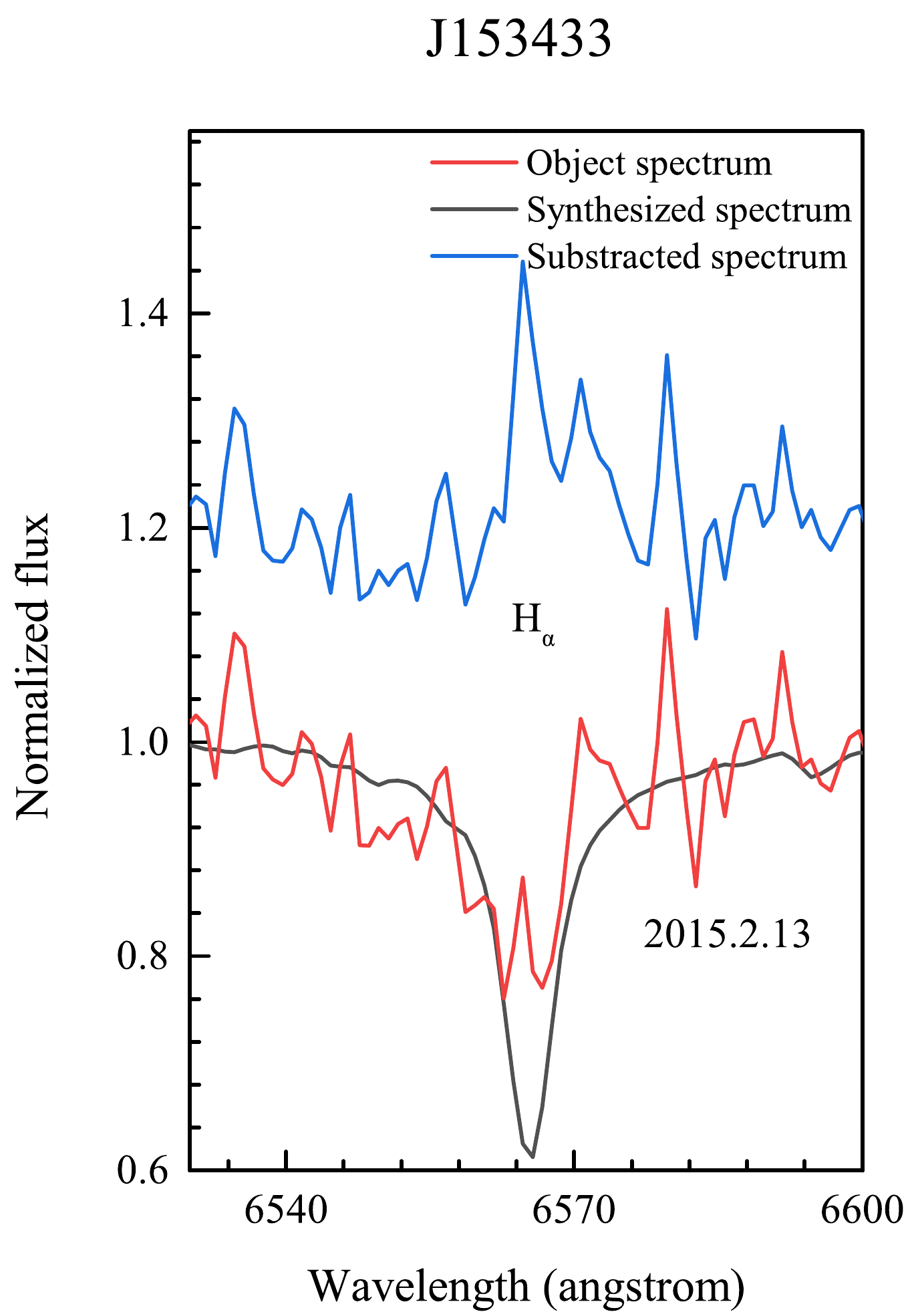}
\plotone{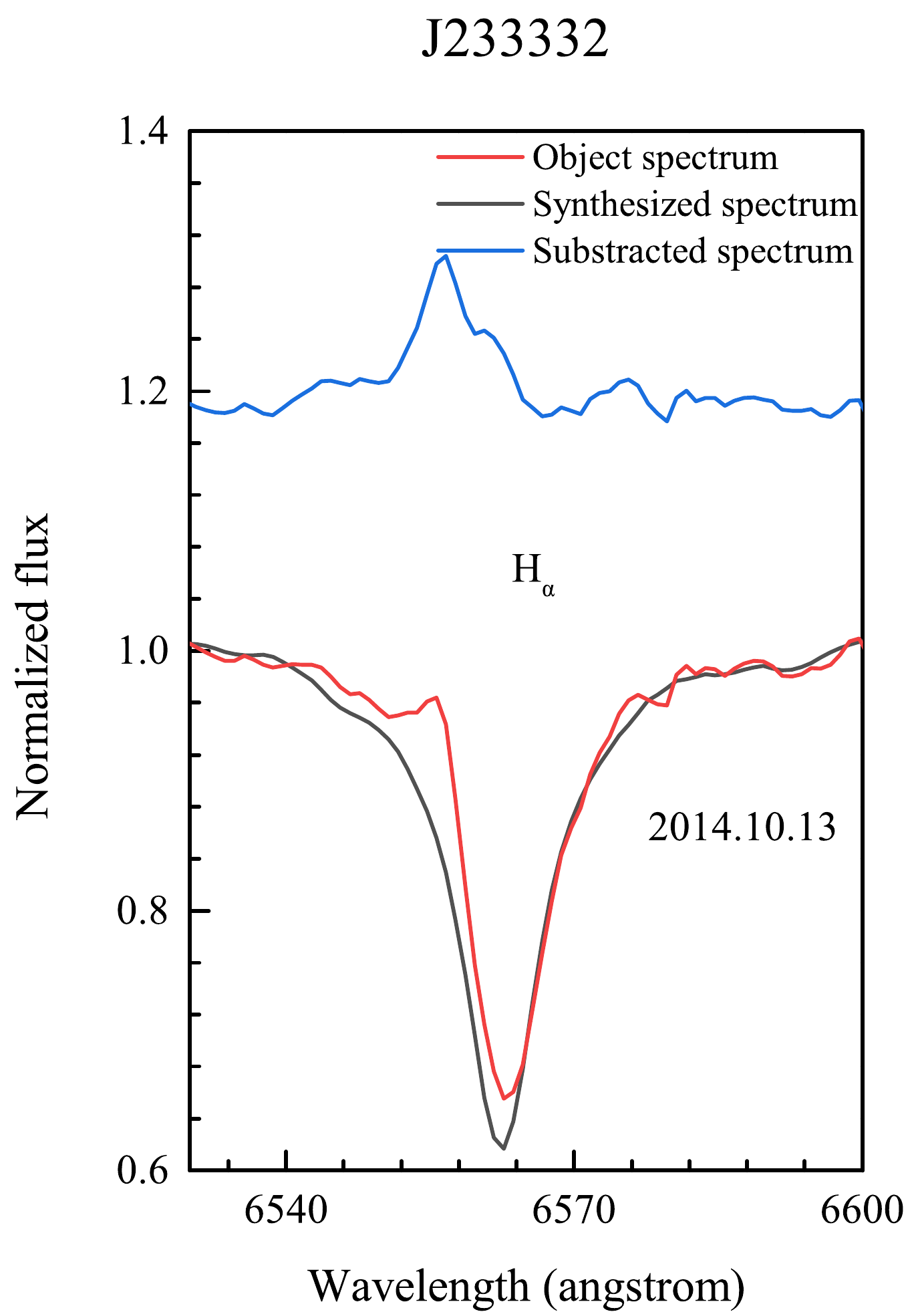}
\caption{The H$_\alpha$ line region of LAMOST spectra of the ten binaries, the red lines represent the spectra of the targets, the black lines refer to the synthesized spectra produced by STARMOD, while the blue ones display the subtracted spectra. All the ten targets exhibit obvious excess H$_\alpha$ emission lines. \label{fig:sp}}
\end{figure}

The chromospheric magnetic activities are usually observed in late-type stars, such as spots, flares, or plages, which are caused by the stellar magnetic dynamo mechanism due to the differential rotation.
The majority of contact binaries are composed of two late-type component stars. Therefore, contact binaries usually exhibit chromospheric emission lines, such as Balmer series (H$_\alpha$, H$_\beta$, H$_\gamma$), Ca II H \& K, Ca II IRT, etc. \citep{2018A&A...615A.120P,2020MNRAS.495.1252Z,2021MNRAS.506.4251Z}. For our ten targets, all the observed LAMOST spectra show absorptions of these lines. Thus, the spectral subtraction technique \citep{1985ApJ...295..162B} was introduced to determine the emission lines. First, two appropriate inactive stars spectra (temperature difference between the inactive star and the corresponding component of the binary is less than 200 K) downloaded from the LAMOST website were selected from the catalog of \citep{2021ApJS..256...14Z} and are treated as the primary and secondary template spectra for each binary system. Secondly, all the spectra were normalized and removed cosmic rays. Thirdly, the STARMOD code developed by \citep{1985ApJ...295..162B} (which has been used by many researchers, e.g. \citealt{2020MNRAS.495.1252Z,2021AJ....161..221P,2022ApJ...927...12P}) was applied to produce the synthetic spectra of the two inactive stars by considering the radial velocity shift, the rotationally broadening, and the weight of the two stars. Finally, the subtracted spectra between the synthetic spectra and the LAMOST spectra were obtained and displayed in Figure \ref{fig:sp} (the H$_\alpha$ line region is shown). From this figure, we can see that all the ten targets exhibit obvious excess H$_\alpha$ emission lines.
Then, we calculated the equivalent width (EW) of the H$_\alpha$ line using the splot package of IRAF, the results are listed in Table \ref{tab:LAMOST}.

\section{Discussions and conclusions}\label{sec:discussion}
We present the first photometric and spectroscopic investigations of ten totally eclipsing contact binaries. Due to our light-curve analysis, the mass ratios of the ten targets are all smaller than 0.15. They are all ELMRCBs. Because of the obvious flat bottom of their secondary eclipsing minima, the photometric physical parameters are reliable \citep{2003CoSka..33...38P,2005Ap&SS.296..221T,2021AJ....162...13L}. Seven of our targets are low mass ratio, deep contact binaries \citep{2005AJ....130..224Q}, two (J093458 and J233332) are medium contact binaries, while only one (J054950) is a shallow contact system. The criteria to classify systems as "deep", "medium" and "shallow" contact systems are as follows: $f\geq50\%$ is deep, $25\%\leq f<50\%$ is medium, while $f<25\%$ is shallow.
Six of them are A-type contact systems, while the others are W-type ones. Five of the ten systems show obvious O'Connell effect, and a dark spot on one of the two components can model this effect very well. A very interesting phenomenon of the TNT light curves of J071924 was found that a flare-like event is seen near phase 0.8. By analyzing the LAMOST spectra, we found that all the ten targets exhibit excess emissions in the H$_\alpha$ line, indicating chromospheric activity of them. By calculating the eclipsing times from the sky surveys and our observational data, we studied the O-C curves of the ten binaries and derived that four of them show long-term period decrease, while the others display continuous period increase. The orbital period change may be resulted from the mass transfer between the two components.

\begin{figure}
\epsscale{0.7}
\plotone{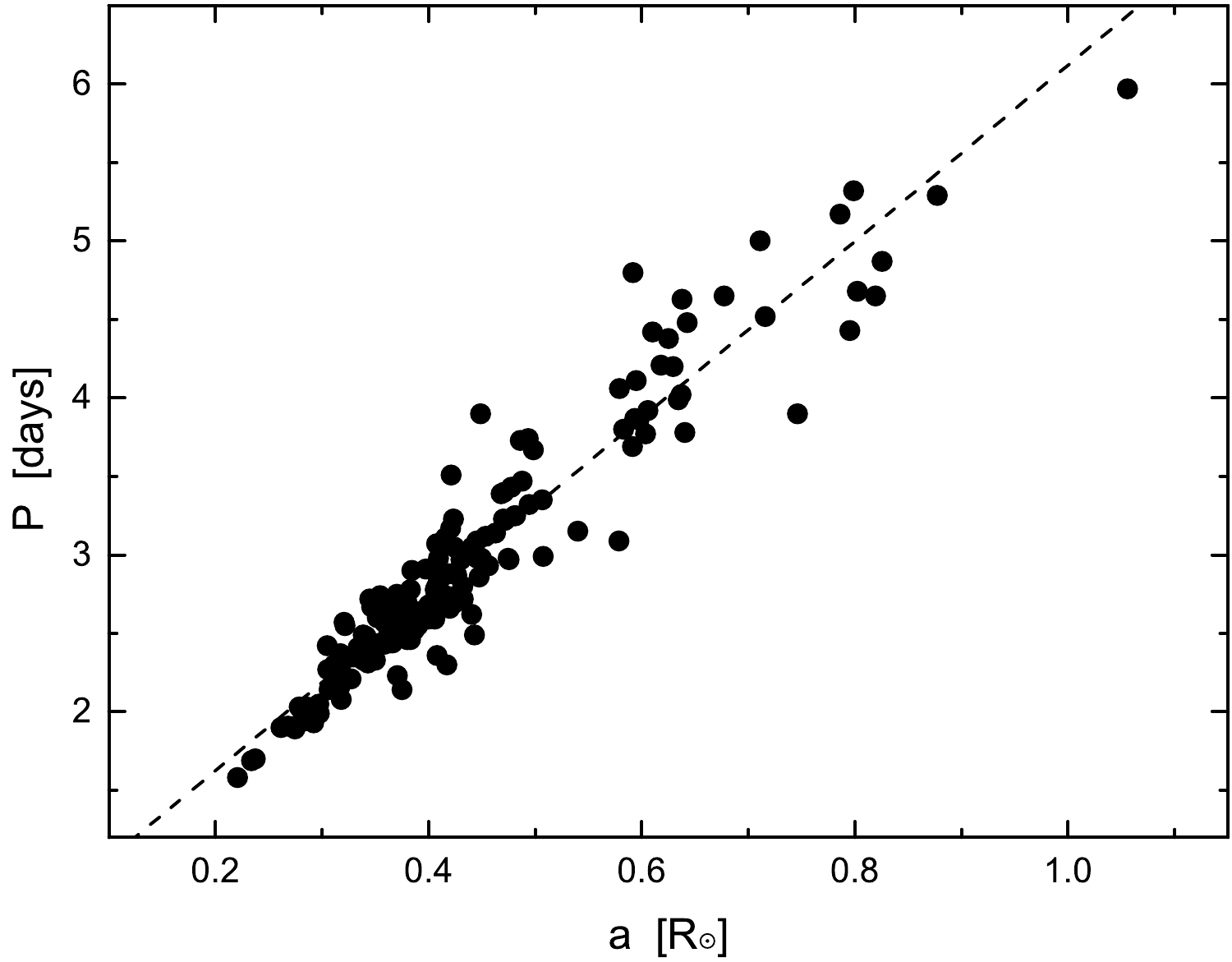}
\caption{The relationship between the period (P) and semi-major axis (a) of the contact binaries whose temperatures are smaller than 10000 K taken from \cite{2021AJ....162...13L}. \label{fig:pa}}
\end{figure}

In order to investigate the the mass transfer rate and the evolutionary status of the ten binaries, we have to estimate their absolute parameters. \cite{2021AJ....162...13L} collected 173 contact binaries whose absolute parameters have been determined by simultaneously analyzing the radial velocity curve and the photometric light curve. Using these contact binaries whose temperatures are smaller than 10000 K (168 targets), we found that they reveal a very good period (P) and semi-major axis (a) relation, which is displayed in Figure \ref{fig:pa}. The relationship between P and a can be described by the following linear equation,
\begin{eqnarray}
P=0.501(\pm0.063)+5.621(\pm0.138)\times a.
\end{eqnarray}
Using this equation and the period listed in Table \ref{tab:information}, we estimated the absolute parameters of the ten binaries which are listed in Table \ref{tab:absolute}. Then,using the following equation,
\begin{eqnarray}
{\dot{P}\over P}=-3\dot{M_1}({1\over M_1}-{1\over M_2}),
\end{eqnarray}
we calculated the mass transfer rate between the two components which are listed in Table \ref{tab:o-cresults}.

\setlength{\tabcolsep}{1.2mm}{
\begin{deluxetable*}{cccccccccc}
\small
\tablecaption{The estimated absolute parameters of the ten targets\label{tab:absolute}}
\tablewidth{0pt}
\tablehead{
\colhead{Star}& \colhead{T$_1$ (K)}&\colhead{T$_2$ (K)} & \colhead{a ($R_\odot$)}&\colhead{M$_1$ ($M_\odot$)}&\colhead{M$_2$ ($M_\odot$)}&\colhead{R$_1$ ($R_\odot$)}&\colhead{R$_2$ ($R_\odot$)}&\colhead{L$_1$ ($L_\odot$)}&\colhead{L$_2$ ($L_\odot$)}\\
}
\startdata
J001422 & $6590\pm13$ & $6463\pm24 $& $2.63\pm0.12$ & $1.49\pm0.20$ & $0.21\pm0.03$ & $1.55\pm0.07$ & $0.72\pm0.04$& $4.05\pm0.40$& $0.80\pm0.09$\\
J022733 & $6494\pm39$ & $6374\pm52 $& $2.83\pm0.12$ & $1.58\pm0.20$ & $0.19\pm0.02$ & $1.65\pm0.07$ & $0.68\pm0.05$& $4.36\pm0.49$& $0.68\pm0.11$\\
J042640 & $6029\pm93$ & $6071\pm116$& $2.54\pm0.11$ & $1.48\pm0.20$ & $0.19\pm0.03$ & $1.47\pm0.07$ & $0.62\pm0.03$& $2.57\pm0.40$& $0.47\pm0.08$\\
J054950 & $7037\pm37$ & $6642\pm55 $& $3.00\pm0.12$ & $1.63\pm0.21$ & $0.20\pm0.03$ & $1.71\pm0.07$ & $0.67\pm0.04$& $6.46\pm0.70$& $0.79\pm0.11$\\
J071924 & $5892\pm35$ & $6059\pm48 $& $2.50\pm0.11$ & $1.51\pm0.20$ & $0.15\pm0.02$ & $1.51\pm0.07$ & $0.59\pm0.03$& $2.47\pm0.28$& $0.41\pm0.06$\\
J093458 & $5822\pm96$ & $5810\pm132$& $2.18\pm0.10$ & $1.38\pm0.20$ & $0.18\pm0.03$ & $1.26\pm0.06$ & $0.52\pm0.04$& $1.63\pm0.27$& $0.28\pm0.07$\\
J110658 & $6231\pm40$ & $6249\pm60 $& $2.75\pm0.12$ & $1.56\pm0.20$ & $0.18\pm0.02$ & $1.65\pm0.07$ & $0.70\pm0.04$& $3.67\pm0.42$& $0.66\pm0.09$\\
J115742 & $5963\pm65$ & $6202\pm107$& $2.12\pm0.10$ & $1.39\pm0.20$ & $0.15\pm0.02$ & $1.25\pm0.06$ & $0.49\pm0.04$& $1.78\pm0.26$& $0.32\pm0.08$\\
J153433 & $6079\pm370$& $6060\pm430$& $2.36\pm0.11$ & $1.47\pm0.20$ & $0.14\pm0.02$ & $1.41\pm0.07$ & $0.53\pm0.03$& $2.44\pm0.83$& $0.33\pm0.13$\\
J233332 & $6770\pm39$ & $6379\pm72 $& $4.04\pm0.15$ & $2.03\pm0.23$ & $0.20\pm0.02$ & $2.40\pm0.09$ & $0.87\pm0.06$& $10.8\pm1.09$& $1.13\pm0.21$\\
\enddata
\end{deluxetable*}}

\begin{figure}
\epsscale{0.47}
\plotone{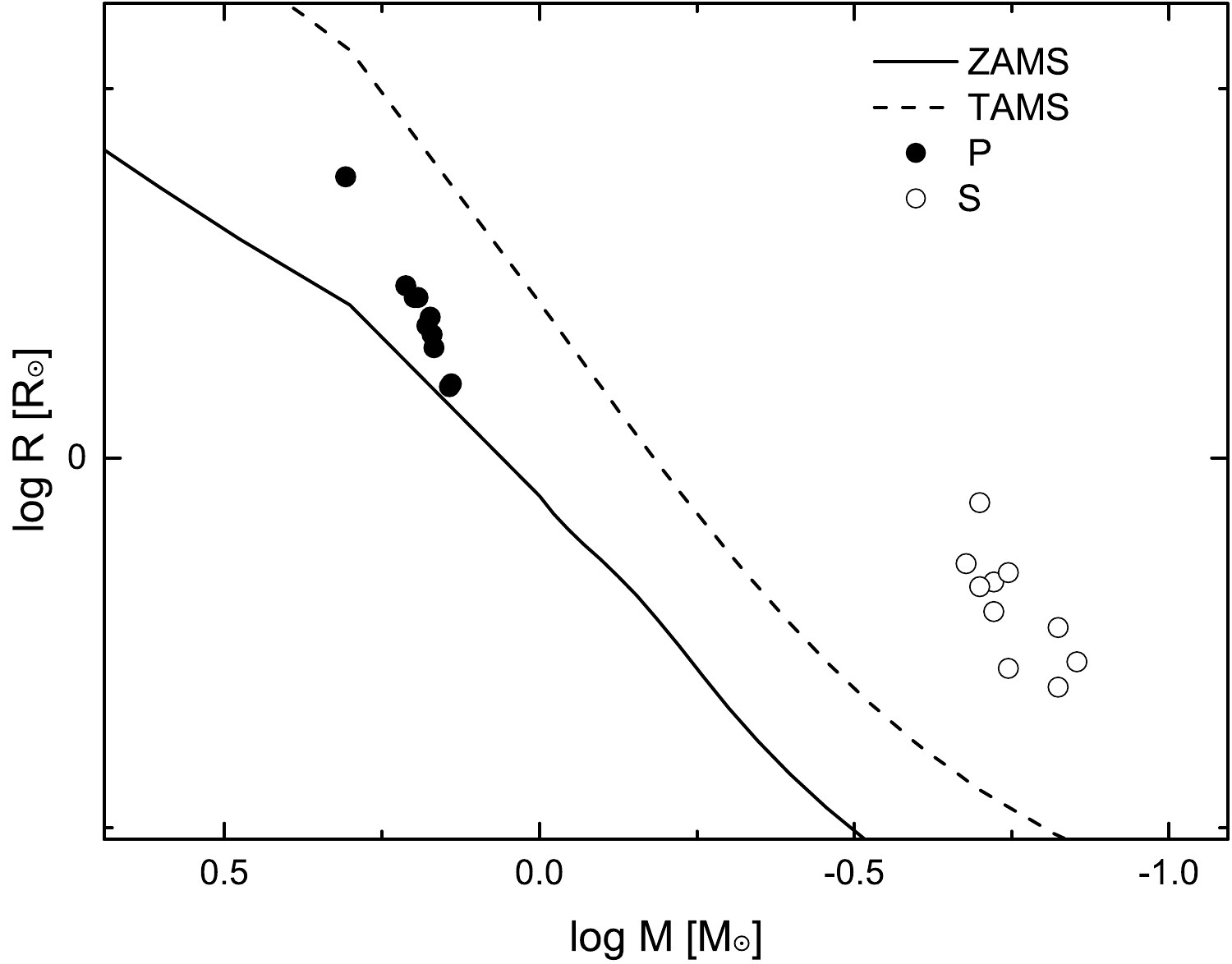}
\plotone{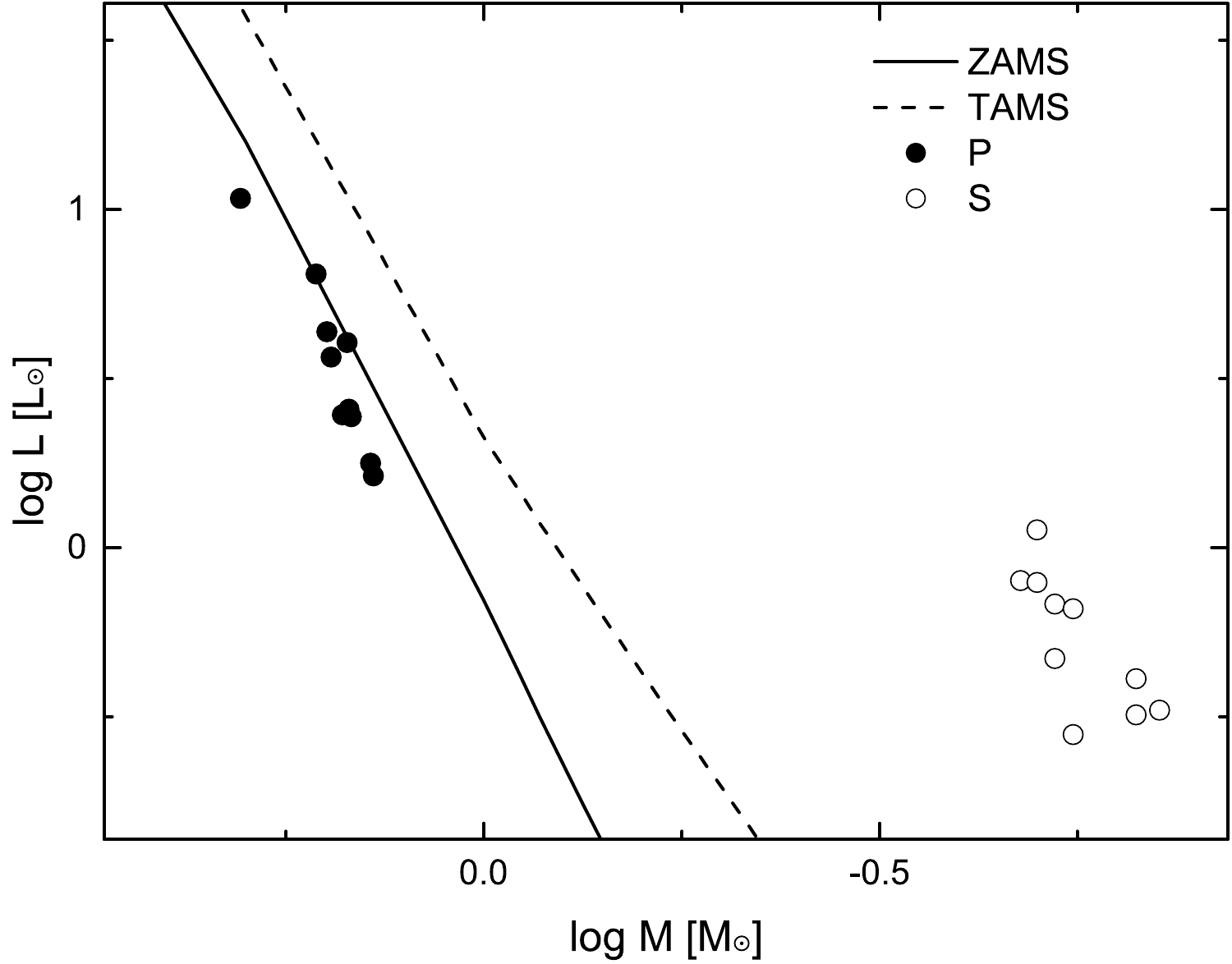}
\caption{The M-R and M-L diagrams for the ten contact binaries, the solid line represents the ZAMS line, while the dashed one refers the TAMS line. The solid circles display the primary components, while the open ones plot the secondary ones. \label{fig:mrml}}
\end{figure}

\begin{figure}
\epsscale{0.7}
\plotone{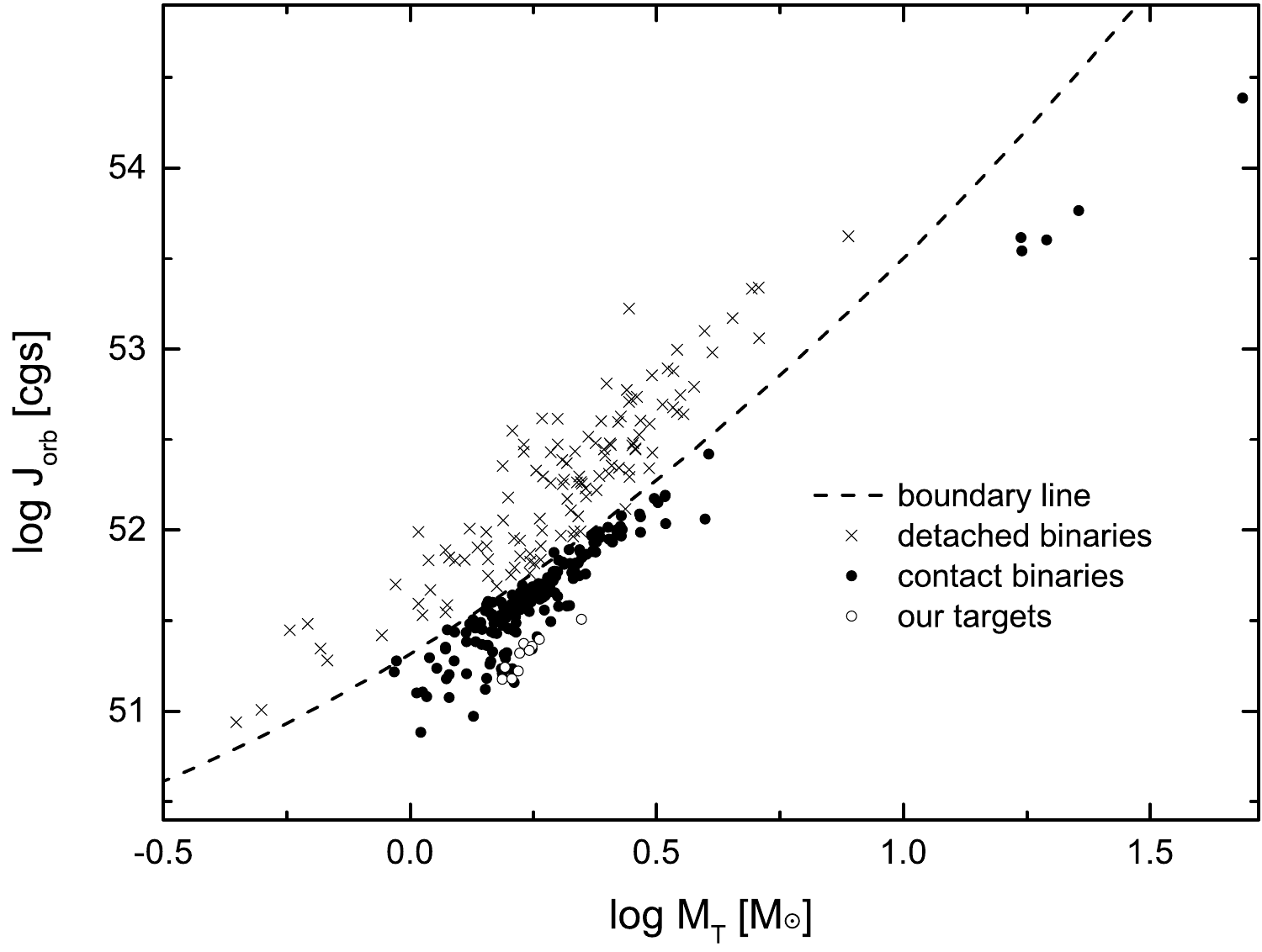}
\caption{The relation between logJ$_{orb}$ and logM$_T$ of the detached binaries from \cite{2006MNRAS.373.1483E} and the contact binaries from \cite{2021AJ....162...13L}. The crosses represent the detached binaries, the solid circles display the contact binaries, and the open circles denote our ten targets. The dashed line refers to the boundary between detached and contact binaries derived by \cite{2006MNRAS.373.1483E}.\label{fig:J-M}}
\end{figure}

In order to analyze the evolutionary states of the ten systems, we draw the mass-radius (M-R) and the mass-luminosity (M-L) diagrams shown in Figure \ref{fig:mrml}. The zero-age main-sequence (ZAMS) and the terminal-age main-sequence (TAMS) which are derived from \cite{2002MNRAS.329..897H} are described with solid and dashed black lines. We can clearly see that the more massive primary components are located around the ZAMS, meaning that they are non-evolved or little-evolved main-sequence stars, and the less massive secondary components are situated above the TAMS, implying that they have evolved away from the main-sequence and are over-sized and over-luminosity comparing the main-sequence star with the same mass which may be due to energy transfer from the primary component to the secondary one. The evolutionary states of the ten binaries are identical to the contact binaries shown in \cite{2021AJ....162...13L}.  The orbital angular momentum, J$_{orb}$, of the ten contact binaries were calculated using the equation provided by \cite{2006MNRAS.373.1483E},
\begin{equation}
	\ J_{orb} = \frac{q}{(1+q)^2}\sqrt[3]{\frac{G^2}{2\pi}M_T^5P}, \\
\end{equation}
where q is the mass ratio, M$_T$ is the total mass, and P is the orbital period. Then, we constructed the diagram of logJ$_{orb}$ and logM$_T$ of the detached binaries from \cite{2006MNRAS.373.1483E} and the contact binaries from \cite{2021AJ....162...13L} and show it in Figure \ref{fig:J-M}. The boundary line between detached and contact binaries derived by \cite{2006MNRAS.373.1483E} is described with a dashed line. The ten binaries are also displayed in this figure. One can see from this figure that the orbital angular momentum of our ten targets is smaller comparing the contact binary having the same mass. This may indicate that the ten ELMRCBs are at the late evolutionary stage of a contact binary.

The initial masses of the two components of contact binaries play an very important role in the their evolution. Then, we computed the initial masses of both components by the method proposed by \cite{2013MNRAS.430.2029Y}. Firstly, the initial mass of the secondary component was calculated using the following equation,
\begin{equation}
\ M_{2i} = M_2+\Delta M= M_2+2.50(M_L-M_2-0.07)^{0.64},
\end{equation}
where M$_{2i}$ is the initial mass of the secondary component, M$_2$ is the mass of the secondary component at present, and M$_L$ is the mass due to mass–luminosity relation defined by M$_L$ =$(\frac{L_2}{1.49})^{\frac{1}{4.216}}$. Secondly, based on the physical constraint of the reciprocal mass ratio 1/$q_i$ ($0<1/q_i<1$), the initial mass of the primary component can be calculated using the following equation,
\begin{equation}
M_{1i} = M_1 - (\Delta M - M_{lost}) = M_1-\Delta M(1-\gamma),
\end{equation}
Where M$_{1i}$ is the initial mass of the primary component, M$_1$ is the mass of the primary component at present, M$_{lost}$ is the mass lost by the system, and $\gamma$ is the ratio $M_{lost}$ to $\Delta M$ and the mean value of $\gamma=0.664$ determined by \cite{2013MNRAS.430.2029Y} was used to calculate M$_{1i}$. Finally, we derived the ages of the ten targets using the following equations \citep{2014MNRAS.437..185Y},
\begin{equation}
\ t \approx t_{MS}(M_{2i})+t_{MS}(\overline{M_{2}}),\\
\end{equation}
\begin{eqnarray}
\ t_{MS}=\frac{10}{(M/M_{\odot})^{4.05}}\times (5.60*10^{-3}(\frac{M}{M_{\odot}}+3.993)^{3.16}+0.042) \quad Gyr,
\end{eqnarray}
where $\overline{M_{2}}$ is the average value of M$_{2i}$ and M$_L$. The derived initial masses of the two components, mass lost, and the ages are listed in Table \ref{tab:instability}. We found that these results are similar to those determined by \cite{2013MNRAS.430.2029Y} and \cite{2014MNRAS.437..185Y}.

All the ten binaries are extremely low mass ratio contact binaries. Several of them have mass ratios that are very close to the cut-off mass ratio of the contact binary ($q_{min}$ is around 0.07 according to  \citealt{1995ApJ...444L..41R,2006MNRAS.369.2001L,2007MNRAS.377.1635A,2009MNRAS.394..501A,2010MNRAS.405.2485J,2021MNRAS.501..229W}). The ratio of the spin angular momentum ($J_{spin}$) to the orbital angular momentum ($J_{orb}$) is very important to examine the dynamic stability of contact binaries \citep{1980A&A....92..167H}. Then, we calculated the $J_{spin}/J_{orb}$ using the following equation \citep{2015AJ....150...69Y} and list them in Table \ref{tab:instability},
{\setlength\abovedisplayskip{0.3cm}
\setlength\belowdisplayskip{0.3cm}
\begin{eqnarray}
{J_{spin}\over J_{orb}}={{1+q}\over q}[(k_1r_1)^2+(k_2r_2)^2q],
\end{eqnarray}}
where $q$ is mass ratio, $k_{1,2}$ is the dimensionless gyration radius, and $r_{1,2}$ is the relative radius. The values of $k^2_{1,2}=0.06$ were adopted from \cite{2006MNRAS.369.2001L}. We found that the values of $J_{spin}/J_{orb}$ for all the ten targets are between 0.17 and 0.25. From the perspective of the $J_{spin}/J_{orb}$, all the ten targets are dynamically stable. Very recently, \cite{2021MNRAS.501..229W} put forward the instability mass ratio, the instability separation, and the instability period of contact binaries. The instability mass ratios of our ten targets were computed using the Equations (14) and (15) of \cite{2021MNRAS.501..229W} and are listed in Table \ref{tab:instability}. The mass ratios of the ten targets are all greater than their instability mass ratios. Then, using Equations (12) and (13) of \cite{2021MNRAS.501..229W} and Kepler's third law, we obtained the instability separations and the instability periods which are also listed in Table \ref{tab:instability}. During the calculations, the gyration radius (k$_1$) of the primary was determined using the relations derived by \cite{2009A&A...494..209L} for rotating
and tidally distorted ZAMS stars, $k_1=0.5391-0.2504\times M_1$ ($0.4M_\odot<M_1<1.4M_\odot$) and $k_1=0.1523+0.0141\times M_1$ ($M_1>1.4M_\odot$). Because the mass of the secondary is very low, it should be a fully convective star, $k^2_2=0.205$ was used \citep{2007MNRAS.377.1635A}. Comparing the present actual values, the instability parameters are smaller, indicating that all the ten ELMRCBs are relatively stable.

In conclusion, ten contact binaries were analyzed for the first time by us and determined that all of them are ELMRCBs. LAMOST spectra of them were studied by the spectral subtraction technique, and the H$_\alpha$ emission line was identified. The O-C diagrams of the ten systems were investigated using all available eclipsing times. The evolutionary states, the initial masses, ages, and the instability parameters were studied, and we found that they are all stable. Future observations, especially the radial velocity observations, are needed to determine very precise absolute parameters and to comprehend the formation and evolution of the ten contact binaries.

\setlength{\tabcolsep}{2mm}{
\begin{deluxetable*}{cccccccccc}
\tablecaption{The initial parameters, the ages, and the instability parameters of the ten targets\label{tab:instability}}
\tablewidth{0pt}
\tablehead{
\colhead{Star}& \colhead{Type}&\colhead{M$_{1i}$ ($M_\odot$)}&\colhead{M$_{2i}$ ($M_\odot$)}&\colhead{M$_{lost}$ ($M_\odot$)}&\colhead{$\tau$ (Gyr)}&\colhead{$J_{spin}/J_{orb}$}&\colhead{$q_{inst}$}&\colhead{$A_{inst}$ ($R_\odot$)}&\colhead{$P_{inst}$ (d)}\\
}
\startdata
J001422 &A & 0.89 &1.98 &1.17 & 3.93 & 0.172 & 0.052& 1.738& 0.203\\
J022733 &A & 0.99 &1.94 &1.16 & 4.27 & 0.196 & 0.045& 1.902& 0.228\\
J042640 &W & 0.94 &1.79 &1.07 & 5.52 & 0.182 & 0.051& 1.669& 0.193\\
J054950 &A & 1.03 &1.99 &1.19 & 3.90 & 0.185 & 0.040& 1.961& 0.235\\
J071924 &W & 0.95 &1.79 &1.09 & 5.66 & 0.247 & 0.050& 1.799& 0.217\\
J093458 &A & 0.89 &1.62 &0.96 & 7.82 & 0.179 & 0.059& 1.563& 0.181\\
J110658 &W & 0.97 &1.93 &1.16 & 4.31 & 0.209 & 0.047& 1.910& 0.231\\
J115742 &W & 0.86 &1.70 &1.03 & 6.71 & 0.216 & 0.059& 1.603& 0.190\\
J153433 &A & 0.94 &1.73 &1.05 & 6.46 & 0.249 & 0.052& 1.683& 0.199\\
J233332 &A & 1.38 &2.13 &1.29 & 3.09 & 0.245 & 0.039& 2.928& 0.389\\
\enddata
\end{deluxetable*}}

\begin{acknowledgments}
Thanks the anonymous referee very much for the constructive and insightful criticisms and suggestions to improve our manuscript. This work is supported by the Joint Research Fund in Astronomy (No. U1931103) under cooperative agreement between National Natural Science Foundation of China (NSFC) and Chinese Academy of Sciences (CAS), and by NSFC (No. 11703016), and by the Qilu Young Researcher Project of Shandong University, and by the Chinese Academy of Sciences Interdisciplinary Innovation Team, and by the Cultivation Project for LAMOST Scientific Payoff and Research Achievement of CAMS-CAS. The calculations in this work were carried out at Supercomputing Center of Shandong University, Weihai.
	
We acknowledge the support of the staff of the Xinglong 80cm, 85cm telescopes, NEXT and WHOT. This work was partially supported by the Open Project Program of the Key Laboratory of Optical Astronomy, National Astronomical Observatories, Chinese Academy of Sciences.
	
The spectral data were provided by Guoshoujing Telescope (the Large Sky Area Multi-Object Fiber Spectroscopic Telescope LAMOST), which is a National Major Scientific Project built by the Chinese Academy of Sciences. Funding for the project has been provided by the National Development and Reform Commission. LAMOST is operated and managed by the National Astronomical Observatories, Chinese Academy of Sciences.
	
Based on observations obtained with the Samuel Oschin 48-inch Telescope at the Palomar Observatory as part of the Zwicky Transient Facility project. ZTF is supported by the National Science Foundation under grant no. AST-1440341 and a collaboration including Caltech, IPAC, the Weizmann Institute for Science, the Oskar Klein Center at Stockholm University, the University of Maryland, the University of Washington, Deutsches Elektronen-Synchrotron and Humboldt University, Los Alamos National Laboratories, the TANGO Consortium of Taiwan, the University of Wisconsin at Milwaukee, and Lawrence Berkeley National Laboratories. Operations are conducted by COO, IPAC, and UW.
	
This paper makes use of data from the DR1 of the WASP data \citep{2010A&A...520L..10B} as provided by the WASP consortium, and the computing and storage facilities at the CERIT Scientific Cloud, reg. no. CZ.1.05/3.2.00/08.0144 which is operated by Masaryk University, Czech Republic.
	
This work includes data collected by the TESS mission. Funding for the TESS mission is provided by NASA Science Mission Directorate. We acknowledge the TESS team for its support of this work.
	
We thank Las Cumbres Observatory and its staff for their continued support of ASAS-SN. ASAS-SN is funded in part by the Gordon and Betty Moore Foundation through grants GBMF5490 and GBMF10501 to the Ohio State University, and also funded in part by the Alfred P. Sloan Foundation grant G-2021-14192.
	
This work has made use of data from the European Space Agency (ESA) mission {\it Gaia} (\url{https://www.cosmos.esa.int/gaia}), processed by the {\it Gaia} Data Processing and Analysis Consortium (DPAC, \url{https://www.cosmos.esa.int/web/gaia/dpac/consortium}). Funding for the DPAC has been provided by national institutions, in particular the institutions participating in the {\it Gaia} Multilateral Agreement.
	
\end{acknowledgments}

\bibliography{sample631}{}

\begin{thebibliography}{}
\expandafter\ifx\csname natexlab\endcsname\relax\def\natexlab#1{#1}\fi
\providecommand{\url}[1]{\href{#1}{#1}}
\providecommand{\dodoi}[1]{doi:~\href{http://doi.org/#1}{\nolinkurl{#1}}}
\providecommand{\doeprint}[1]{\href{http://ascl.net/#1}{\nolinkurl{http://ascl.net/#1}}}
\providecommand{\doarXiv}[1]{\href{https://arxiv.org/abs/#1}{\nolinkurl{https://arxiv.org/abs/#1}}}

\bibitem[{{Arbutina}(2007)}]{2007MNRAS.377.1635A}
{Arbutina}, B. 2007, \mnras, 377, 1635,
  \dodoi{10.1111/j.1365-2966.2007.11723.x}

\bibitem[{{Arbutina}(2009)}]{2009MNRAS.394..501A}
---. 2009, \mnras, 394, 501, \dodoi{10.1111/j.1365-2966.2008.14332.x}

\bibitem[{{Barden}(1985)}]{1985ApJ...295..162B}
{Barden}, S.~C. 1985, \apj, 295, 162, \dodoi{10.1086/163361}

\bibitem[{{Bellm} {et~al.}(2019){Bellm}, {Kulkarni}, {Graham}, {Dekany},
  {Smith}, {Riddle}, {Masci}, {Helou}, {Prince}, {Adams}, {Barbarino},
  {Barlow}, {Bauer}, {Beck}, {Belicki}, {Biswas}, {Blagorodnova}, {Bodewits},
  {Bolin}, {Brinnel}, {Brooke}, {Bue}, {Bulla}, {Burruss}, {Cenko}, {Chang},
  {Connolly}, {Coughlin}, {Cromer}, {Cunningham}, {De}, {Delacroix}, {Desai},
  {Duev}, {Eadie}, {Farnham}, {Feeney}, {Feindt}, {Flynn}, {Franckowiak},
  {Frederick}, {Fremling}, {Gal-Yam}, {Gezari}, {Giomi}, {Goldstein},
  {Golkhou}, {Goobar}, {Groom}, {Hacopians}, {Hale}, {Henning}, {Ho}, {Hover},
  {Howell}, {Hung}, {Huppenkothen}, {Imel}, {Ip}, {Ivezi{\'c}}, {Jackson},
  {Jones}, {Juric}, {Kasliwal}, {Kaspi}, {Kaye}, {Kelley}, {Kowalski},
  {Kramer}, {Kupfer}, {Landry}, {Laher}, {Lee}, {Lin}, {Lin}, {Lunnan},
  {Giomi}, {Mahabal}, {Mao}, {Miller}, {Monkewitz}, {Murphy}, {Ngeow},
  {Nordin}, {Nugent}, {Ofek}, {Patterson}, {Penprase}, {Porter}, {Rauch},
  {Rebbapragada}, {Reiley}, {Rigault}, {Rodriguez}, {van Roestel}, {Rusholme},
  {van Santen}, {Schulze}, {Shupe}, {Singer}, {Soumagnac}, {Stein}, {Surace},
  {Sollerman}, {Szkody}, {Taddia}, {Terek}, {Van Sistine}, {van Velzen},
  {Vestrand}, {Walters}, {Ward}, {Ye}, {Yu}, {Yan}, \&
  {Zolkower}}]{2019PASP..131a8002B}
{Bellm}, E.~C., {Kulkarni}, S.~R., {Graham}, M.~J., {et~al.} 2019, \pasp, 131,
  018002, \dodoi{10.1088/1538-3873/aaecbe}

\bibitem[{{Bradstreet} \& {Guinan}(1994)}]{1994ASPC...56..228B}
{Bradstreet}, D.~H., \& {Guinan}, E.~F. 1994, in Astronomical Society of the
  Pacific Conference Series, Vol.~56, Interacting Binary Stars, ed. A.~W.
  {Shafter}, 228

\bibitem[{{Butters} {et~al.}(2010){Butters}, {West}, {Anderson}, {Collier
  Cameron}, {Clarkson}, {Enoch}, {Haswell}, {Hellier}, {Horne}, {Joshi},
  {Kane}, {Lister}, {Maxted}, {Parley}, {Pollacco}, {Smalley}, {Street},
  {Todd}, {Wheatley}, \& {Wilson}}]{2010A&A...520L..10B}
{Butters}, O.~W., {West}, R.~G., {Anderson}, D.~R., {et~al.} 2010, \aap, 520,
  L10, \dodoi{10.1051/0004-6361/201015655}

\bibitem[{{Caton} {et~al.}(2019){Caton}, {Gentry}, {Samec}, {Chamberlain},
  {Robb}, {Faulkner}, \& {Hill}}]{2019PASP..131e4203C}
{Caton}, D., {Gentry}, D.~R., {Samec}, R.~G., {et~al.} 2019, \pasp, 131,
  054203, \dodoi{10.1088/1538-3873/aafb8f}

\bibitem[{{Christopoulou} \& {Papageorgiou}(2013)}]{2013AJ....146..157C}
{Christopoulou}, P.~E., \& {Papageorgiou}, A. 2013, \aj, 146, 157,
  \dodoi{10.1088/0004-6256/146/6/157}

\bibitem[{{Cui} {et~al.}(2012){Cui}, {Zhao}, {Chu}, {Li}, {Li}, {Zhang}, {Su},
  {Yao}, {Wang}, {Xing}, {Li}, {Zhu}, {Wang}, {Gu}, {Luo}, {Xu}, {Zhang},
  {Liu}, {Zhang}, {Yang}, {Cao}, {Chen}, {Chen}, {Chen}, {Chen}, {Chu}, {Feng},
  {Gong}, {Hou}, {Hu}, {Hu}, {Hu}, {Jia}, {Jiang}, {Jiang}, {Jiang}, {Jin},
  {Li}, {Li}, {Li}, {Liu}, {Liu}, {Lu}, {Mao}, {Men}, {Qi}, {Qi}, {Shi},
  {Tang}, {Tao}, {Wang}, {Wang}, {Wang}, {Wang}, {Wang}, {Wang}, {Wang},
  {Wang}, {Wang}, {Wang}, {Wang}, {Wang}, {Xu}, {Xu}, {Yang}, {Yu}, {Yuan},
  {Yuan}, {Zhai}, {Zhang}, {Zhang}, {Zhang}, {Zhao}, {Zhou}, {Zhou}, {Zhu}, \&
  {Zou}}]{2012RAA....12.1197C}
{Cui}, X.-Q., {Zhao}, Y.-H., {Chu}, Y.-Q., {et~al.} 2012, Research in Astronomy
  and Astrophysics, 12, 1197, \dodoi{10.1088/1674-4527/12/9/003}

\bibitem[{{Cutri} {et~al.}(2003){Cutri}, {Skrutskie}, {van Dyk}, {Beichman},
  {Carpenter}, {Chester}, {Cambresy}, {Evans}, {Fowler}, {Gizis}, {Howard},
  {Huchra}, {Jarrett}, {Kopan}, {Kirkpatrick}, {Light}, {Marsh}, {McCallon},
  {Schneider}, {Stiening}, {Sykes}, {Weinberg}, {Wheaton}, {Wheelock}, \&
  {Zacarias}}]{2003yCat.2246....0C}
{Cutri}, R.~M., {Skrutskie}, M.~F., {van Dyk}, S., {et~al.} 2003, VizieR Online
  Data Catalog, II/246

\bibitem[{{Drake} {et~al.}(2009){Drake}, {Djorgovski}, {Mahabal}, {Beshore},
  {Larson}, {Graham}, {Williams}, {Christensen}, {Catelan}, {Boattini},
  {Gibbs}, {Hill}, \& {Kowalski}}]{2009ApJ...696..870D}
{Drake}, A.~J., {Djorgovski}, S.~G., {Mahabal}, A., {et~al.} 2009, \apj, 696,
  870, \dodoi{10.1088/0004-637X/696/1/870}

\bibitem[{{Eastman} {et~al.}(2010){Eastman}, {Siverd}, \&
  {Gaudi}}]{2010PASP..122..935E}
{Eastman}, J., {Siverd}, R., \& {Gaudi}, B.~S. 2010, \pasp, 122, 935,
  \dodoi{10.1086/655938}

\bibitem[{{Eggleton} \& {Kiseleva-Eggleton}(2002)}]{2002ApJ...575..461E}
{Eggleton}, P.~P., \& {Kiseleva-Eggleton}, L. 2002, \apj, 575, 461,
  \dodoi{10.1086/341215}

\bibitem[{{Eker} {et~al.}(2006){Eker}, {Demircan}, {Bilir}, \&
  {Karata{\c{s}}}}]{2006MNRAS.373.1483E}
{Eker}, Z., {Demircan}, O., {Bilir}, S., \& {Karata{\c{s}}}, Y. 2006, \mnras,
  373, 1483, \dodoi{10.1111/j.1365-2966.2006.11073.x}

\bibitem[{{Guinan} \& {Bradstreet}(1988)}]{1988ASIC..241..345G}
{Guinan}, E.~F., \& {Bradstreet}, D.~H. 1988, in NATO Advanced Study Institute
  (ASI) Series C, Vol. 241, Formation and Evolution of Low Mass Stars, ed.
  A.~K. {Dupree} \& M.~T.~V.~T. {Lago}, 345

\bibitem[{{Hu} {et~al.}(2014){Hu}, {Han}, {Guo}, \& {Du}}]{2014RAA....14..719H}
{Hu}, S.-M., {Han}, S.-H., {Guo}, D.-F., \& {Du}, J.-J. 2014, Research in
  Astronomy and Astrophysics, 14, 719, \dodoi{10.1088/1674-4527/14/6/010}

\bibitem[{{Huang} {et~al.}(2012){Huang}, {Li}, {Wang}, {Shang}, {Zhang}, {Hu},
  {Qiu}, \& {Jiang}}]{2012RAA....12.1585H}
{Huang}, F., {Li}, J.-Z., {Wang}, X.-F., {et~al.} 2012, Research in Astronomy
  and Astrophysics, 12, 1585, \dodoi{10.1088/1674-4527/12/11/012}

\bibitem[{{Hurley} {et~al.}(2002){Hurley}, {Tout}, \&
  {Pols}}]{2002MNRAS.329..897H}
{Hurley}, J.~R., {Tout}, C.~A., \& {Pols}, O.~R. 2002, \mnras, 329, 897,
  \dodoi{10.1046/j.1365-8711.2002.05038.x}

\bibitem[{{Hut}(1980)}]{1980A&A....92..167H}
{Hut}, P. 1980, \aap, 92, 167

\bibitem[{{Jayasinghe} {et~al.}(2018){Jayasinghe}, {Kochanek}, {Stanek},
  {Shappee}, {Holoien}, {Thompson}, {Prieto}, {Dong}, {Pawlak}, {Shields},
  {Pojmanski}, {Otero}, {Britt}, \& {Will}}]{2018MNRAS.477.3145J}
{Jayasinghe}, T., {Kochanek}, C.~S., {Stanek}, K.~Z., {et~al.} 2018, \mnras,
  477, 3145, \dodoi{10.1093/mnras/sty838}

\bibitem[{{Jiang} {et~al.}(2010){Jiang}, {Han}, {Wang}, {Jiang}, \&
  {Li}}]{2010MNRAS.405.2485J}
{Jiang}, D., {Han}, Z., {Wang}, J., {Jiang}, T., \& {Li}, L. 2010, \mnras, 405,
  2485, \dodoi{10.1111/j.1365-2966.2010.16615.x}

\bibitem[{{Kochanek} {et~al.}(2014){Kochanek}, {Adams}, \&
  {Belczynski}}]{2014MNRAS.443.1319K}
{Kochanek}, C.~S., {Adams}, S.~M., \& {Belczynski}, K. 2014, \mnras, 443, 1319,
  \dodoi{10.1093/mnras/stu1226}

\bibitem[{{Kochanek} {et~al.}(2017){Kochanek}, {Shappee}, {Stanek}, {Holoien},
  {Thompson}, {Prieto}, {Dong}, {Shields}, {Will}, {Britt}, {Perzanowski}, \&
  {Pojma{\'n}ski}}]{2017PASP..129j4502K}
{Kochanek}, C.~S., {Shappee}, B.~J., {Stanek}, K.~Z., {et~al.} 2017, \pasp,
  129, 104502, \dodoi{10.1088/1538-3873/aa80d9}

\bibitem[{{Kwee} \& {van Woerden}(1956)}]{1956BAN....12..327K}
{Kwee}, K.~K., \& {van Woerden}, H. 1956, \bain, 12, 327

\bibitem[{{Landin} {et~al.}(2009){Landin}, {Mendes}, \&
  {Vaz}}]{2009A&A...494..209L}
{Landin}, N.~R., {Mendes}, L.~T.~S., \& {Vaz}, L.~P.~R. 2009, \aap, 494, 209,
  \dodoi{10.1051/0004-6361:20078403}

\bibitem[{{Li} {et~al.}(2020){Li}, {Kim}, {Xia}, {Michel}, {Hu}, {Gao}, {Guo},
  \& {Chen}}]{2020AJ....159..189L}
{Li}, K., {Kim}, C.-H., {Xia}, Q.-Q., {et~al.} 2020, \aj, 159, 189,
  \dodoi{10.3847/1538-3881/ab7cda}

\bibitem[{{Li} {et~al.}(2021{\natexlab{a}}){Li}, {Xia}, {Kim}, {Hu}, {Guo},
  {Jeong}, {Chen}, \& {Gao}}]{2021ApJ...922..122L}
{Li}, K., {Xia}, Q.-Q., {Kim}, C.-H., {et~al.} 2021{\natexlab{a}}, \apj, 922,
  122, \dodoi{10.3847/1538-4357/ac242f}

\bibitem[{{Li} {et~al.}(2019){Li}, {Xia}, {Michel}, {Hu}, {Guo}, {Gao}, {Chen},
  \& {Gao}}]{2019MNRAS.485.4588L}
{Li}, K., {Xia}, Q.-Q., {Michel}, R., {et~al.} 2019, \mnras, 485, 4588,
  \dodoi{10.1093/mnras/stz715}

\bibitem[{{Li} {et~al.}(2021{\natexlab{b}}){Li}, {Xia}, {Kim}, {Gao}, {Hu},
  {Guo}, {Gao}, {Chen}, \& {Guo}}]{2021AJ....162...13L}
{Li}, K., {Xia}, Q.-Q., {Kim}, C.-H., {et~al.} 2021{\natexlab{b}}, \aj, 162,
  13, \dodoi{10.3847/1538-3881/abfc53}

\bibitem[{{Li} \& {Zhang}(2006)}]{2006MNRAS.369.2001L}
{Li}, L., \& {Zhang}, F. 2006, \mnras, 369, 2001,
  \dodoi{10.1111/j.1365-2966.2006.10462.x}

\bibitem[{{Liao} {et~al.}(2019){Liao}, {Qian}, \&
  {Sarotsakulchai}}]{2019AJ....157..207L}
{Liao}, W.~P., {Qian}, S.~B., \& {Sarotsakulchai}, T. 2019, \aj, 157, 207,
  \dodoi{10.3847/1538-3881/ab17d4}

\bibitem[{{Liao} {et~al.}(2022){Liao}, {Qian}, {Shi}, {Li}, {Liu}, {He},
  {Zang}, \& {Li}}]{2022ApJ...927..183L}
{Liao}, W.~P., {Qian}, S.~B., {Shi}, X.~D., {et~al.} 2022, \apj, 927, 183,
  \dodoi{10.3847/1538-4357/ac5038}

\bibitem[{{Luo} {et~al.}(2015){Luo}, {Zhao}, {Zhao}, {Deng}, {Liu}, {Jing},
  {Wang}, {Zhang}, {Shi}, {Cui}, {Chu}, {Li}, {Bai}, {Wu}, {Cai}, {Cao}, {Cao},
  {Carlin}, {Chen}, {Chen}, {Chen}, {Chen}, {Chen}, {Chen}, {Chen},
  {Christlieb}, {Chu}, {Cui}, {Dong}, {Du}, {Fan}, {Feng}, {Fu}, {Gao}, {Gong},
  {Gu}, {Guo}, {Han}, {He}, {Hou}, {Hou}, {Hou}, {Hu}, {Hu}, {Hu}, {Huo},
  {Jia}, {Jiang}, {Jiang}, {Jiang}, {Jin}, {Kong}, {Kong}, {Lei}, {Li}, {Li},
  {Li}, {Li}, {Li}, {Li}, {Li}, {Li}, {Li}, {Li}, {Li}, {Li}, {Liang}, {Lin},
  {Liu}, {Liu}, {Liu}, {Liu}, {Lu}, {Luo}, {Mao}, {Newberg}, {Ni}, {Qi}, {Qi},
  {Shen}, {Shi}, {Song}, {Song}, {Su}, {Su}, {Tang}, {Tao}, {Tian}, {Wang},
  {Wang}, {Wang}, {Wang}, {Wang}, {Wang}, {Wang}, {Wang}, {Wang}, {Wang},
  {Wang}, {Wang}, {Wang}, {Wang}, {Wang}, {Wang}, {Wang}, {Wang}, {Wang},
  {Wang}, {Wei}, {Wei}, {Wu}, {Wu}, {Wu}, {Wu}, {Xing}, {Xu}, {Xu}, {Xu},
  {Yan}, {Yang}, {Yang}, {Yang}, {Yang}, {Yao}, {Yu}, {Yuan}, {Yuan}, {Yuan},
  {Yuan}, {Zhai}, {Zhang}, {Zhang}, {Zhang}, {Zhang}, {Zhang}, {Zhang},
  {Zhang}, {Zhang}, {Zhao}, {Zhou}, {Zhou}, {Zhu}, {Zhu}, {Zou}, \&
  {Zuo}}]{2015RAA....15.1095L}
{Luo}, A.~L., {Zhao}, Y.-H., {Zhao}, G., {et~al.} 2015, Research in Astronomy
  and Astrophysics, 15, 1095, \dodoi{10.1088/1674-4527/15/8/002}

\bibitem[{{Masci} {et~al.}(2019){Masci}, {Laher}, {Rusholme}, {Shupe}, {Groom},
  {Surace}, {Jackson}, {Monkewitz}, {Beck}, {Flynn}, {Terek}, {Landry},
  {Hacopians}, {Desai}, {Howell}, {Brooke}, {Imel}, {Wachter}, {Ye}, {Lin},
  {Cenko}, {Cunningham}, {Rebbapragada}, {Bue}, {Miller}, {Mahabal}, {Bellm},
  {Patterson}, {Juri{\'c}}, {Golkhou}, {Ofek}, {Walters}, {Graham}, {Kasliwal},
  {Dekany}, {Kupfer}, {Burdge}, {Cannella}, {Barlow}, {Van Sistine}, {Giomi},
  {Fremling}, {Blagorodnova}, {Levitan}, {Riddle}, {Smith}, {Helou}, {Prince},
  \& {Kulkarni}}]{2019PASP..131a8003M}
{Masci}, F.~J., {Laher}, R.~R., {Rusholme}, B., {et~al.} 2019, \pasp, 131,
  018003, \dodoi{10.1088/1538-3873/aae8ac}

\bibitem[{{O'Connell}(1951)}]{1951PRCO....2...85O}
{O'Connell}, D.~J.~K. 1951, Publications of the Riverview College Observatory,
  2, 85

\bibitem[{{Panchal} \& {Joshi}(2021)}]{2021AJ....161..221P}
{Panchal}, A., \& {Joshi}, Y.~C. 2021, \aj, 161, 221,
  \dodoi{10.3847/1538-3881/abea0c}

\bibitem[{{Panchal} {et~al.}(2022){Panchal}, {Joshi}, {De Cat}, \&
  {Tiwari}}]{2022ApJ...927...12P}
{Panchal}, A., {Joshi}, Y.~C., {De Cat}, P., \& {Tiwari}, S.~N. 2022, \apj,
  927, 12, \dodoi{10.3847/1538-4357/ac45fb}

\bibitem[{{Pavlenko} {et~al.}(2018){Pavlenko}, {Evans}, {Banerjee},
  {Southworth}, {Shahbandeh}, \& {Davis}}]{2018A&A...615A.120P}
{Pavlenko}, Y.~V., {Evans}, A., {Banerjee}, D.~P.~K., {et~al.} 2018, \aap, 615,
  A120, \dodoi{10.1051/0004-6361/201832717}

\bibitem[{{Pojmanski}(1997)}]{1997AcA....47..467P}
{Pojmanski}, G. 1997, \actaa, 47, 467.
\newblock \doarXiv{astro-ph/9712146}

\bibitem[{{Pribulla} {et~al.}(2003){Pribulla}, {Kreiner}, \&
  {Tremko}}]{2003CoSka..33...38P}
{Pribulla}, T., {Kreiner}, J.~M., \& {Tremko}, J. 2003, Contributions of the
  Astronomical Observatory Skalnate Pleso, 33, 38

\bibitem[{{Qian}(2003)}]{2003MNRAS.342.1260Q}
{Qian}, S. 2003, \mnras, 342, 1260, \dodoi{10.1046/j.1365-8711.2003.06627.x}

\bibitem[{{Qian} {et~al.}(2006){Qian}, {Yang}, {Zhu}, {He}, \&
  {Yuan}}]{2006Ap&SS.304...25Q}
{Qian}, S., {Yang}, Y., {Zhu}, L., {He}, J., \& {Yuan}, J. 2006, \apss, 304,
  25, \dodoi{10.1007/s10509-006-9114-z}

\bibitem[{{Qian} {et~al.}(2017){Qian}, {He}, {Zhang}, {Zhu}, {Shi}, {Zhao}, \&
  {Zhou}}]{2017RAA....17...87Q}
{Qian}, S.-B., {He}, J.-J., {Zhang}, J., {et~al.} 2017, Research in Astronomy
  and Astrophysics, 17, 087, \dodoi{10.1088/1674-4527/17/8/87}

\bibitem[{{Qian} {et~al.}(2005{\natexlab{a}}){Qian}, {Yang}, {Soonthornthum},
  {Zhu}, {He}, \& {Yuan}}]{2005AJ....130..224Q}
{Qian}, S.~B., {Yang}, Y.~G., {Soonthornthum}, B., {et~al.} 2005{\natexlab{a}},
  \aj, 130, 224, \dodoi{10.1086/430673}

\bibitem[{{Qian} {et~al.}(2005{\natexlab{b}}){Qian}, {Zhu}, {Soonthornthum},
  {Yuan}, {Yang}, \& {He}}]{2005AJ....130.1206Q}
{Qian}, S.~B., {Zhu}, L.~Y., {Soonthornthum}, B., {et~al.} 2005{\natexlab{b}},
  \aj, 130, 1206, \dodoi{10.1086/432544}

\bibitem[{{Rasio}(1995)}]{1995ApJ...444L..41R}
{Rasio}, F.~A. 1995, \apjl, 444, L41, \dodoi{10.1086/187855}

\bibitem[{{Ricker} {et~al.}(2015){Ricker}, {Winn}, {Vanderspek}, {Latham},
  {Bakos}, {Bean}, {Berta-Thompson}, {Brown}, {Buchhave}, {Butler}, {Butler},
  {Chaplin}, {Charbonneau}, {Christensen-Dalsgaard}, {Clampin}, {Deming},
  {Doty}, {De Lee}, {Dressing}, {Dunham}, {Endl}, {Fressin}, {Ge}, {Henning},
  {Holman}, {Howard}, {Ida}, {Jenkins}, {Jernigan}, {Johnson}, {Kaltenegger},
  {Kawai}, {Kjeldsen}, {Laughlin}, {Levine}, {Lin}, {Lissauer}, {MacQueen},
  {Marcy}, {McCullough}, {Morton}, {Narita}, {Paegert}, {Palle}, {Pepe},
  {Pepper}, {Quirrenbach}, {Rinehart}, {Sasselov}, {Sato}, {Seager},
  {Sozzetti}, {Stassun}, {Sullivan}, {Szentgyorgyi}, {Torres}, {Udry}, \&
  {Villasenor}}]{2015JATIS...1a4003R}
{Ricker}, G.~R., {Winn}, J.~N., {Vanderspek}, R., {et~al.} 2015, Journal of
  Astronomical Telescopes, Instruments, and Systems, 1, 014003,
  \dodoi{10.1117/1.JATIS.1.1.014003}

\bibitem[{{Rucinski}(2001)}]{2001AJ....122.1007R}
{Rucinski}, S.~M. 2001, \aj, 122, 1007, \dodoi{10.1086/321153}

\bibitem[{{Rucinski}(2007)}]{2007MNRAS.382..393R}
---. 2007, \mnras, 382, 393, \dodoi{10.1111/j.1365-2966.2007.12377.x}

\bibitem[{{Shappee} {et~al.}(2014){Shappee}, {Prieto}, {Grupe}, {Kochanek},
  {Stanek}, {De Rosa}, {Mathur}, {Zu}, {Peterson}, {Pogge}, {Komossa}, {Im},
  {Jencson}, {Holoien}, {Basu}, {Beacom}, {Szczygie{\l}}, {Brimacombe},
  {Adams}, {Campillay}, {Choi}, {Contreras}, {Dietrich}, {Dubberley},
  {Elphick}, {Foale}, {Giustini}, {Gonzalez}, {Hawkins}, {Howell}, {Hsiao},
  {Koss}, {Leighly}, {Morrell}, {Mudd}, {Mullins}, {Nugent}, {Parrent},
  {Phillips}, {Pojmanski}, {Rosing}, {Ross}, {Sand}, {Terndrup}, {Valenti},
  {Walker}, \& {Yoon}}]{2014ApJ...788...48S}
{Shappee}, B.~J., {Prieto}, J.~L., {Grupe}, D., {et~al.} 2014, \apj, 788, 48,
  \dodoi{10.1088/0004-637X/788/1/48}

\bibitem[{{Sriram} {et~al.}(2016){Sriram}, {Malu}, {Choi}, \& {Vivekananda
  Rao}}]{2016AJ....151...69S}
{Sriram}, K., {Malu}, S., {Choi}, C.~S., \& {Vivekananda Rao}, P. 2016, \aj,
  151, 69, \dodoi{10.3847/0004-6256/151/3/69}

\bibitem[{{St{\c{e}}pie{\'n}}(2011)}]{2011AcA....61..139S}
{St{\c{e}}pie{\'n}}, K. 2011, \actaa, 61, 139.
\newblock \doarXiv{1105.2645}

\bibitem[{{Stepien}(2006)}]{2006AcA....56..347S}
{Stepien}, K. 2006, \actaa, 56, 347.
\newblock \doarXiv{astro-ph/0701529}

\bibitem[{{Terrell}(2022)}]{2022Galax..10....8T}
{Terrell}, D. 2022, Galaxies, 10, 8, \dodoi{10.3390/galaxies10010008}

\bibitem[{{Terrell} \& {Wilson}(2005)}]{2005Ap&SS.296..221T}
{Terrell}, D., \& {Wilson}, R.~E. 2005, \apss, 296, 221,
  \dodoi{10.1007/s10509-005-4449-4}

\bibitem[{{Tylenda} {et~al.}(2011){Tylenda}, {Hajduk}, {Kami{\'n}ski},
  {Udalski}, {Soszy{\'n}ski}, {Szyma{\'n}ski}, {Kubiak}, {Pietrzy{\'n}ski},
  {Poleski}, {Wyrzykowski}, \& {Ulaczyk}}]{2011A&A...528A.114T}
{Tylenda}, R., {Hajduk}, M., {Kami{\'n}ski}, T., {et~al.} 2011, \aap, 528,
  A114, \dodoi{10.1051/0004-6361/201016221}

\bibitem[{{van Hamme}(1993)}]{1993AJ....106.2096V}
{van Hamme}, W. 1993, \aj, 106, 2096, \dodoi{10.1086/116788}

\bibitem[{{von Zeipel}(1924)}]{1924MNRAS..84..665V}
{von Zeipel}, H. 1924, \mnras, 84, 665, \dodoi{10.1093/mnras/84.9.665}

\bibitem[{{Wadhwa} {et~al.}(2021){Wadhwa}, {De Horta}, {Filipovi{\'c}},
  {Tothill}, {Arbutina}, {Petrovi{\'c}}, \&
  {Djura{\v{s}}evi{\'c}}}]{2021MNRAS.501..229W}
{Wadhwa}, S.~S., {De Horta}, A., {Filipovi{\'c}}, M.~D., {et~al.} 2021, \mnras,
  501, 229, \dodoi{10.1093/mnras/staa3637}

\bibitem[{{Wang} {et~al.}(2020){Wang}, {Fu}, {Zong}, {Smith}, {De Cat}, {Shi},
  {Luo}, {Zhang}, {Frasca}, {Corbally}, {Molenda-{\.Z}akowicz}, {Catanzaro},
  {Gray}, {Wang}, \& {Pan}}]{2020ApJS..251...27W}
{Wang}, J., {Fu}, J.-N., {Zong}, W., {et~al.} 2020, \apjs, 251, 27,
  \dodoi{10.3847/1538-4365/abc1ed}

\bibitem[{{Wilson}(1979)}]{1979ApJ...234.1054W}
{Wilson}, R.~E. 1979, \apj, 234, 1054, \dodoi{10.1086/157588}

\bibitem[{{Wilson}(1990)}]{1990ApJ...356..613W}
---. 1990, \apj, 356, 613, \dodoi{10.1086/168867}

\bibitem[{{Wilson} \& {Devinney}(1971)}]{1971ApJ...166..605W}
{Wilson}, R.~E., \& {Devinney}, E.~J. 1971, \apj, 166, 605,
  \dodoi{10.1086/150986}

\bibitem[{{Wu} {et~al.}(2014){Wu}, {Du}, {Luo}, {Zhao}, \&
  {Yuan}}]{2014IAUS..306..340W}
{Wu}, Y., {Du}, B., {Luo}, A., {Zhao}, Y., \& {Yuan}, H. 2014, in Statistical
  Challenges in 21st Century Cosmology, ed. A.~{Heavens}, J.-L. {Starck}, \&
  A.~{Krone-Martins}, Vol. 306, 340--342, \dodoi{10.1017/S1743921314010825}

\bibitem[{{Wu} {et~al.}(2011){Wu}, {Luo}, {Li}, {Shi}, {Prugniel}, {Liang},
  {Zhao}, {Zhang}, {Bai}, {Wei}, {Dong}, {Zhang}, \&
  {Chen}}]{2011RAA....11..924W}
{Wu}, Y., {Luo}, A.~L., {Li}, H.-N., {et~al.} 2011, Research in Astronomy and
  Astrophysics, 11, 924, \dodoi{10.1088/1674-4527/11/8/006}

\bibitem[{{Yakut} \& {Eggleton}(2005)}]{2005ApJ...629.1055Y}
{Yakut}, K., \& {Eggleton}, P.~P. 2005, \apj, 629, 1055, \dodoi{10.1086/431300}

\bibitem[{{Yang} \& {Qian}(2015)}]{2015AJ....150...69Y}
{Yang}, Y.-G., \& {Qian}, S.-B. 2015, \aj, 150, 69,
  \dodoi{10.1088/0004-6256/150/3/69}

\bibitem[{{Y{\i}ld{\i}z}(2014)}]{2014MNRAS.437..185Y}
{Y{\i}ld{\i}z}, M. 2014, \mnras, 437, 185, \dodoi{10.1093/mnras/stt1874}

\bibitem[{{Yildiz} \& {Do{\u{g}}an}(2013)}]{2013MNRAS.430.2029Y}
{Yildiz}, M., \& {Do{\u{g}}an}, T. 2013, \mnras, 430, 2029,
  \dodoi{10.1093/mnras/stt028}

\bibitem[{{Zhang} {et~al.}(2021){Zhang}, {Li}, {Yang}, {Xiong}, {Fu}, {Liu},
  {Tian}, {Li}, {Wang}, {Liang}, {Zhou}, {Zong}, {Yang}, {Liu}, \&
  {Hou}}]{2021ApJS..256...14Z}
{Zhang}, B., {Li}, J., {Yang}, F., {et~al.} 2021, \apjs, 256, 14,
  \dodoi{10.3847/1538-4365/ac0834}

\bibitem[{{Zhang} {et~al.}(2017){Zhang}, {Qian}, {Han}, \&
  {Wu}}]{2017MNRAS.466.1118Z}
{Zhang}, J., {Qian}, S.-B., {Han}, Z.-T., \& {Wu}, Y. 2017, \mnras, 466, 1118,
  \dodoi{10.1093/mnras/stw3153}

\bibitem[{{Zhang} {et~al.}(2019){Zhang}, {Qian}, {Wu}, \&
  {Zhou}}]{2019ApJS..244...43Z}
{Zhang}, J., {Qian}, S.-B., {Wu}, Y., \& {Zhou}, X. 2019, \apjs, 244, 43,
  \dodoi{10.3847/1538-4365/ab442b}

\bibitem[{{Zhang} {et~al.}(2020){Zhang}, {Long}, {Shi}, {Lu}, {Gao}, {Han},
  {Wang}, {Prabhakar}, \& {Lamost Mrs Collaboration}}]{2020MNRAS.495.1252Z}
{Zhang}, L.-Y., {Long}, L., {Shi}, J., {et~al.} 2020, \mnras, 495, 1252,
  \dodoi{10.1093/mnras/staa942}

\bibitem[{{Zheng} {et~al.}(2021){Zheng}, {Li}, \& {Xia}}]{2021MNRAS.506.4251Z}
{Zheng}, S.-Y., {Li}, K., \& {Xia}, Q.-Q. 2021, \mnras, 506, 4251,
  \dodoi{10.1093/mnras/stab1829}

\bibitem[{{Zhou} {et~al.}(2016){Zhou}, {Qian}, {Zhang}, {Zhang}, \&
  {Kreiner}}]{2016AJ....151...67Z}
{Zhou}, X., {Qian}, S.~B., {Zhang}, J., {Zhang}, B., \& {Kreiner}, J. 2016,
  \aj, 151, 67, \dodoi{10.3847/0004-6256/151/3/67}

\bibitem[{{Zwitter} {et~al.}(2003){Zwitter}, {Munari}, {Marrese}, {Pr{\v{s}}a},
  {Milone}, {Boschi}, {Tomov}, \& {Siviero}}]{2003A&A...404..333Z}
{Zwitter}, T., {Munari}, U., {Marrese}, P.~M., {et~al.} 2003, \aap, 404, 333,
  \dodoi{10.1051/0004-6361:20030446}

\end{thebibliography}
\bibliographystyle{aasjournal}



\end{document}